\documentclass[a4paper,clock]{article}
\usepackage[dvips]{epsfig}
\usepackage{amssymb}
\usepackage{amsmath}
\usepackage{bm}
\usepackage{dcolumn}

\catcode`\@=11
\def\marginnote#1{}
\newcount\hour
\newcount\minute
\newtoks\amorpm
\hour=\time\divide\hour by60
\minute=\time{\multiply\hour by60 \global\advance\minute
by-\hour}\edef\standardtime{{\ifnum\hour<12
\global\amorpm={am}%
        \else\global\amorpm={pm}\advance\hour by-12 \fi
        \ifnum\hour=0 \hour=12 \fi
        \number\hour:\ifnum\minute<10
0\fi\number\minute\the\amorpm}}
\edef\militarytime{\number\hour:\ifnum\minute<10
0\fi\number\minute}

\def\draftlabel#1{{\@bsphack\if@filesw {\let\thepage\relax
   \xdef\@gtempa{\write\@auxout{\string
      \newlabel{#1}{{\@currentlabel}{\thepage}}}}}\@gtempa
   \if@nobreak \ifvmode\nobreak\fi\fi\fi\@esphack}
        \gdef\@eqnlabel{#1}}
\def\@eqnlabel{}
\def\@vacuum{}
\def\draftmarginnote#1{\marginpar{\raggedright\scriptsize\tt#1}}
\def\draft{\oddsidemargin -.5truein
        \def\@oddfoot{\sl preliminary draft \hfil
        \rm\thepage\hfil\sl\today\quad\militarytime}
        \let\@evenfoot\@oddfoot \overfullrule 3pt
        \let\label=\draftlabel
        \let\marginnote=\draftmarginnote

\def\@eqnnum{(\theequation)\rlap{\kern\marginparsep\tt\@eqnlabel}%
\global\let\@eqnlabel\@vacuum}  }

\def\numberbysection{\@addtoreset{equation}{section}
        \def\theequation{\thesection.\arabic{equation}}}

\def\underline#1{\relax\ifmmode\@@underline#1\else
 $\@@underline{\hbox{#1}}$\relax\fi}

\catcode`@=12
\relax

\numberbysection

\advance \topmargin by -\headheight
\advance \topmargin by -\headsep

\evensidemargin \oddsidemargin
\def\br{\begin{eqnarray}}
\def\er{\end{eqnarray}}
\def\be{\begin{equation}}
\def\ee{\end{equation}}

\def\({\left(}
\def\){\right)}

\relax

\def\a{\alpha}

\def\d{\delta}
\def\D{\Delta}

\def\g{\gamma}
\def\G{\Gamma}

\def\l{\lambda}

\def\pa{\partial}

\def\tp0{\Theta_{+}^{(0)}}
\def\tm0{\Theta_{-}^{(0)}}

\def\f#1#2#3 {f^{#1#2}_{#3}}

\def\win1{{\sf w_{1+\infty}}}

\def\Win1{{\sf W_{1+\infty}}}

\def\rlx{\relax\leavevmode}
\def\inbar{\vrule height1.5ex width.4pt depth0pt}
\def\IZ{\rlx\hbox{\sf Z\kern-.4em Z}}
\def\IR{\rlx\hbox{\rm I\kern-.18em R}}
\def\IT{\rlx\hbox{\rm I\kern-.18em T}}
\def\IC{\rlx\hbox{\,$\inbar\kern-.3em{\rm C}$}}
\def\IN{\rlx\hbox{\rm I\kern-.18em N}}
\def\IO{\rlx\hbox{\,$\inbar\kern-.3em{\rm O}$}}
\def\IP{\rlx\hbox{\rm I\kern-.18em P}}
\def\IQ{\rlx\hbox{\,$\inbar\kern-.3em{\rm Q}$}}
\def\IF{\rlx\hbox{\rm I\kern-.18em F}}
\def\IG{\rlx\hbox{\,$\inbar\kern-.3em{\rm G}$}}
\def\IH{\rlx\hbox{\rm I\kern-.18em H}}
\def\II{\rlx\hbox{\rm I\kern-.18em I}}
\def\IK{\rlx\hbox{\rm I\kern-.18em K}}
\def\IL{\rlx\hbox{\rm I\kern-.18em L}}
\def\one{\hbox{{1}\kern-.25em\hbox{l}}}
\def\0#1{\relax\ifmmode\mathaccent"7017{#1}%
B        \else\accent23#1\relax\fi}

\def\PRL#1#2#3{{\sl Phys. Rev. Lett.} {\bf#1} (#2) #3}
\def\NPB#1#2#3{{\sl Nucl. Phys.} {\bf B#1} (#2) #3}

\def\CMP#1#2#3{{\sl Commun. Math. Phys.} {\bf #1} (#2) #3}

\def\PRE#1#2#3{{\sl Phys. Rev.} {\bf E#1} (#2) #3}

\def\PLA#1#2#3{{\sl Phys. Lett.} {\bf #1A} (#2) #3}

\def\PLB#1#2#3{{\sl Phys. Lett.} {\bf #1B} (#2) #3}

\def\RMP#1#2#3{{\sl Rev. Mod. Phys.} {\bf #1} (#2) #3}

\def\IJMPA#1#2#3{{\sl Int. J. Mod. Phys.} {\bf A#1} (#2) #3}

\def\JPA#1#2#3{{\sl J. Physics} {\bf A#1} (#2) #3}

\def\PHSD#1#2#3{{\sl Physica D: Nonlinear Phenomena} {\bf D#1} (#2) #3}

\def\JHEP#1#2#3{{\sl JHEP} {\bf #1} (#2) #3}

\def\JCP#1#2#3{{\sl Journal of Computational Physics} {\bf #1} (#2) #3}

\def\Nonl#1#2#3{{\sl Nonlinearity} {\bf #1} (#2) #3}

\def\EPL#1#2#3{{\sl Europhysics Letters} {\bf #1} (#2) #3}
\def\RQE#1#2#3{{\sl Radiophysics and Quantum Electronics} {\bf #1} (#2) #3}
\def\IJNMF#1#2#3{{\sl Int.   J.   Numer.   Methods   Fluids} {\bf #1} (#2) #3} 
\def\CNSNS#1#2#3{{\sl Commun Nonlinear Sci Numer Simulat} {\bf #1} (#2) #3}
\def\ScR#1#2#3{{\sl Sci. Rep.} {\bf #1} (#2) #3}
\def\JNS#1#2#3{{\sl J. Nonlinear Sci} {\bf #1} (#2) #3}
\begin{document}

\begin{titlepage}

\vspace{.2in}
\begin{center}
{\large\bf Quasi-integrable KdV models, towers of infinite number of anomalous charges and soliton collisions}
\end{center}

\vspace{.2in}

\begin{center}

H. Blas$^{(a)}$, R. Ochoa$^{(b)}$ and D. Suarez$^{(b)}$

\par \vskip .2in \noindent

$^{(a)}$Instituto de F\'{\i}sica\\
Universidade Federal de Mato Grosso\\
Av. Fernando Correa, $N^{0}$ \, 2367\\
Bairro Boa Esperan\c ca, Cep 78060-900, Cuiab\'a - MT - Brazil \\
$^{(b)}$ Facultad de Ciencias\\
Universidad Nacional de Ingenier\'{\i}a\\
Av. Tupac Amaru, $N^{0}$ \, 210, Rimac, Lima-Per\'u 
\normalsize
\end{center}

\vspace{.3in}

\begin{abstract}
\vspace{.3in}
We found, through analytical and numerical methods, new towers of infinite number of asymptotically conserved charges for deformations of the Korteweg-de Vries equation (KdV). It is shown analytically that the standard KdV also exhibits some towers of infinite number of anomalous charges, and that their relevant anomalies vanish for $N-$soliton solution. Some deformations of the KdV model are performed through the Riccati-type pseudo-potential approach, and infinite number of exact non-local conservation laws is provided using a linear formulation of the deformed model. In order to check the degrees of modifications of the charges around the soliton interaction regions, we compute numerically some representative anomalies, associated to the lowest order quasi-conservation laws, depending on the deformation parameters $\{\epsilon_1, \epsilon_2\}$, which include the standard KdV ($\epsilon_1=\epsilon_2=0$), the regularized long-wave (RLW)  ($\epsilon_1=1,\epsilon_2=0$), the  modified regularized long-wave (mRLW)  ($\epsilon_1=\epsilon_2=1$) and the KdV-RLW (KdV-BBM) type ($\epsilon_2=0,\,\epsilon \neq \{0,1\}$) equations, respectively. Our numerical simulations show the elastic scattering of two and three solitons for a wide range of values of the set $\{\epsilon_1, \epsilon_2\}$, for a variety of amplitudes and relative velocities. The KdV-type equations are quite ubiquitous in several areas of non-linear science, and they find relevant applications in the study of General Relativity on $AdS_{3}$, Bose-Einstein condensates, superconductivity and soliton gas and turbulence in fluid dynamics. 
\end{abstract} 

\end{titlepage}

\section{Introduction}

The soliton solutions  and the existence of infinite number of conserved charges are among the main properties of the integrable models; however, certain non-linear field theory models with important physical applications and solitary wave solutions are not integrable. Recently, there have  been performed certain deformations of integrable models such that they possess soliton-like waves (solitary waves) with approximately similar properties to their counterparts of the true soliton theories. In this context, it has been put forward the quasi-integrability concept related to the anomalous zero-curvature approach to modifications of integrable models \cite{jhep1, jhep2}. For earlier discussions  on non-linear field theories with solitary waves and the study of their collisions, see e.g. \cite{hietarinta}. There are different approaches regarding the deformations of integrable theories, see e.g. \cite{malomed, arnaudon} and references therein. 

The quasi-integrability concept has recently been developed and certain deformations of the  sine-Gordon, Toda, Bullough-Dodd, KdV, non-linear Schr\"odinger (NLS) and supersymmetric sine-Gordon models \cite{jhep1, jhep2, jhep6, jhep3, toda, npb, susy} have been studied using their relevant anomalous zero-curvature representations. The main developments have been focused on the construction of infinite number of asymptotically conserved charges and the study of their relevant properties. The asymptotically conserved charges exhibit the same form as the ones from the relevant undeformed theories, and their quasi-conservation properties hold provided the vanishing of the space-time integral of the corresponding anomaly densities. The space-time integration of the anomalies are shown to vanish in special cases; i.e. for two or three-soliton configurations with definite parity under a special space-time inversion symmetry. Remarkably, the presence of several new towers of infinite number of asymptotically conserved charges was recently uncovered in the context of deformed sine-Gordon models \cite{arxiv2}. These new charges differ in form from the relevant charges corresponding to the undeformed model. As it has been mentioned in \cite{arxiv2}, an infinite subset of those new charges turned out to be anomalous even for the standard sine-Gordon model.    

The complete understanding of the dynamics underlying the quasi-soliton behavior of the soliton-like configurations are, so far, largely unknown. The main features can be summarized as follows. First, the one-soliton sectors of those theories have an infinite number of exact conservation laws since the so-called anomalies of the quasi-conservation laws vanish for the one-soliton like solutions. Second, the anomalies also vanish for configurations in which one-soliton like solutions are well separated from each other. The anomalies are significant only when the solitons are close together and they interact with each other. Third, the observed phenomenon seems to occur when the multi-soliton solutions of the equation of motion possess special symmetry properties under a space-time parity transformation. The two or three-soliton configurations possess definite parity, either odd or even, under a space-time reflection around a point in space–time that depends on the individual parameters of the solitons, i.e. velocity, width, initial position, deformation parameters, etc. When the anomaly densities are odd under this parity transformation, one has that the space-time integration, in a rectangle centered at the point around which the reflection is performed, provides a vanishing anomaly, and consequently an asymptotically conserved charge. The presence of that mirror-like symmetry is argued to be a sufficient condition in order to have quasi-integrability \cite{npb}. Fourth, some deformed 
models possess a subset of infinite number of exactly conserved charges for two-soliton 
field configurations being eigenstates of solely the space-reflection parity symmetry. The deformed defocusing (focusing) NLS model with dark (bright) solitons for a variety of two-soliton configurations \cite{jhep4, jhep5} and the deformed sine-Gordon model with kink-kink, kink-antikink and breather solutions \cite{cnsns} have been shown to exhibit this property. 

Several new towers of infinite number of anomalous conservation laws for deformed sine-Gordon models have been uncovered by direct construction \cite{arxiv2}. Remarkably, it has been observed that even the standard sine-Gordon model possesses those types of anomalous charges for soliton configurations satisfying the special space-time inversion symmetry properties. So,  one is lead to think that a truly integrable system inherits to its deformed counterpart that novel property. Moreover, in \cite{arxiv2} it has been developed the so-called Riccati-type pseudo-potential approach to quasi-integrability, and shown that the anomalous conservation laws of \cite{jhep1, jhep2} are, in fact, exact conservation laws, i.e. they become simply the higher order derivatives of the energy-momentum charges. In addition, it has been uncovered an infinite set of exact non-local conservation laws associated to a linear system formulation of the deformed sine-Gordon model \cite{arxiv2}. The above results have been obtained by combining analytical and numerical methods.   

The models considered in \cite{arxiv2} were deformed sine-Gordon models; i.e. relativistic  models with topological solitons. So, it is worth to search for new anomalous charges and perform the pseudo-potential approach to deformations of models with different symmetries, such as the non-relativistic KdV and non-linear Schr\"odinger models. These models stand on the same level of importance as the SG model in their applications, which are abundant in all areas of nonlinear science. Let us mention some applications. The SG, NLS and KdV type models have been applied to the study of Bose-Einsten condensates and superconductivity \cite{frantzeskakis, tanaka1}, General Relativity on $AdS_{3}$ \cite{grav}, soliton gas and soliton turbulence in fluid dynamics \cite{pla1, prlgas} and in the Alice-Bob physics \cite{alice}.    

In this paper we will examine some of the patterns mentioned above in the context of deformations of the KdV model. We examine carefully the anomalous conservation laws presented in \cite{npb}, and demonstrate that each of them hides a trivial conservation law, since  the relevant anomalies can be written as a sum of the type $[\pa_t (\,...\,) + \pa_x (\, ... \,)]$ which, in turn, cancels their similar terms in each conservation law. We search for additional quasi-conservation laws, different from  the ones related to the anomalous zero-curvature approach of \cite{npb}, and study the role played by them in the phenomenon of quasi-integrability. As a byproduct of our constructions we have found that even the standard KdV model exhibits some towers of infinite number of anomalous conservation laws with analogous properties to their counterparts in the quasi-integrable KdV theory. It is shown analytically the quasi-conservation of the infinite towers of anomalous charges for $N-$soliton solution satisfying a special parity symmetry. In particular, some of the lowest order anomalous charges, e.g. the so-called  statistical moments of the KdV model, have been argued to play a fundamental role in the undertanding of the phenomena of soliton gas and soliton turbulence, see e.g. \cite{pla1, prlgas} and references therein. Moreover, we perform the deformation of the KdV model in the framework of  the Riccati-type pseudo-potential approach. In this context, we obtain a linear system formulation of a general deformation of the KdV model and provide an infinite set of exact non-local conservation laws for the deformed model. 

We numerically simulate the various two-soliton and three-soliton interactions of the deformed model by numerically evolving
linear superpositions of two or three (initially well-separated) single-soliton exact solutions of the deformed model. The collisions 
were shown to be very elastic (i.e. the solitons preserved their initial shapes and velocities and there was no appreciable loss of radiation). This property holds for integrable models and the fact that it holds also for the deformed KdV 
equation, which is not an integrable system, characterizes the quasi-integrability of the deformed model. By numerical simulations of 2-soliton and 3-soliton collisions we verify our analytical expectations for the new set of quasi-conservation laws, and prove the vanishing of the lowest order anomalies associated to the relevant new towers of anomalous conservation laws of the deformed KdV model for a variety of values of the deformation parameters $\{\epsilon_1, \epsilon_2\}$.  In order to perform the numerical simulations we follow the methods discussed by J.C. Eilbeck and G.R. McGuire \cite{eil1, eil2} and by L.A. Ferreira, et.al. \cite{npb}.     

This paper is organized as follows: The next section examines the particular deformation introduced in \cite{npb}. The exact one-soliton solutions of the deformed KdV are discussed, and the special parity symmetry, i.e. a shifted space-reflection and time-delayed inversion, is discussed. The properties of the quasi-conservation laws of \cite{npb}, which have been found in the anomalous zero-curvature approach, are examined and discussed. In the section \ref{sec:newasy} we obtain new towers of infinite number of anomalous conservation laws of the model. In subsection \ref{sec:mom} we obtain the higher order moments of the model as anomalous charges. In subsection \ref{sec:mix}, new asymptotically conserved charges with mixed scale dimensions are discussed. These asymptotically conserved charges with mixed scale dimensions are composed by local and non-local terms of their charge densities. In \ref{sec:mrlw1} it is discussed the mRLW model and its quasi-integrability. In section \ref{sec:standard}, we show by direct construction that even the standard KdV model possesses some towers of infinite number of anomalous conservation laws. It is shown analytically the quasi-conservation of the towers of infinite number of anomalous charges for $N-$soliton solution. The section \ref{sec:num} presents some results of our numerical simulations. These simulations
were performed using the LU decomposition method to solve a linear system of equations. The time evolution of various soliton field configurations, corresponding to two- or three-soliton systems, initially located far away, are performed and then verified whether the observed results supported the vanishing of the integrated anomalies of the quasi-conservation laws for several values of the deformation parameters. 

The last section \ref{sec:riccati} considers a general deformation of the KdV model in the context of the Riccati-type pseudo-potential approach. In subsection \ref{sec:nonlocal}, it is found a linear system formulation of the deformed model and constructed an infinite set of non-local conserved charges. Finally, we present our conclusions and three short appendices presenting more details about our numerical techniques and providing some additional results on the construction of the quasi-conserved quantities.

\section{A particular deformation of the KdV model}
\label{sec:sim}

In this section we will consider the model studied in \cite{npb} as  a particular deformation of the KdV equation. It  involves the real scalar field $u$ and the auxiliary fields $w$ and $v$ with equation of motion
\br
\label{mrkdv}
u_t + u_x +\Big[\frac{\a}{2} u^2 + \epsilon_2 \frac{\a}{4} w_x v_t + u_{xx} - \epsilon_1 (u_{xt} + u_{xx} )\Big]_x
= 0,
\er
such that the auxiliary fields satisfy 
\br
\label{wv1}
u &=& w_ t \\
u &=& v_x . \label{wv2}
\er
The real parameters $\epsilon_1$  and  $\epsilon_2$ plays the role of deformation parameters   away from the standard KdV and $\a$ is an arbitrary real parameter. The model (\ref{mrkdv}) encompasses a variety of sub-models. In fact, for $\epsilon_1=  \epsilon_2=0$ one has the integrable KdV model. The case  $\epsilon_1= 1,  \epsilon_2=0$ corresponds to the so-called
regularized long wave equation (RLW). It is not integrable and possesses one-soliton solution. The two and three-soliton solutions for the RLW  model have been constructed numerically and their analytic expressions are not known.
Whereas, the case  $\epsilon_1= \epsilon_2=1$ corresponds to the modified  regularized long wave equation (mRLW). The mRLW equations presents the remarkable property of possessing  analytical  two-soliton solutions. Moreover, for  $\epsilon_2=0,\,\epsilon \neq \{0,1\}$ one has the KdV-RLW or Korteweg-de Vries-Benjamin-Bona-Mahony (KdV-BBM) type equations. We will consider below in sec. \ref{sec:riccati} a more general deformation of KdV in the Riccati-type pseudo-potential approach.

A suitable parametrization of the model (\ref{mrkdv}) is available in order to construct  analytical or numerical soliton solutions  of the model. So, let us consider
\br
\label{uqxt}
u = -\frac{8}{\a} q_{xt}.
\er
In addition, for soliton-type solutions in the context of the tau function Hirota constructions one can make the following parameterizations \cite{npb}
\br
\label{wxvt}
 w_x= -\frac{8}{\a} q_{xx} \,\,\, \mbox{and}  \,\,\,v_t= -\frac{8}{\a} q_{tt}.
\er 
So, substituting the expressions of $u$,  $w_x$ and $v_t$  from (\ref{uqxt})-(\ref{wxvt}), respectively,  into (\ref{mrkdv}) one gets an equation for $q$ as the $x-$derivative of the following equation
\br
\label{eqq}
q_{tt} + q_{xt} - 4 q^2_{xt} - 2 \epsilon_2  q_{xx} q_{tt} + q_{xxxt} - \epsilon_1 (q_{xxtt} + q_{xxxt} ) = 0.
\er
For later purposes we write the next identities. Let us  define 
\br
\label{x1}
X \equiv
\frac{\a}{6} \Big[ \frac{\a}{4} \epsilon_2 w_x v_t - \epsilon_1 \(u_{xt} + u_{xx} \)\Big].
\er
Using the system of eqs. of motion (\ref{mrkdv})-(\ref{wv2}) and the definition of the field  $X$ in (\ref{x1}) one can write
\br
\label{dx}
 u_t + (\frac{\a}{2} u^2 + u_{xx})_x = - \frac{6}{\a} ( X + \frac{\a}{6} u)_x,
\er
which shows on the l.h.s. the usual terms of the  KdV model.  Notice that for $\epsilon_1=\epsilon_2=0$ the  field  $X$ vanishes, so the effect of the deformation is completely encoded in this field. The eq. (\ref{dx})  can  further be written as
\br
\label{dxv}
X_x = - \frac{\a}{6} \Big[ v_t + u + (\frac{\a}{2} u^2 + u_{xx})\Big]_x ,
\er
where we have used the eq.  (\ref{wv2}) which introduces the field $v$ as $u=v_x$. So, the last eq. once integrated in $x$,  can be written as
\br
\label{xvt}
X = - \frac{\a}{6} \Big[ v_t + u + (\frac{\a}{2} u^2 + u_{xx})\Big] + f(t),
 \er
with $f(t)$ being an arbitrary real function of $t$. This field can be set to zero provided that suitable boundary conditions are assumed for the fields. 

Next, let us discuss some space-time symmetries related to soliton-type solutions of the model. So, consider the space-time reflection around a given fixed point $(x_{\Delta},t_{\Delta})$
\br
\label{parity1}
{\cal P}:  (\widetilde{x},\widetilde{t}) \rightarrow (-\widetilde{x},-\widetilde{t});\,\,\,\,\,\,\,\,\widetilde{x} = x - x_{\Delta},\,\,\widetilde{t} = t- t_{\Delta}. 
\er 
In fact, the transformation ${\cal P}$ defines a shifted parity ${\cal P}_{s}$ for the spatial variable  and the  delayed time reversal ${\cal T}_d$ for the time variable. When $x_{\Delta}=0$ ($t_\Delta=0$), ${\cal P}_{s}$ (${\cal T}_d$) is reduced back to the pure parity ${\cal P}$ (pure time reversal ${\cal T}$).  
  
As in the quasi-integrability approach \cite{npb} let us assume that the $u-$field solution of the deformed KdV model evaluated on the N-soliton solution, viz. $u_{N_{-}sol}$, is even under the transformation (\ref{parity1})
\br
\label{paritys1}
{\cal P} (u_{N-sol})=  u_{N-sol}. 
\er 
This implies, according to (\ref{wv1})-(\ref{wv2}), that 
\br
\label{vwtr}
{\cal P} (v_{N-sol})= -v_{N-sol},\,\,\,\,\,{\cal P} (w_{N-sol})= - w_{N-sol},\,\,\,\,\, {\cal P} (q_{N-sol})=  q_{N-sol}.
\er
Therefore, one has 
\br
\label{paritys2}
{\cal P} (X)=  X.
\er
Two and three-soliton solutions of the standard KdV satisfying the above parity symmetries have been constructed in \cite{npb}. Moreover, the analytical two-soliton $u_{2-sol}$
solution of the mRLW model ($\epsilon_1=\epsilon_2=1$) possesses an even parity under the above transformation. In \cite{npb} it has been presented an analytical proof of the quasi-integrability of the non-integrable mRLW theory showing that the relevant charges are asymptotically conserved in the scattering of two solitons. In fact, this has been the first analytical, not only numerical, proof of the quasi-integrability of a (non-integrable)  field theory in 1 + 1 dimensions.
 
Two types of 1-soliton solutions of (\ref{eqq}) have been provided in  \cite{npb} using the Hirota and a direct method, respectively. Below we provide, by direct method, a general 1-soliton solution of the model for any set of values of the parameters $\{\epsilon_1, \epsilon_2\}$, such that the known solutions appear as particular solutions of that general 1-soliton solution. 

\subsection{Two types of 1-soliton solutions}
\label{1soli}

The Hirota method furnishes the first type  of solution of (\ref{eqq})\cite{npb} 
\br
q_{I} = \frac{3}{(2+\epsilon_2)\left(1+(1-\epsilon_1) k^2\right)} \Big\{ \log{2}+ \frac{\G}{2}+ \log{\cosh{(\frac{\G}{2})}}  \Big\},
\er
with
\br
\G = k x - w_1 t + \d; \,\,\,\,\,\, w_1 = \frac{k+ (1-\epsilon_1) k^3}{1- \epsilon_1 k^2}.
\er
So, the eq.  (\ref{uqxt}) provides  the first type of 1-soliton solution for the field $u$
\br
u_{I}= \frac{6}{\a} \frac{k^2}{(2+\epsilon_2)\( 1-\epsilon_1 k^2\)} \mbox{sech}^2\Big[\frac{1}{2} \(k x - w_1 t + \d\)\Big].
\er
A direct method provides a general 1-soliton solution of  (\ref{eqq}) by assuming the form
\br
\label{qii}
q_{II} &=& q_0 \Big\{\log{ \cosh{[\frac{\zeta}{2 a}] } }+ b \zeta + c \Big\},\,\,\,\,\,\zeta =  k x - w_2 t + \d.
\er
A direct substituion of $q_{II}$ into (\ref{eqq})  provides the relationships
\br
\label{disp}
w_2 = \frac{a^2 k + (1-\epsilon_1) k^3}{a^2- \epsilon_1 k^2};\,\,\,\,\, q_0 =\frac{3 a^2}{(a^2+(1-\epsilon_1) k^2)(2 + \epsilon_2)},
\er
such that $a,\,b$ and $c$ are arbitrary real parameters. So, through (\ref{uqxt}) one has the second type of 1-soliton solution for $u$
\br 
\label{solgeral}
u_{II}= \frac{6}{\a} \frac{k^2}{(2+\epsilon_2) ( a^2- \epsilon_1 k^2 )} \mbox{sech}^2\Big[\frac{1}{2 a} (k x - w_2 t + \d)\Big].
\er
This is a new general form of 1-soliton solution which can not be found by the usual Hirota method. Clearly, the two types of solutions become the same for $a^2 =1$ and for arbitrary values of the set $\{\epsilon_1, \epsilon_2\}$. Moreover, for the case $\epsilon_1 \neq 1$ and $w_2 = \frac{k}{1-k^2}$ one has  $a^2=1-(1-\epsilon_1) k^2 $,  and the  1-soliton solution takes the form
\br 
u'_{II} = \frac{6}{\a} \frac{k^2}{(2+\epsilon_2) ( 1- k^2 )} \mbox{sech}^2\Big[\frac{k x - w_2 t + \d}{2  \sqrt{1-(1-\epsilon_1) k^2}}\Big].
\er
This particular case has been reported in \cite{npb}, and this type of solution $u'_{II}$ coincides with $u_I$ for $\epsilon_1=1$.
 
\subsection{2-soliton type solution: the case $\epsilon_1 =\epsilon_2 =1$}
 
The $2-$soliton solution exists for the particular case $\epsilon_1 =\epsilon_2 =1$. The field $q$ takes the form\cite{npb, gibbon}
\br
\label{qq22}
q &=& \log{\Big[1+ e^{\G_1}+e^{\G_2}+ A_{12} e^{\G_1} e^{\G_2}\Big]},\,\,\,\, \G_i = k_i x - w_i t + \delta_i,\,\,\,w_i= \frac{k_i}{1-k_i^2},\,\,i=1,2.\\
A_{12}&=& -\frac{(w_1 - w_2)^2 (k_1 - k_2)^2 + (w_1 - w_2)(k_1 - k_2) - (w_1 - w_2)^2}{(w_1 + w_2)^2 (k_1 + k_2)^2 + (w_1 + w_2)(k_1 + k_2) - (w_1 + w_2)^2}.
\er
In order to implement the parity transformation (\ref{parity1}) and check the space-time parity inversion symmetry of the 2-soliton solution we will derive a new expression for $q$ in (\ref{qq22}), such that  $u_{2-sol}$ in (\ref{uqxt}) becomes a manifestly ${\cal P}$ invariant function. So, let us define a new parameter $\D$, as $A_{12} = e^{\D}$, and
\br
\label{etas1}
 \G_j &=& k_j \widetilde{x} - w_j \widetilde{t} + \eta_{0j} - \frac{\D}{2} \equiv \eta_j - \frac{\D}{2}, \,\,\,\,\, j=1,2  
\er
where 
\br \label{deltas1}
\delta_j  = -k_j x_{\D} + w_j t_{\D} + \eta_{0j} - \frac{\D}{2}, \,\,\,j=1,2.
\er
Therefore, $q$ can be rewritten as 
\br
q = \log{\Big[2 e^{-\D/4} \, e^{(\eta_1+\eta_2)/2} \(e^{\D/4} \cosh{(\frac{\eta_1+\eta_2}{2})} + e^{-\D/4}  \cosh{(\frac{\eta_1-\eta_2}{2})}  \)\Big]}.
\er
So, using (\ref{uqxt}) one has 
\br
\label{u2ee}
u_{2-sol} &=& -\frac{8}{\a} \pa_x \pa_t   \log{\Big[ e^{\D/4} \cosh{(\frac{\eta_1+\eta_2}{2})} + e^{-\D/4}  \cosh{(\frac{\eta_1-\eta_2}{2})} \Big]}.
\er
Therefore, the parity invariant 2-soliton becomes 
\br
\label{usolxt}
{\bf u}_{2-sol} = u_{2-sol} \Big|_{\eta_{01}=\eta_{02}=0}.
\er
Taking into account the condition $\eta_{01}=\eta_{02}=0$, one gets the next relationships for the coordinates of the special point $(x_{\D},t_{\D})$
\br
x_{\D} &\equiv & \frac{w_2 \widetilde{\theta}_1 - w_1 \widetilde{\theta}_2}{k_2 w_1 - k_1 w_2}\\
t_{\D} & \equiv & \frac{k_2 \widetilde{\theta}_1 - k_1 \widetilde{\theta}_2}{k_2 w_1 - k_1 w_2},\,\,\,\,\,\widetilde{\theta}_j \equiv \frac{\D}{2} + \delta_j,\,\,\,j=1,2.  
\er 
Therefore, one has 
\br
\label{parity2}
{\cal P}({\bf u}_{2-sol}) = { \bf u}_{2-sol},
\er
and using (\ref{wv1})-(\ref{wv2}) one has that
\br
\label{parity3}
{\cal P}({\bf v}_{2-sol}) = - {\bf v}_{2-sol},\,\,\,\,\,{\cal P}({\bf w}_{2-sol}) = - {\bf w}_{2-sol}.
 \er
In \cite{npb} it was provided a different procedure to construct  ${\bf u}_{2-sol}$, and it  has also been shown that the exact Hirota three-soliton solutions of the standard KdV equation possess the relevant parity properties when their solitons collide at the same point in space.  

\section{KdV-type asymptotically conserved charges}
\label{sec:kdvtype}

In the   context of  deformations of the sine-Gordon model, the higher order quasi-conservation laws obtained in the anomalous zero-curvature approach \cite{jhep1, jhep2}, when conveniently rewritten them as exact  conservation laws, simply become the higher order derivatives of the energy-momentum conservation law \cite{arxiv2}. Then, the higher order charges have been written as $\frac{d^n}{dt^n} (E\pm P)$, where $E$ stands for energy and $P$ for momentum. So, it is interesting to examine the similar quasi-conservation laws related to the deformations of the KdV model in the approach of \cite{npb}. 

In this section we will examine the properties of the anomalous  conservation laws considered  in \cite{npb} for the particular  deformation of the  KdV model (\ref{mrkdv})-(\ref{wv2}).  

The anomalous conservation laws of the particular  deformation of the  KdV model (\ref{mrkdv}) have been defined as \cite{npb}
\br
\label{curr1}
\pa_t a_x^{(-2n-1)} - \pa_x a_t^{(-2n-1)} =  -  X \g^{(-2n-1)}, \,\,\,\,\,\,n \in \IZ^{+}_{0}
\er
where $X$ has been defined in (\ref{x1}) and the first components of $ a_x^{(-2n-1)}$ and $\g^{(-2n-1)}$ are provided  in (\ref{axn}) and (\ref{gdef})-(\ref{gn}), respectively. So, one can write
\br
\label{qa1}
\frac{d}{dt}  Q^{ (-2n-1)}_a = \a^{(-2n-1)},\,\,\,\,\,\,n \in \IZ^{+}_{0}
\er
such that the asymptotically conserved charges and their associated anomalies are defined as 
\br
\label{can}
Q^{ (-2n-1)}_a \equiv \int_{-\infty}^{+\infty}  a_x^{(-2n-1)}\,\,\,\,  \mbox{and} \,\,\,\, \a^{(-2n-1)} \equiv - \int_{-\infty}^{+\infty}  dx X \g^{(-2n-1)}.
\er 
In \cite{npb} it has been defined the asymptotically conserved charges
\br
\label{asymp1}
Q^{ (-2n-1)}_{a\,(t\rightarrow -\infty)} = Q^{ (-2n-1)}_{a\, (t \rightarrow +\infty)}.
\er 
In fact, for soliton configurations satisfying  the parity symmetries (\ref{paritys1})-(\ref{paritys2}) the time integrated anomalies vanish, i.e.
\br
\label{avan}
\int_{t=-\infty}^{t=+\infty}\, dt \a^{(-2n-1)} &=& - \int_{t=-\infty}^{t=+\infty}\, dt  \int_{x = -\infty}^{x= +\infty} \, dx X \g^{(-2n-1)}\\
&=&0.
\er
In that approach, the form of the charge densities $ a_x^{(-2n-1)}$ are the same as the relevant ones corresponding to the usual KdV model at each order. Notice that the deformation parameters enter only on the r.h.s., $ \a^{(-2n-1)}$,  of the eq. (\ref{qa1}). 

Next, instead of assuming the quasi-conservation laws (\ref{qa1}), which define the relevant asymptotically conserved charges and anomalies,  we inquire about the properties  of the anomalous r.h.s.'s; in particular, if they would directly be rewritten in the form 
\br
\label{rhs}
 -  X \g^{(-2n-1)} \equiv \pa_{t} \(j_{x}^{(-2n-1)}\) + \pa_{x} \( j_{t}^{(-2n-1)}\),
\er
by explicitly obtaining the relevant current components $j_{x}^{(-2n-1)}$\, and \, $j_{t}^{(-2n-1)}$. So, let us rewrite the r.h.s.'s of the first four equations of (\ref{curr1}) for $n=0,1,2$ and $3$. 

{\bf Zeroth order} ($n=0$)
\br
\label{n0}
\pa_t \(\frac{\a}{2^2 3} u\) - \pa_x \( -\frac{\a^2}{72} u^2 +\frac{\a}{12} (v_{t} + \frac{\a}{6} u^2) \) =0,
\er
where the trivially  vanishing  term $\g^{(-1)}$ (\ref{gdef})-(\ref{gn})  has been used in the r.h.s. of (\ref{n0}). Then, one can define the charge
\br
\label{mass}
Q^{(-1)} = \frac{\a}{12} \int_{-\infty}^{+\infty} dx \, u.  
\er
This charge is generally associated with the ``mass".  Since it emerges from an exact conservation law,  this charge is conserved even in the deformed KdV model (\ref{mrkdv}). 

{\bf First order} ($n=1$)
\br
\label{n1}
\pa_t [\frac{\a^2}{2^5 3^2} u^2] - \pa_x a_t^{(-3)} = - X \g^{(-3)}.
\er
A remarkable fact is that the r.h.s. of (\ref{n1}), using the expression for $\g^{(-3)}$ in  (\ref{gdef})-(\ref{gn}) and the eq. of motion (\ref{mrkdv}),  can be written as 
\br
- X \g^{(-3)} =
\frac{\a^2}{2^5 3^2}  \pa_t[ u^2]  +\pa_x\Big[ \frac{\a}{2^3 3}X u  +  \frac{\a^2}{2^5 3^2}   u^2 + \frac{\a^2}{2^4 3^2}   u  (\frac{\a}{2} u^2 + u_{xx}) -  \frac{\a^3}{2^5 3^3}   u^3 -  \frac{\a^2}{2^5 3^2}  u_x^2\Big].\label{xm3}
\er
Substituting the last identity into the eq. (\ref{n1}) one gets a trivial identity, and then a vanishing charge density. Therefore, one gets a trivial charge at this order
\br
q^{(-3)} =0.\label{q3}
\er
However,  following \cite{npb},  at this order one can define the quasi-conservation law
\br
\label{qc3}
\frac{d Q^{(-3)}_a}{dt} = \a^{(-3)}
\er
where
\br
\label{qan1}
Q^{(-3)}_a \equiv  \frac{\a^2}{2^5 3^2} \int_{-\infty}^{+\infty} dx \, u^2 ,\,\,\,\,\,\,\, \a^{(-3)}  \equiv  - \int_{-\infty}^{+\infty} dx  \, X \g^{(-3)},
\er 
is the asymptotically conserved charge $Q_a^{(-3)}$, with $ \a^{(-3)}$ being its relevant anomaly. 

{\bf Second order}($n=2$)
 \br
\label{n2}
\pa_t [\frac{\a^3}{2^7 3^3} u^3+\frac{\a^2}{2^7 3^2} u u_{xx} ] - \pa_x a_t^{(-5)} = - X \g^{(-5)}.
\er
Similarly, the r.h.s. of (\ref{n2}) can be written as
\br\nonumber
 - X \g^{(-5)} = \frac{\a^3}{2^7 3^3} \pa_t u^3 - \frac{\a^2}{2^7 3^2} \pa_t (u_x)^2 + \pa_x\Big[ \frac{\a^2}{2^6 3^2} u^3 +\frac{\a}{2^6 3} u_x^2 + \frac{\a^3}{2^8 3} u^4 + \frac{\a^2}{2^6 3} u_{xx} u^2 + \frac{\a}{2^6 3} u_{xx}^2 + \frac{\a^2}{2^6 3^2} u_ t u_x\Big]
\er
Substituting the last identity into the r.h.s. of eq. (\ref{n2}) and collecting the charge density terms one has
\br
q^{(-5)} &=&  \frac{\a^2}{2^7 3^2} \int_{-\infty}^{+\infty} dx \, \pa_x \( u_x u \) ,\\
&=&0.
\er
Therefore, one has a trivially vanishing charge also at this order. Following \cite{npb} one can define the quasi-conservation law
\br
\frac{d Q^{(-5)}_a}{dt} = \a^{(-5)}
\er
where
\br
\label{qan2}
Q^{(-5)}_a \equiv \frac{\a^2}{2^7 3^2} \int_{-\infty}^{+\infty} dx  [ \frac{\a}{3} u^3 - (u_{x})^2],\,\,\, \a^{(-5)}  \equiv  - \int_{-\infty}^{+\infty} dx  \, X \g^{(-5)},
\er 
are the asymptotically conserved charge $Q^{(-5)}_a$ and its relevant anomaly $ \a^{(-5)}$. This charge maintains the same form as in the usual KdV at this order.
 
{\bf Third order} ($n=3$)
 \br
\label{n3}
\pa_t [\frac{5 \a^4}{2^{11} 3^4} u^4+\frac{\a^3}{2^7 3^3} u^2 u_{xx} + \frac{\a^2}{2^9 3^2} (\a u u_{x}^2+ u u_{xxxx})] - \pa_x a_t^{(-7)} = - X \g^{(-7)}.
\er
Next, the r.h.s. of (\ref{n3}) can be rewritten as \footnote{The correct form of $\g^{(-7)}$ has been presented in (\ref{gdef})-(\ref{gn}) of the present paper. In fact, the term $\frac{5\a^2}{2^7 3^2} u_{xx}$ inside $-\pa_x[....]$ appearing in the expression of $\g^{(-7)}$ in the fourth line of (2.20) of \cite{npb} should be replaced by  $\frac{5\a^2}{2^7 3^2} u u_{xx}$. }
\br
\nonumber
 - X \g^{(-7)}&=&  \pa_t \Big[  \frac{5 \a^2}{2^9 3^2} \(\frac{\a^2}{36} u^4 - \frac{\a}{3} u (u_{x})^2 + \frac{1}{5} (u_{xx})^2 \)\Big] +  \\
&&\frac{5 \a^2}{2^9 3^2}  \pa_x \Big[\frac{2\a}{3} uu_xu_t-\frac{2}{5}u_{xx}u_{xt}+\frac{2}{5} u_{xxx}u_t-\frac{\a^2}{3^2}u^4-\nonumber \\
&&\frac{1}{6} (\frac{\a}{2} u^2+u_{xx})^2-
\frac{2}{5} u_xu_{xxx}+\frac{1}{5} u_{xx}^2-\frac{\a^3}{45}u^5-\frac{1}{30} u_{xxx}^2 -\frac{\a}{15} uu_xu_{xxx} -\frac{\a^2}{9}u^3u_{xx}-\nonumber\\
&&\frac{\a}{15} uu_{xx}^2 +\frac{\a}{15} u_x^2u_{xx} -\frac{\a}{15} uu_{xx}^2\Big].
\label{Xg7}
\er
Notice that the charge density in the l.h.s. of (\ref{n3}) and the term inside the bracket $\pa_t[...]$ of (\ref{Xg7}) are the same up to an expression of the form $\pa_x[...]$. With this observation in mind and substituting (\ref{Xg7}) into the r.h.s. of eq. (\ref{n3}) and collecting the charge density terms one has
\br
q^{(-7)} &=& 0.
\er
So, one gets a trivial charge also at this order. However, related to (\ref{n3}) the next asymptotically-conserved charge has been defined \cite{npb}
\br
\label{qan3}
Q^{(-7)}_a =\frac{5 \a^2}{2^9 3^2} \int_{-\infty}^{+\infty} dx  [ \frac{\a^2}{36} u^4 - \frac{\a}{3} u (u_{x})^2 + \frac{1}{5} (u_{xx})^2].
\er   
Therefore, from (\ref{n3})  the quasi-conservation law for $Q^{(-7)}_a$ can be written as
\br
\label{n31}
\frac{d}{dt} Q^{(-7)}_a &=& - \int_{-\infty}^{+\infty} dx \, X \g^{(-7)},
\er
where the r.h.s. of (\ref{n3}) provides the so-called anomaly.

In summary, what we have done in the computations which follow the eqs. (\ref{n1}), (\ref{n2}) and  (\ref{n3})  is to rewrite  the relevant r.h.s.'s in the form (\ref{rhs}) and shown that the relevant quasi-conservation laws reduce to trivial identities.

Some comments are in order here. First,  the charges of the sequence $Q^{(-2n-1)}_a, \,n=0,1,2,... ,$ in the anomalous zero-curvature approach, maintain the same form as the relevant charges of the usual KdV.
Second, in the ordinary KdV, i.e. when the anomaly $X=0$,   the  charges  of (\ref{n0}), (\ref{n1}) and (\ref{n2}) are usually associated with the
``mass", ``momentum" and ``energy" conservation, respectively. The quantity $u$ inside the time-derivative of (\ref{n0}) can be interpreted as the mass density, while the terms inside the $x-$derivative represent the mass flux. However,  the charges of (\ref{n2})  and (\ref{n3})  do not  have a direct interpretation and their relationships to the relevant physical quantities have recently been considered (see e.g. \cite{mme}). 
Third, in order to achieve the trivial charges $q^{(-n)},\,n=3,5,7$ we have removed the non-homogeneous terms, which were dubbed as ``anomalies'' in \cite{npb}, and  conveniently rewritten the relevant equations as exact conservation laws. So, one can argue that the conservation laws (\ref{qa1}) become trivially satisfied, since the r.h.s.'s of the relevant quasi-conservation laws (\ref{curr1}) can be expressed as time-derivatives of the relevant KdV-type charges. Fourth, the behavior above is in contradistinction to the deformed sine-Gordon models, in which the analogous higher order quasi-conservation laws, in the  anomalous zero-curvature approach, become simply the higher derivatives of the non-trivial energy-momentum conservation law \cite{arxiv2}.  

\section{New asymptotically conserved charges and scale dimensions}
\label{sec:newasy}

In the context of the integrable KdV,  modified KdV (mKdV) and Gardner (mixed KdV-mKdV) models there have been analyzed the behavior of the so-called statistical moments defined by the integrals of type \cite{pla1, physd}
\br
\label{smom}
M_n(t) = \int_{\infty}^{+\infty} u^n \, dx,\,\,\,\,\,\,n=1,2,3,4.
 \er
For the above integrals it has been examined the two-soliton interactions which are thought to play an important role in the formations of the structures of soliton turbulence and soliton gas in integrable systems. The soliton gas associated to the quasi-integrable model (\ref{mrkdv}) for $\epsilon_2=0,\epsilon_1 \neq \{0, 1\}$ (the so-called KdV-RLW or the Korteweg-de Vries-Benjamin-Bona-Mahony (KdV-BBM) model), has been examined \cite{pla2}. 

It is interesting to notice that  the two-soliton interaction in the framework of the KdV equation leads to the decrease of the 3rd and 4th moments, $M_{3,4}$, respectively,  around the interaction region \cite{pla1}, revealing a qualitatively analogous behavior to the asymptotically conserved charges reported in quasi-integrable models \cite{npb}. While the first two moments  $M_{1,2}$ are integrals of the KdV and mKdV evolutions, respectively, the 3rd and 4th moments, corresponding to the KdV and mKdV systems, undergo significant variations in the dominant interaction region, resembling to the behavior of asymptotically conserved charges of quasi-integrable KdV models \cite{npb}. In fact, the charge $Q_a^{(-3)}$ in the quasi-integrable KdV model, see (\ref{qan1}), which has the same form as $M_2$, is an asymptotically conserved charge.  

We believe that those types of charges will play an important role in the study of soliton gases and formation of certain structures in (quasi-)integrable systems, such as soliton turbulence, soliton gas dynamics and rogue waves \cite{pla1, prlgas}. It has recently been achieved the experimental realization of a hydrodynamic soliton gas in a wave flume in a
shallow water regime \cite{prlgas}. In this scenario a pure integrable dynamics and  the two-soliton interaction are the basic ingredients in the formation of the soliton gas.

In addition, the above type of moment functions have been used to implement a numerical solution of KdV-type models in the context of the so-called general lattice Boltzmann model \cite{chai}.  

The question arises whether one could find  other independent sets, different from the above set $Q_a^{(-2n-1)}$ in (\ref{qa1})-(\ref{can}), of  asymptotically conserved charges in the quasi-integrable KdV model. So, we will search for new infinite towers of asymptotically conserved charges of the deformed KdV (\ref{mrkdv}), or equivalently (\ref{dx}). 

Let us examine the scaling dimensions and some symmetries of the model. By inspecting the scaling (inverse length dimension) of the relevant fields in the l.h.s. of  eq. (\ref{dx}) one notices that the fields and derivatives can be associated with the scale dimensions\footnote{ Note that all terms  in the l.h.s. of (\ref{dx}) should be of the same scale dimension with $\pa_x$ being $L^{-1}$, which we define as $\deg(\pa_x)=1$. The $u_x$ term in the r.h.s. can be removed by field redefinition, as in the usual KdV.}
\br
\deg(\pa_t) = 3;\,\,\,\,\deg(\pa_x) = 1;\,\,\,\, \deg(u)=2; \,\,\,\, \deg(v) =1;\,\,\,\,\deg(w) =-1,
\er 
where the scaling dimensions of the fields $w, v$ have been defined by using the relationships (\ref{wv1})-(\ref{wv2}), respectively. The standard KdV eq. is symmetric under scaling transformation; so, its conservation laws, generalized symmetries, and recursion operator inherit the same scaling property of the KdV system \cite{hereman}.  According to the above definition, the terms of the r.h.s. of (\ref{dx}) exhibits non-homogeneous scaling dimensions, i.e. the components of  $X$ comprises fourth and sixth order scale dimensions, while the remaining term $u_x$ possesses third order.  We will systematically use this concept below to identify the higher scale dimensions of different quantities, such as the relevant charge densities.  
In fact, the KdV-type asymptotically conserved charges defined in section \ref{sec:kdvtype} exhibits homogeneous scale dimensions in each term of their relevant charge densities; i.e. $\deg{\{a_x^{(-2n-1)}\}} = 2n + 2$.

\subsection{Higher order moments as asymptotically conserved charges}
\label{sec:mom}

One can get an infinite tower of quasi-conservation laws by multiplying (\ref{dx}) on the both sides by $n u^{n-1}$ and using the form of $X$ in (\ref{xvt}) one gets  the non-homogeneous conservation laws
 \br
\label{bbmn1}
\pa_t [u^n] + \pa_x[ \frac{\a}{2} n u^{n+1} + n u^n + n u^{n-1} u_{xx}+  \frac{6 n}{\a} u^{n-1} X ] = - n(n-1) u^{n-2} u_x v_t; \,\,\,\, n=1,2,3....
\er
Notice that the $n=1$ case corresponds to the eq. of motion (\ref{dx}) itself; since the r.h.s. of (\ref{bbmn1}) for $n=1$ vanishes the relevant conservation law defines, up to the constant factor $\a/12$, the mass  $Q^{(-1)}$ in (\ref{mass}). Taking  $n=2$ in  (\ref{bbmn1}) and using the expression  of $v_t$ in (\ref{xvt}) one gets the quasi-conservation law
\br
\pa_t [u^2] + \pa_x[ \frac{2 \a}{3} u^{3} + u^2 + 2 u u_{xx} +  \frac{12}{\a} u X -u_x^2] = \frac{12}{\a} u_x X,
\er
which is, up to the overall constant factor $\frac{\a^2}{2^5 3^2}$, the anomalous conservation law (\ref{n1}). So, we define the following quasi-conservation laws for $n\geq 3$
\br
\label{mom1}
\frac{d}{dt}  \widetilde{q}_a^{(n)} &=& \widetilde{\beta}^{(n)},\,\,\,\,n=3,4,5,...\\
  \widetilde{q}_a^{(n)}  &\equiv& \frac{\a^n}{2^{2n+1} 3^{n}} \int_{-\widetilde{x}}^{+\widetilde{x}} dx \,\, u^n,\,\,\,\,\,  \widetilde{\beta}^{(n)} \equiv  \frac{\a^n}{2^{2n+1} 3^{n}} \int_{-\widetilde{x}}^{+\widetilde{x}} dx \,\, [  - n(n-1) u^{n-2} u_x v_t ].\label{mom11}
\er  
Since we can assume  the parity symmetry (\ref{paritys1})-(\ref{paritys2}) for the corresponding fields, the  time integrated anomalies $\widetilde{\beta}^{(n)}$ vanish for $\widetilde{t} \rightarrow+\infty,\, \widetilde{x} \rightarrow +\infty$
\br
\label{anoint}
\int_{-\widetilde{t}}^{+ \widetilde{t}} dt \widetilde{\beta}^{(n)} &=& \frac{\a^n}{2^{2n+1} 3^{n}} \int_{-\widetilde{t}}^{+ \widetilde{t}} dt  \int_{-\widetilde{x}}^{+\widetilde{x}} dx \,\, [  - n(n-1) u^{n-2} u_x v_t ].\\
&=& 0\nonumber
\er
Integrating in time (\ref{mom1})  and making  $\widetilde{x} \rightarrow +\infty $ one can write
\br
\label{mom11t}
\widetilde{q}_a^{(n)}(+ \widetilde{t}) = \widetilde{q}_a^{(n)}(-\widetilde{t}),\,\,\,\,\,\,n=2,3,4,...
\er
provided that the vanishing of the time-integrated anomaly (\ref{anoint}) is taken into account. So,  the higher order moments defined in (\ref{mom1}) become asymptotically conserved charges, in analogy to the relevant moments  (\ref{smom}) of the standard KdV as mentioned above. 

Notice that the relevant charge densities exhibit the scale dimension $\deg{u^{n}} = 2n$. So, this tower of asymptotically conserved charges stands as a different set  from the previous set constructed using the KdV-type conservation laws. The last results  beg the question of whether it could be possible to construct new quasi-conservation laws directly from the equations of motion. In the next constructions we will show the appearance of other sets of asymptotically conserved charges with mixed scale dimensions.

In sec. \ref{sec:num}, we will numerically simulate the anomaly $\widetilde{\beta}^{(3)}$ in (\ref{mom1}) of the quasi-conservation laws  (\ref{bbmn1}) for 2-soliton and 3-soliton scatterings.  

\subsection{Asymptotically conserved charges and mixed scale dimensions}
\label{sec:mix}
In addition to the above charges, we can also define another set of charges with corresponding  charge density terms  possessing mixed scale dimensions, i.e. in contradistinction to the charge densities $ a_x^{(-2n-1)}$ in (\ref{axn}) with $\deg( a_x^{(-2n-1)}) = 2n+2$ for each of their terms, and the above  higher order moment charges (\ref{mom1}) with $\deg{u^n}=2 n$. 

\subsubsection{Local asymptotically conserved charges and mixed scale dimensions}
\label{sec:local}

Let us examine local asymptotically conserved charges with  mixed scale dimensions composing their charge densities. Taking into account the form of $X$ in (\ref{x1}) and multiplying (\ref{dx}) on the both hand sides by $2 u$,  the next non-homogeneous conservation law can directly be obtained  
\br
\label{bbm1}
\pa_t [u^2 + \epsilon_1 u_x^2 ] + \pa_x[u^2+ \frac{2 \a}{3} u^3 +2 u u_{xx} - u_x^2 -2 \epsilon_1 u u_{xt} + \frac{\a \epsilon_2}{2} u w_x v_t ] = \frac{\a \epsilon_2}{2} w_x v_ t u_x.
\er
Notice that for $\epsilon_2 =0$ the r.h.s. of (\ref{bbm1}) vanishes; so,  this eq. turns out to be an exact conservation law; and it will give rise to the second conserved charge of the KdV-RLW model \cite{bbm, pla2, pla3, junior} defined for $\epsilon_2 =0, \epsilon_1 \neq \{0, 1\}$. For the particular choice  $\epsilon_2 =0, \epsilon_1=1$  (\ref{bbm1}) defines the second conservation law of the RLW model \cite{olver}. For  $\epsilon_2 \neq 0$ one can define an asymptotically conserved charge from the quasi-conservation law (\ref{bbm1}) as
\br
\label{qq2}
\frac{d}{dt} \widetilde{Q}^{(2)}_a & = & \widetilde{\alpha}_2 \\
 \widetilde{Q}_a^{(2)}  & \equiv & \frac{\a^2}{2^5 3^2}\int_{-\widetilde{x}}^{+\widetilde{x}} dx  [u^2 + \epsilon_1 u_x^2 ]; \,\,\,\,  \widetilde{\alpha}_2  \equiv \frac{\a^3 \epsilon_2}{2^6 3^2} \int_{-\widetilde{x}}^{+\widetilde{x}} w_x v_ t u_x \, dx,\label{qq21}
\er
where the r.h.s. defines the relevant anomaly $ \widetilde{\alpha}_2$, and an overall normalization factor has been introduced for later convenience. Notice that the terms in the charge density possess different scale dimensions; i.e.  $\deg(u^2)=4,\,\,\deg(u_x^2) = 6$. The time-integrated anomaly vanishes provided that the fields satisfy the parity transformations (\ref{paritys1})-(\ref{paritys2}), i.e.
\br
\label{tia}
\int_{-\widetilde{t}}^{+\widetilde{t}}  \widetilde{\alpha}_2 \, dt& = & \frac{\a^3 \epsilon_2}{2^6 3^2} \int_{-\widetilde{t}}^{+\widetilde{t}} dt \int_{-\widetilde{x}}^{+\widetilde{x}} dx \,\, w_x v_ t u_x \\
 &=& 0.
\er
So, integrating in time (\ref{qq2})  and making  $\widetilde{x} \rightarrow +\infty $ one can write for $\widetilde{t} \rightarrow +\infty$
\br
\label{asym2}
\widetilde{Q}^{(2)}_a(+ \widetilde{t}) = \widetilde{Q}^{(2)}_a(-\widetilde{t}),
\er
provided that the vanishing of the time-integrated anomaly (\ref{tia}) is taken into account. 

Similarly, from $X$ in (\ref{x1}) and multiplying (\ref{dx}) on the both sides by $3 u^2$,  the following non-homogeneous conservation law can be obtained  
\br
\nonumber
\pa_t [u^3 + 3 \epsilon_1 u u_x^2 ] + \pa_x[\frac{3 \a}{4} u^4 + u^3 +\frac{3 \a}{4} \epsilon_2 w_x v_t u^2 -3 \epsilon_ 1 u_{xt} u^2] &=& 3 (\epsilon_ 1 -1) u_{xxx} u^2 +\\
\label{bbm2} && \frac{3 \a \epsilon_2}{2} w_x v_ t u u_x +3 \epsilon_ 1  u_{x}^2 u_t.
\er
So, the generalization of this sequence for higher order quasi-conservation laws become
\br
 \label{bbmn}
\pa_t \Big[u^n + \frac{n(n-1) \epsilon_1}{2} u^{n-2} u_x^2 \Big] &+& \pa_x\Big[ n u^{n-1}\(\frac{\a}{n+1} u^{2} + \frac{u}{n} +\frac{\a  \epsilon_2}{4 } w_x v_t  - \epsilon_ 1 u_{xt}\)\Big] = {\cal A}_n,\,\,\,\,\,\,\,\,\,\,\,n\geq 3  \\
\nonumber
{\cal A}_n & \equiv &  nu^{n-3}\Big[ (\epsilon_ 1 -1) u^2 u_{xxx} + \frac{ \a \epsilon_2 (n-1)}{4} w_x v_ t  u u_x +  \frac{(n-1)(n-2)}{2} \epsilon_ 1  u_{x}^2 u_t\Big].
\er
From (\ref{bbmn}) one can define the asymptotically conserved charges
\br
\label{qqn}
\frac{d}{dt} \widetilde{Q}^{(n)}_a & = & \widetilde{\alpha}_n ,\,\,\,\,\,\,\,\,\,\,\,\,\,\,\,n\geq 3 \\
 \widetilde{Q}_a^{(n)}  & \equiv & \frac{\a^n}{2^{2n+1} 3^n}\int_{-\widetilde{x}}^{+\widetilde{x}} dx  [u^n + \frac{n(n-1) \epsilon_1}{2} u^{n-2} u_x^2]; \,\,\,\,  \widetilde{\alpha}_n  \equiv \frac{\a^{n+1}}{2^{2n+2} 3^n} \int_{-\widetilde{x}}^{+\widetilde{x}} {\cal A}_n   \, dx.\label{qqn1}
\er
One can show that the anomaly density terms of ${\cal A}_n$ are odd functions under the parity transformation (\ref{paritys1})-(\ref{paritys2}); so, following similar arguments as above one can conclude that the charges  $\widetilde{Q}_a^{(n)}$ are asymptotically conserved.

In sec. \ref{sec:num}, we will numerically simulate the anomaly $\widetilde{\alpha}_2$ in (\ref{qq21}) of the quasi-conservation law (\ref{qq2}) for 2-soliton and 3-soliton scatterings.  

Next, let us search for another set of charges. So, write the eq. (\ref{dx})  in the form of an evolution equation
\br
\label{evol}
u_t &=& -\frac{6}{\a}  F_x,\\ \nonumber
F &\equiv& X + \frac{\a}{6} u + \frac{\a}{6} (\frac{\a}{2} u^2+u_{xx}).
\er
Multiplying the both sides of (\ref{evol}) by $F$ and using the expression for $X$ in (\ref{x1}) one can get the quasi-conservation law
\br
\label{evoq1}
\pa_t [\frac{\a}{3} u^3 + u^2+  (\epsilon_1-1)  u_x^2] + \pa_x [\frac{36}{\a^2} F^2 -\epsilon_1 u_t^2- 2(\epsilon_1-1) u_t u_x ] = -\frac{\a \epsilon_2}{2} w_x v_t u_ t.
\er
Clearly, for $\epsilon_2 =0$ it provides an exact conservation law. In fact, for  $\epsilon_2 =0,  \epsilon_1 =1$ it provides the third conserved charge of the RLW model \cite{olver}.  In addition, it will give the third conserved charge of the KdV-RLW model defined for $\epsilon_2 =0, \epsilon_1 \neq \{0, 1\}$. In fact, the eq. (\ref{mrkdv}) for $\epsilon_2=0, \epsilon_1 \neq \{0, 1\}$ can be written as
\br
\label{kdvrlw}
u_t + \pa_x [R]=0,\,\,\,\,R\equiv u +\frac{\a}{2} u^2 + u_{xx} - \epsilon_1 (u_{xt} + u_{xx} ). 
\er
So, multiplying by $R$ the eq. (\ref{kdvrlw}) on can rewrite it as
\br
\pa_t\Big[\frac{1}{2} u^2 + \frac{\alpha}{3} u^3 - (1-\epsilon_1)(u_x)^2 \Big] +\pa_x\Big[ (1-\epsilon_1) u_t u_x -\epsilon_1 (u_t)^2 +\frac{1}{2} R^2\Big] =0.
\er
The last construction provides the third conserved charge of the KdV-RLW model. For other choices of the set of parameters $\{\epsilon_1, \epsilon_2 \}$, from (\ref{evoq1}) one can define the asymptotically conserved charge 
\br
\label{qq3}
\frac{d}{dt}  {\cal Q}_{a}^{(3)} & = & {\cal \alpha}_{3} \\
  {\cal Q}_{a}^{(3)} & \equiv &
\frac{\a^2}{2^5 3^2}\int_{-\widetilde{x}}^{+\widetilde{x}} [ \frac{\a}{3} u^3 + u^2 +(\epsilon_1-1)  u_x^2] dx ,\,\,\,\,\,\,  {\cal \alpha}_{3}  \equiv -\frac{\a^3 \epsilon_2}{2^6 3^2} \int_{-\widetilde{x}}^{+\widetilde{x}} w_x v_ t u_t \, dx.\label{qq31}
\er
Since the anomaly $  {\cal \alpha}_{3} $ vanishes when integrated in space-time and evaluated on fields satisfying the parity symmetry  (\ref{paritys1})-(\ref{paritys2}) , following similar steps as above one can define the asymptotically conserved charge
\br
\label{asym21}
 {\cal Q}_{a}^{(3)}(\widetilde{t}\rightarrow \infty) = {\cal Q}_{a}^{(3)}(\widetilde{t} \rightarrow -\infty).
\er
Next, let us construct the sequence of quasi-conservation laws containing the eq. (\ref{evoq1}). So, substitute  $u=v_x$ into the l.h.s. of the evolution eq. (\ref{evol}) and perform a $x$ integration once, the outcome will be  
\br
\label{evol1}
v_t = -\frac{6}{\a}  F.
\er
Multiplying on the both hand sides of the eq. above by $u_t$ and performing some algebraic manipulations one can get
\br
\label{evoq11}
\pa_t [\frac{\a}{3} u^3 + u^2+  (\epsilon_1-1)  u_x^2]+ \pa_x[v_t^2 +2 (1-\epsilon_1) u_t u_x - \epsilon_1 u_t^2] = - \frac{\a}{2} \epsilon_2 w_x v_t u_t.  
\er
One can show that this eq. is the same as the quasi-conservation law obtained above in (\ref{evoq1}). Next, let us multiply by $u^n u_t$ on the both hand sides of the eq. (\ref{evol1}) and perform some algebraic manipulations in order to get
\br
\label{bbmn2}
\pa_t [\frac{\a}{2(n+1)} u^{n+2} + \frac{u^{n+1}}{n+1}+  (1-\epsilon_1)  u^n u_{xx}]+ \frac{1}{2}\pa_x[v_x^{n-1}v_t^2 -\epsilon_1 u_t^2 u^{n-1} ] = {\cal C}_n,\,\,\,\,\,n\geq 2,\\
\label{bbmn21}
{\cal C}_n \equiv - \frac{\a}{4} \epsilon_2 w_x v_t u^{n-1}u_t +\frac{n-1}{2} v_x^{n-2}v_{xx}v_t^2-\frac{\epsilon_1 (n-1)}{2} u^{n-2} u_x u_t^2+n (1-\epsilon_1) u^n u_{xxt}. 
\er 
For $n=1$ the last eq. can be rewritten as (\ref{evoq11}) after some algebraic manipulations. Then, for $n\geq 2$ from (\ref{bbmn2}) one can define the asymptotically conserved charges
\br
\label{qqn22}
\frac{d}{dt} {\cal Q}_{a}^{(n+2)} & = & \tau_n ,\,\,\,\,\,\,\,\,\,\,\,\,\,\,\,n\geq 2 \\
 {\cal Q}_{a}^{(n+2)} & \equiv & \frac{\a^{n+1}}{2^{2n+1} 3^n}\int_{-\widetilde{x}}^{+\widetilde{x}} dx  [\frac{\a}{2(n+1)} u^{n+2} + \frac{u^{n+1}}{n+1}+  (1-\epsilon_1)  u^n u_{xx}]; \\
 \tau_n  &\equiv& \frac{\a^{n+1} }{2^{2n+2} 3^n} \int_{-\widetilde{x}}^{+\widetilde{x}} {\cal C}_n   \, dx.\label{qqn221}
\er
The anomaly density terms of ${\cal C}_n$ are odd functions under the parity transformation (\ref{paritys1})-(\ref{paritys2}); so, following similar arguments as above one can conclude that the charges  ${\cal Q}_a^{(n)}$ are asymptotically conserved.

In sec. \ref{sec:num}, we will numerically simulate the anomaly $ {\cal \alpha}_{3} $ in (\ref{qq31})  of the quasi-conservation law (\ref{qq3}) for 2-soliton and 3-soliton scatterings.  

Next, let us discuss some relationships between the lowest order charges. The density of KdV-type charge $Q_{a}^{(-5)}$ defined in (\ref{qan2})  does not depend explicitly on the deformation parameters $\{\epsilon_1, \epsilon_2\}$; however, it can be related to the charges  $\widetilde{Q}^{(2)}_a$ and $ {\cal Q}_{a}^{(3)}$ of  (\ref{qq21}) and (\ref{qq31}), respectively, and they satisfy the relationship 
\br
\label{relation}
Q_{a}^{(-5)} = \frac{1}{2^2} [ {\cal Q}_{a}^{(3)} - \widetilde{Q}^{(2)}_a].
\er
Let us compute the  first conserved charges $\widetilde{Q}_{a}^{(2)}$, ${\cal Q}_{a}^{(3)}$ and  $Q_{a}^{(-5)}$  defined in (\ref{qq21}), (\ref{qq31}) and (\ref{qan2}), respectively,  for the general 1-soliton solution presented in (\ref{solgeral}). So, one has
\br
\label{1charges}
\widetilde{Q}_{a}^{(2)} &=&\frac{k^3(5 a^2+\epsilon_1 k^2)}{15 a (a^2-\epsilon_1 k^2)^2 (2+\epsilon_2)^2} ;\\
\,\,\,\, \,\,
{\cal Q}_{a}^{(3)} &=& \frac{5 a^4 k^3 (2+\epsilon_2)-k^7 (\epsilon_1-1) \epsilon_1 (2+\epsilon_2)+ a^2 k^5 (6-\epsilon_2 - 4\epsilon_1 (2+\epsilon_2))}{15 a (a^2-k^2 \epsilon_1)^3 (2+\epsilon_2)^3}\\
Q_{a} ^{(-5)} &=&\frac{ a^2 k^5 (6-\epsilon_2) + k^7 \epsilon_1 (2+\epsilon_2)}{60 a (a^2-k^2 \epsilon_1)^3 (2+\epsilon_2)^3}.
\er
For these values one can perform a direct  verification of the relationship (\ref{relation}) for 1-soliton. In sec. \ref{sec:num} we will verify numerically the relationship (\ref{relation}) for 1-soliton.

The RLW model ($\epsilon_2 =0,  \epsilon_1 =1$) possesses only three independent conserved charges, defined by the following charge densities \cite{olver}: $u$, $u^2 + u_x^2$ and $u^2 + \frac{\a}{3} u^3$, which we have identified above. Moreover, we have found the three conserved charges of the KdV-RLW model ( $\epsilon_2 =0, \epsilon_1 \neq \{0, 1\}$ ) with charge densities: $u$, $u^2 + \epsilon_1 u_x^2$ and $u^2 + \frac{\a}{3} u^3 + (\epsilon_1-1)  u_x^2$. The above towers of infinite number of asymptotically conservation laws, (\ref{mom1}), (\ref{qqn}) and (\ref{qqn22}), respectively,  can be considered as generalizations of these charges such that each conserved charge constitutes the lowest order charge of a family of infinite tower of higher order asymptotically conserved charges. So, it seems to be that for each exact conservation law of the deformed model one can construct  a tower of related family of higher order infinite number of quasi-conservation laws.

The above charge densities and anomalies show different degrees of dependence on the parameters $\epsilon_{1,2}$, and since each tower of quasi-conservation law defines an infinite set of asymptotically conserved charges, we may argue that they probe the degree of deformation away form the usual KdV-type charges. In fact, in comparison to the KdV-type   charges, $Q^{(-2n-1)}_a$ in (\ref{can}), the charges  $\widetilde{Q}^{(n)}_a$ \, in (\ref{qqn}) and  $ {\cal Q}_{a}^{(n+2)}$ in (\ref{qqn22}) encode more accurately the property of being nearly conserved, since their densities incorporate additional terms possessing explicit dependence on the deformation parameters. Each  tower  encapsulates different degree of deformation away from KdV-type charges, so we could argue that they probe more accurately the regions of interactions of the solitons. 

\subsection{Non-local charges and mixed scale dimensions}
\label{sec:nonlocal1}

The asymptotically conserved charges we have found so far incorporate only local expressions in their charge densities; so, since the deformed  KdV (\ref{mrkdv}) possesses the nonlocal terms $v_t,\, w_x$,  one could inquire about the existence of non-local charges, e.g. those incorporating these terms and their $x$ and  $t$ derivatives. Then, in the next steps we construct nonlocal asymptotically conserved charges.     

Let us consider the evolution eq. (\ref{evol}) and  multiply it by $(n+1) F^n$ on the both sides; so, after some algebraic computations one can define the following anomalous conservation laws 
\br
\label{fn}
 \pa_t [ u  F^n ] + \pa_x [\frac{6}{\a (n+1)} F^{n+1} ] = u \pa_t F^n,\,\,\,\,\,\,n=1,2,3,....
\er
where $F$ has been defined in (\ref{evol}). Therefore, one can define
\br
\label{qnonl}
\frac{d}{dt} \hat{Q}_{nonl, a}^{(n)} &=& {\cal B}_n, \,\,\,\,\,\,\,\,n=1,2,3,...\\
  \hat{Q}_{nonl, a}^{(n)} & \equiv & \int dx u  F^n,\,\,\,\,\,\,\,{\cal B}_n \equiv \int dx  u \pa_t F^n
\label{qnonl1}
\er

The case $n=1$ of the eq. (\ref{fn}), after some algebraic manipulations, can be rewritten as (\ref{evoq1}) or (\ref{evoq11}). The anomaly $ u \pa_t F^n$ on the r.h.s. of (\ref{fn})  is an odd function  provided that the fields satisfy the parity transformations (\ref{parity1}) and the properties (\ref{paritys1})-(\ref{paritys2}). In general, one can show the vanishing of the space-time integrated anomaly $u \pa_t F^n$ on the r.h.s. of (\ref{fn}), and so, making  $|\widetilde{x}| \rightarrow +\infty $ in (\ref{qnonl})-(\ref{qnonl1}) one can write
\br
 \hat{Q}_{nonl, a}^{(n)}(\widetilde{t} \rightarrow \infty) =  \hat{Q}_{nonl, a}^{(n)}(\widetilde{t}\rightarrow - \infty).
\er
Next, let us examine the conserved charges of the deformed KdV written in terms of the $q$ field. In fact, the deformed KdV can be written as a local equation of motion in terms of this field (\ref{eqq}). So, take the $x-$derivative of the eq. (\ref{wv1}) and use the first eq. in (\ref{wxvt}) in order to get
\br
u_x = w_{xt} \Rightarrow  \frac{8}{\a} \pa_t (-q_{xx} ) - u_x =0.  \label{qxx}
\er
This eq. allows us to define the next conserved charge
\br
\bar{Q}_1= -\frac{8}{\a} \int_{-\infty}^{+\infty} dx\,  q_{xx}.  \label{Q0} 
\er
It can directly be verified by taking  
\br
\frac{d}{dt} \bar{Q}_1 &=& -\frac{8}{\a} \int_{-\infty}^{+\infty} dx\,  q_{xxt} \\ 
                             &=& u(+\infty)-u(-\infty)\\
&=&0, \er
where in the second line  the relationship (\ref{uqxt}) has been used. It is a remarkable fact that this charge is conserved for the deformed KdV model, for any set of values of the parameters $\epsilon_1$ and  $\epsilon_2$. In addition, the conservation law in (\ref{qxx}) and the charge (\ref{Q0}) can easily be constructed by taking the $x-$derivative of the eq. (\ref{eqq}). The next order quasi-conservation law can be constructed by taking firstly the $x-$derivative of (\ref{eqq}) and then by multiplying the resulting eq.  by $2 q_{xx}$. So, one can write   
\br
\label{Q2}
\pa_t \Big[q_{xx}^2 - 2 q_t q_{xxx} - (1-\epsilon_1) q_{xxx}^2 + 2 \epsilon_1 q_{xxx} q_{xxt} \Big] + \pa_x\Big[ 2 q_{xx} H + 2 q_t q_{txx} -  q_{xt}^2 - \epsilon_1 q_{xxt}^2\Big] = {\cal H}_1 \\
 {\cal H}_1 \equiv  - 4 q_{xxx} \(2 q_{xt}^2 + \epsilon_2 q_{xx} q_{tt}\),\nonumber \\
H\equiv -q_{tt} + 4 q^2_{xt} + 2 \epsilon_2  q_{xx} q_{tt} - q_{xxxt} + \epsilon_1 (q_{xxtt} + q_{xxxt}). \label{hh}
\er
Therefore, one can define the asymptotically conserved charge
\br
\frac{d}{dt} \bar{Q}_2 &=& h_1, \label{hh0}\\
 \bar{Q}_2 & = & \frac{2^{4}}{\a^2}\int\, dx \Big[q_{xx}^2 - 2 q_t q_{xxx} - (1-\epsilon_1) q_{xxx}^2 + 2 \epsilon_1 q_{xxx} q_{xxt} \Big],\,\,\,\,h_1 \equiv \int\, dx  {\cal H}_1.\label{hh1}
\er
Notice that, taking into account the symmetry property of $q$ (\ref{vwtr}), the anomaly $ {\cal H}_1$ exhibits odd parity under (\ref{parity1}).
 
The higher order quasi-conservation laws of this sequence can be constructed similarly, taking firstly the $x-$derivative of (\ref{eqq}) and then by multiplying the resulting eq.  by $(n+1) q_{xx}^{n}$. Afterwards, one can write   
\br
\nonumber
\pa_t\Big[q_{xx}^{n+1} -n(n+1) q_t q_{xxx} q_{xx}^{n-1} +\frac{n(n+1)(\epsilon_1-1)}{2} q_{xx}^{n-1} q_{xxx}^2 + n(n+1)\epsilon_1 q_{xx}^{n-1} q_{xxx} q_{xxt}  \Big]-\\
\label{hnl}
\pa_x\Big[ (n+1) H q_{xx}^n\Big] =  {\cal H}_n,\,\,\,\,\,\,\,\,\,\,\,\,\,\,\,\,\,\,\,\,n \geq 2\\
 {\cal H}_n \equiv -n(n+1) q_t \pa_t(q_{xxx} q_{xx}^{n-1})+\frac{(n-1)n(n+1)(\epsilon_1-1)}{2} q_{xx}^{n-2} q_{xxt} q_{xxx}^2 - \nonumber\\
n(n+1) q_{xx}^{n-1} q_{xxx}(4 q^2_{xt} + 2 \epsilon_2  q_{xx} q_{tt})+(n+1)\epsilon_1 \pa_x\pa_t(q_{xx}^n) q_{xxt},\label{ah}
\er
where $H$ is defined in (\ref{hh}). Notice that the anomaly $ {\cal H}_n$ exhibits odd parity under (\ref{parity1}). Therefore, the quasi-conservation law (\ref{Q2}) and the infinite tower of eqs. (\ref{hnl}), following similar discussions as above, will present asymptotically conserved charges. Moreover, when expressed in terms of the $u$ field, using the relationship (\ref{uqxt}), they will become highly non-local charges. 

In sec. \ref{sec:num}, we will numerically simulate the anomaly $ h_{1} $ in (\ref{hh1}) of the quasi-conservation law (\ref{hh0}) for 2-soliton and 3-soliton scatterings. 

\section{The mRLW theory and the quasi-conservation laws}
\label{sec:mrlw1}

For the non-integrable modified regularized long-wave (mRLW)  model ($\epsilon_1=\epsilon_2=1$) we have an analytical form of a 2-soliton solution, whose ${\cal P}$ invariant representation was obtained in (\ref{usolxt}). So, taking into account  this symmetry of the 2-soliton in (\ref{usolxt}), which satisfies the parity symmetry (\ref{parity2}), and the symmetries of the auxiliary fields in (\ref{parity3}), one can show analytically the vanishing of the anomalies belonging to the various infinite towers of new quasi-conservation laws presented above. Thus, it is an analytical proof of the quasi-integrability of the mRLW theory. Notice that similar argument has been used in order to present this proof for the KdV-like quasi-conservation laws in \cite{npb}. Here, we are generalizing this proof for the new quasi-conservation laws presented in the last section. Then, it is 
worth to mention that this adds a new strong result on the analytical proof, not only numerical,  of the quasi-integrability of a (non-integrable) theory, first discussed in \cite{npb}. 

\section{Standard KdV: quasi-conservations and anomalies for $N-$soliton}
\label{sec:standard}

Next, we show by direct construction that the standard KdV model possesses some towers of infinite number of anomalous conservation laws. In  this way, higher order analogs to the series (\ref{bbmn1}) and (\ref{fn}), respectively, will be constructed for the standard KdV model. Subsequently, it will be shown analytically the quasi-conservation of the infinite towers of anomalous charges for $N-$soliton solution satisfying the  special parity symmetry (\ref{parity1})-(\ref{paritys1}).

So, let us rewrite the eq. (\ref{mrkdv})  for $\epsilon_1=\epsilon_2=0$, as 
\br
\label{k1}
\pa_t u + \pa_x K=0,\,\,\,\,K\equiv u + \frac{\alpha}{2} u^2 + u_{xx}.
\er 
This KdV equation is not written in the standard form. The standard form of the KdV model can be written as \cite{hirota, alice}
\br
\label{k2}
\pa_t u + \alpha u u_x + u_{xxx} =0,\,\,\,\,\,\alpha=6,
\er
which can be obtained from (\ref{k1}) provided the transformation $u \rightarrow u - \frac{1}{\alpha}$ is performed.

Multiplying by $u^n$ the eq. (\ref{k1}) and rewritten conveniently one has
\br
\label{kdvmom}
\pa_t [\frac{u^{n+1}}{n+1} ] +\pa_{x} [u^n K ] = n u^{n-1} \pa_x K,\,\,\,\,\,\,n=2,3,...
\er
So, one can define the higher order moment charges and anomalies as
\br
\label{moman}
\frac{d}{dt} M_{n+1} &=& {\cal A}^{n}_{kdv}\\
M_{n+1} &=& \int_{-\infty}^{+\infty} \frac{u^{n+1}}{n+1},\,\,\,\,{\cal A}^{n}_{kdv} \equiv \int_{-\infty}^{+\infty} n u^{n-1} \pa_x K,\,\,\,\,n=2,3,...
\er 
Then, these eqs.  are the quasi-conservation laws satisfied by the higher order moments of the usual KdV model defined in (\ref{smom}). Then, following  the discussion above,  one has that for the field $u$ satisfying the property (\ref{paritys1}) under the parity transformation (\ref{parity1}) one can show  that the anomaly $[n u^{n-1} \pa_x K]$ is an odd function. So, the vanishing of these anomalies upon  integration in space and time will provide an infinite series of asymptotically conserved charges, even for the standard KdV model. We will find  below other type of  anomalous charges for the standard KdV. These new kind of  anomalous charges are expected to appear in the other quasi-integrable theories considered in the literature \cite{arxiv2}. 

Moreover, an analogous series to the  asymptotically conservation laws in (\ref{fn}) can be constructed even for the standard KdV model. So, multiplying by $K$ the eq. (\ref{k1}) and after some algebraic manipulations, one has
\br
\pa_t \Big[ \frac{1}{2} u^2 + \frac{\a}{6} u^3 -\frac{1}{2} (u_x)^2\Big] + \pa_x[\frac{1}{2} K^2] =0.
\er
This eq. is simply the linear combination of the exact conservation laws for the KdV charges $Q^{(-3)}$ and $Q^{(-5)}$ in (\ref{n1}) and (\ref{n2}), respectively, provided that one sets $X=0$ for the standard KdV.

Non-trivial anomalies appear for the higher order constructions of this sequence. Let us multiply by $\frac{1}{n+1} K^{n},\,(n\geq 2),$ the eq. (\ref{k1}); so, rewritten conveniently one gets  
\br
\label{k11}
\pa_t\Big[ \frac{1}{n+1} u K^n \Big] + \pa_x\Big[ K^{n+1} \Big] = \frac{u}{n+1} \pa_t (K^n),\,\,\,\,\,\,\,n\geq 2.  
\er
So, one can define the next anomalous charges and their relevant anomalies as
\br
\label{kdvan}
\frac{d}{dt} {\cal V}_{n} &=& {\cal B}_{kdv}^{n}\\
{\cal V}_{n} &=& \int_{-\infty}^{+\infty} \frac{1}{n+1} u K^n,\,\,\,\,{\cal B}_{kdv}^{n} \equiv \int_{-\infty}^{+\infty} \frac{1}{n+1} u \pa_t (K^{n}),\,\,\,\,n=2,3,...
\er 
By inspecting the anomaly density $[\frac{u}{n+1} \pa_t (K^n)]$ on the r.h.s. of (\ref{k11}) one  can see that is is an odd function provided that the field $u$ satisfies the parity transformations (\ref{parity1}) and (\ref{paritys1}). As discussed above, these series of quasi-conservation laws will provide a tower of anomalous conserved charges for the standard KdV model. So, these kind of anomalous charges and the higher order moments, we have described above in (\ref{kdvmom}), appear for the standard KdV model. We have mentioned above some physical consequences and certain patterns that might be expected for the higher order moments of the undeformed KdV  when evaluated for the two-soliton collisions. 

The dynamical mechanism responsible for the behavior of the anomalous charges from (\ref{k11}) for general solutions of the integrable KdV, to our knowledge, has not been studied in the literature yet. However, we will use the symmetry argument to advance the, so far,  only plausible  explanation for the presence of those set of anomalous charges and the $N-$soliton collision of the integrable KdV. So, let us construct N-soliton solutions of KdV satisfying the symmetry property (\ref{parity1}) and (\ref{paritys1}). The relevant construction of the solutions for the cases $N = 1,2 $ and $ 3$ have been discussed in \cite{npb}; however, for the case $N=3$ it has been discussed a particular case $x_\Delta=t_\Delta= 0$. Here, we follow the approach of \cite{alice} in order to construct a general $N-$soliton solution possessing the space-time parity symmetry (\ref{parity1})-(\ref{paritys1}), for any shifted point and delayed time ($x_\Delta,\,t_\Delta$) in space-time.

The Hirota's tau function for the eq. (\ref{k2}) is introduced as
\br
 u  = \frac{12}{\alpha} \pa_{x}^2 \log{\tau}. 
\er  
The $N-$soliton solutions of the equation (\ref{k2}) possesses the well known form\cite{hirota}
\br
\label{Nsol}
\tau_N = \sum_\mu \exp\(\sum_{j=1}^{N} \mu_j \G_{j} + \sum_{1\leq j< l}^{N} \mu_j \mu_l \theta_{jl}\) 
\er
where the $\mu-$summation is performed in all the permutations of $\mu_i = 0,1$, for $i=1,2,...N$, and
\br
\G_j = k_i x - w_i t + \xi_{0j},\,\,\,\,e^{\theta_{ij}} = \(\frac{k_i - k_j}{k_i + k_j}\)^2,\,\,\,\,w_i = k_i^3.
\er  
Notice that the $\xi_{0j}$ are arbitrary constants revealing the space-time translation invariance of the KdV equation, such that  each $j-$soliton component of the $N-$soliton can be located anywhere $\xi_{0j}$. In order to construct a subset of solutions possessing the space-time symmetry (\ref{parity1})-(\ref{paritys1}) we follow the method of \cite{alice}. The key idea is to make a convenient choice of the set of parameters  $\xi_{0j}$, such that the space-time translation symmetry of the solution (\ref{Nsol}) is broken. So, let us consider \cite{alice}
\br
\G_j = k_j (x-x_{\Delta}) - w_j (t-t_\D) + \eta_{0j} - \frac{1}{2} \sum_{i=1}^{j-1} \theta_{ij} -  \frac{1}{2} \sum_{i=j+1}^{N} \theta_{ji} \equiv \eta_j -  \frac{1}{2} \sum_{i=1}^{j-1} \theta_{ij} -  \frac{1}{2} \sum_{i=j+1}^{N} \theta_{ji}. 
\er
With the redefinitions above, the $N-$soliton solutions (\ref{Nsol}) take the equivalent form
\br
\label{Nsol1}
 u_N  = \frac{12}{\alpha} \pa^2_{x} \Big[\log{ \sum_{\nu} k_\nu \cosh{\(\frac{1}{2} \sum_{j=1}^{N} \nu_j \eta_j \)}}   \Big],
\er
where the summation in $\nu$ is performed in all the permutations of $\nu_i = 1, -1$, $i=1,2,...N$, and $K_\nu = \Pi_{i>j} (k_i - \nu_i \nu_j k_j)$.

Therefore, the shifted parity and delayed time inversion symmetric $N-$soliton solution is directly obtained from (\ref{Nsol1}) as
\br
\label{usym}
{\bf u}_N = u|_{\eta_{0j}=0}. 
\er
It is clear that this solution ${\bf u}_N$ will exhibit the symmetry (\ref{parity1})-(\ref{paritys1}), i.e.
\br
\label{sym0}
{\cal P} ({\bf u}_N)= {\bf u}_N.
\er
Next, we follow the above construction to write explicitly the tau functions for the cases $N=1, 2, 3$, and obtain their associated solitons and describe their main properties. 

The case $N=1$, $\tau_1 = 1 + e^{\G_1}$,  becomes
\br
\tau_1 = 2 e^{-\frac{\eta_1}{2}}\Big[\cosh{\frac{\eta_1}{2}}\Big],\,\,\,\,\eta_1 = k_1 (x-x_{\Delta}) - w_1 (t-t_\D) + \eta_{01}.
\er
Therefore, using (\ref{Nsol1}) and (\ref{usym}) one can get 
\br
{\bf u}_1 = \frac{3}{\alpha} k_1^2\, \mbox{sech}^2\Big[\frac{k_1 (x-x_\Delta) - w_1 (t-t_\D)}{2}\Big].
\er
In fact, it is an even  parity 1-soliton under ${\cal P}$, i.e. ${\cal P}({\bf u}_1)= {\bf u}_1$.

The case $N=2$ tau function becomes 
\br
\tau_2 = 1 + e^{\G_{1}} + e^{\G_2} +  e^{\G_{1} + \G_2 + \theta_{12}}, 
\er
which, taking into account the above construction, can be rewritten as 
\br
\tau_2 &=& \frac{2}{k_1-k_2} e^{(\eta_1+\eta_2)/2}\Big[ (k_1-k_2) \cosh{\(\frac{\eta_1+\eta_2}{2}\)} + (k_1+k_2)  \cosh{\(\frac{\eta_1-\eta_2}{2}\)} \Big].\\
\eta_i &=& k_i (x-x_\D) - w_i (t-t_\D) + \eta_{0i},\,\,\,\,i=1,2.
\er    
Then, it is straightforward to construct a ${\cal P}$ invariant 2-soliton as 
\br
{\bf u}_2  = \frac{12}{\alpha} \pa^2_{x} \log{ \Big[ (k_1-k_2) \cosh{\(\frac{\eta_1+\eta_2}{2}\)} + (k_1+k_2)  \cosh{\(\frac{\eta_1-\eta_2}{2}\)} \Big]}\Big|_{\eta_{01}=\eta_{02}=0}.
\er
Thus, this KdV 2-soliton solution is even under the parity transformation ${\cal P} : (\widetilde{x} , \widetilde{t} ) \rightarrow (-\widetilde{x} , -\widetilde{t})$. It is interesting to determine (from the conditions $\eta_{01}=\eta_{02}=0$) the coordinates of the special point $(x_{\D}, t_{\D})$ provided by
\br
\label{xd}
x_{\D} &=& \frac{w_1(\xi_{02} + \frac{1}{2} \theta_{12}) - w_2 (\xi_{01} + \frac{1}{2} \theta_{12})}{k_1 w_2 - k_2 w_1},\\
t_{\D} &=& \frac{k_1(\xi_{02} + \frac{1}{2} \theta_{12}) - k_2 (\xi_{01} + \frac{1}{2} \theta_{12})}{k_1 w_2 - k_2 w_1}.\label{td}
\er
Notice that this point depends on the initial space coordinate positions $x_{0i} = - \frac{\xi_{0i}}{k_i}, (i=1,2)$, assumed at initial time $t_o = 0$, of the 2-soliton components, as well as on the wave numbers $k_i,\,i=1,2$.
  
The $N=3$ case follows similarly. So, the tau function $\tau_3$ 
\br
\tau_3 =  1 + e^{\G_{1}} + e^{\G_2} +  e^{\G_{3}} + e^{\G_1 + \G_{2} + \theta_{12}} + e^{\G_1 + \G_{3} + \theta_{13}} + e^{\G_2 + \G_{3} + \theta_{23}} + e^{\G_1 + \G_{2} + \G_3 + \theta_{12}+ \theta_{13}+\theta_{23}}, 
\er 
can be rewritten as 
\br
\tau_3 &=& \frac{2 e^{(\eta_1+\eta_2+\eta_3)/2}}{(k_1-k_2)(k_1-k_3)(k_2-k_3)}\Big[ C(x,t)\Big]\\ \nonumber
    C(x,t)  &\equiv& (k_1-k_2)(k_1-k_3)(k_2-k_3) \cosh{[(\eta_1+\eta_2+\eta_3)/2]} + \nonumber\\
&&(k_1+k_2)(k_1+k_3)(k_2-k_3) \cosh{[(-\eta_1+\eta_2+\eta_3)/2]} + \nonumber\\
&&
(k_1+k_2)(k_1-k_3)(k_2+k_3) \cosh{[(\eta_1-\eta_2+\eta_3)/2]}+\nonumber\\
&& (k_1-k_2)(k_1+k_3)(k_2+k_3) \cosh{[(\eta_1+\eta_2-\eta_3)/2]},\nonumber\\
\eta_{i}&=& k_i (x-x_\D) - w_i (t-t_\D) + \eta_{0i},\,\,\,\,i=1,2,3. \nonumber  
\er
Similarly, it is straightforward to construct a ${\cal P}$ invariant 3-soliton as 
\br
{\bf u}_3  = \frac{12}{\alpha} \pa^2_{x} \log{ C(x,t)} \Big|_{\eta_{01}=\eta_{02}=\eta_{03}=0}.
\er
Clearly, this KdV 3-soliton solution is even under the parity transformation ${\cal P}({\bf u}_3 ) = {\bf u}_3 $. The point $(x_{\D}, t_{\D})$, upon imposing $\eta_{01}=\eta_{02}=\eta_{03}=0$, becomes
\br
x_{\D} &=& \frac{{\cal E}_1 w_2 - {\cal E}_2 w_1}{k_1 w_2 - k_2 w_1}\\
t_{\D} &=& \frac{{\cal E}_1 k_2 - {\cal E}_2 k_1}{k_1 w_2 - k_2 w_1}\\
{\cal E}_i &\equiv&  -(\xi_{0i} + \Theta_{i}),\,\,\,i=1,2,3\\
\Theta_{1} &\equiv& \frac{1}{2}(\theta_{12}+\theta_{13}),\,\,\Theta_{2} \equiv \frac{1}{2}(\theta_{12}+\theta_{23}),\,\,\Theta_{3} \equiv \frac{1}{2}(\theta_{13}+\theta_{23}).
\er
For the solution above the initial positions ($x_{0i}=-\frac{\xi_{0i}}{k_i},\,i=1,2$) of the first two solitons are assumed to be fixed a priori for initial time $t_0 = 0$. Then, the position of the third soliton ($x_{03}$, at $t_0 =0$) must be fixed as 
\br
x_{03} &=&-\frac{\xi_{03}}{k_3}\\
\xi_{03} &=& \frac{w_3 ({\cal E}_1 k_2 - {\cal E}_2 k_1 ) - k_3 ({\cal E}_1 w_2 - {\cal E}_2 w_1 )}{k_1 w_2 - k_2 w_1}.
\er 
Notice that the initial position of the third soliton is not completely arbitrary, but depends on the initial positions of the other two solitons $x_{01}$ and  $x_{02}$ previously fixed, as well as on the wave numbers $k_i,\, i=1,2,3$. 

Then, one must conclude that the above anomalies ${\cal A}_{kdv}^{n}$ in (\ref{moman}) and  ${\cal B}_{kdv}^{n}$ in (\ref{kdvan}) will vanish upon integration in space-time for all the N-soliton configurations (\ref{usym}), since the relevant anomaly densities possess odd parities for soliton configurations satisfying the parity symmetry (\ref{sym0}). Consequently, the quantities $M_{n}, (n=3,4,...),$ and ${\cal V}_n, (n=2,3...),$ in (\ref{moman}) and (\ref{kdvan}), respectively, are asymptotically conserved charges of the standard KdV model. Thus, the above results show the first example of an analytical, and not only numerical, demonstration of the vanishing of the anomalies for $N-$soliton, associated to an infinite series of quasi-conservation laws in soliton theory.  
 
We believe that, for deformed models, the existence of asymptotically conserved charges associated to several towers of  
infinite number of quasi-conservation laws reflects, as in the integrable soliton theories, in the special behavior of the dynamics of the deformed  model in such a way that the soliton-like solutions emerge from the scattering region basically as they have entered it. 

The above patterns will be qualitatively reproduced below in our numerical simulations of the relevant anomalies for the 2-soliton and 3-soliton interactions of the deformed KdV model, for a variety of soliton configurations and a wide range of values of the set of deformation parameters $\{\epsilon_1, \epsilon_2\}$.  

The Liouville's integrability criterion stated for a system with a finite number of degrees of freedom goes through for a system with an infinite number of degrees of freedom with convenient modifications \cite{das}. Some qualitative features of the Liouville's theorem remain true for continuous non-linear systems admitting a zero-curvature formulation (Lax representation). So, one must have an infinite number of conservation laws whose conserved charges are in involution \cite{das, faddeev}. In this context, the appearance of the novel towers of asymptotically conserved charges as above, even in the standard KdV model, are restricted to special field configurations satisfying the symmetry property (\ref{parity1}) and (\ref{paritys1}). Therefore, one can not use these types of charges, even though they are infinitely many, in order to match to the number of degrees of freedom of the KdV model. Of course, the true conserved charges hold for general field configurations, i.e. being solitonic or not. So, the relationships between the anomalous charges and the set of true conserved charges of the standard KdV model remains to be investigated. Moreover, since the quasi-conservation laws give rise to asymptotically conserved charges for the N-soliton sector of the models, it would be interesting to study the solitons of the deformed model as an $N-$body problem, in analogy to the approach followed in \cite{babelon} for the restricted sine-Gordon model. Interestingly, the relationship between the restricted SG model and the KdV model has been described in the last reference. 

In \cite{tao}, in the harmonic analysis approach, it has been introduced the method of {\sl almost conservation laws} ($I$-method) for integrable systems. In particular, they considered the KdV model and provided a rigorous proof on how the so-called {\sl almost conservation laws} can be used to recover infinitely many conserved charges that make the KdV model an integrable system. We think that our results will be useful for some analysts in order to tackle these problems and try to establish more definitive statements about the role played by the above anomalous charges for general field configurations and, then, provide some clarifications on the quasi-integrability approach to deformations of integrable systems.  

\section{Numerical treatment of the anomalies}
\label{sec:num}

We have found an analytical expression for a general one soliton solution of the model for any set of values of the deformation parameters $\{\epsilon_1, \epsilon_2\}$ in (\ref{solgeral}), and it involves itself a general dispersion relation (\ref{disp}). It reduces to the two different expressions provided in \cite{npb} when appropriate limits for the parameters  $\{\epsilon_{1}, \epsilon_2, a\}$ are chosen. The  simulations of \cite{npb} started with a slightly incorrect initial condition for the individual solitons (e.g. solitons that do not solve the RLW equation analytically for any choice of the parameters ). However, the justification to such an initial condition would be that the radiation effects are small and restricted to the initial re-adjustments. They argued that that initial radiation has not interfered considerably and the results were in good agreement with the quasi-integrability expectations. Here, we will use general exact one soliton solutions located far apart as the initial conditions in order to simulate two and three-soliton collisions for the deformed model. So, our initial conditions will avoid the emissions of radiation and we expect to provide a strong result for the quasi-conservation of the charges. 

An analytical solution for 2-soliton, for any values of $\epsilon_1$ and $\epsilon_2$, of (\ref{mrkdv}) is not known; so, we will take as an initial condition the superposition of two solitons of the general type (\ref{solgeral}) located far away from each other. So, let us consider the next linear superposition of two expressions of type (\ref{qii}) 
\br
\label{q2s}
q_{2s}(x, t) &=& q_1 \Big\{\log{ \cosh{[\frac{\zeta_1}{2 a_1}] } }+ b_1 \zeta_1 + c_1 \Big\} +  q_2 \Big\{\log{ \cosh{[\frac{\zeta_2}{2 a_2}] } }+ b_2 \zeta_2 + c_2 \Big\} ,\\
w_j &=& \frac{a_j^2 k_j + (1-\epsilon_1) k_j^3}{a_j^2- \epsilon_1 k_j^2};\,\,\,\,\, q_j =\frac{3 a_j^2}{[a_j^2+(1-\epsilon_1) k_j^2](2 + \epsilon_2)};\,\,\,\zeta_j =  k_j x - w_j t + \d_j,\,\,\,\, j =1,2;
\er
such that $a_j,\,b_j$ and $c_j, \, j=1,2,$ are arbitrary real parameters. The relevant form of each component of the field $q_{2s}(x,t)$ is presented in (\ref{qii}). Let us define the field $p(x,t)$ as
\br
\label{pdef}
p(x,t) = \frac{\pa}{\pa t} q(x,t).
\er
We plot the functions  $q_{2s}(x, t_i)$, $p_{2s}(x, t_i)$ and $u_{2s}(x, t_i) = -\frac{8}{\alpha}  \frac{\pa^2}{\pa x \pa t} q_{2s}(x, t)|_{t=t_i} $ for initial time $t_i = -17$ in the Fig. 1. The field $u_{2s}(x, t_i)$ (green) represents the initial configuration of our numerical solution of the model (\ref{mrkdv}) for two-soliton collision. Notice that the function $q_{2s}(x, t_i)$ (gray)  undergoes significant changes only around the soliton regions  from an approximately linear behavior in regions far away from the solitons, whereas the field $p_{2s}(x, t_i)$ (brown) behaves as a kink-like function around each soliton and approaches approximately constant values outside those regions, and tends to constant values asymptotically for $x \rightarrow \pm \infty$. These patterns and properties will be useful when imposing the relevant initial and boundary conditions of our numerical simulations. In the appendix \ref{app:num} we provide more details on the numerical techniques we have used. 

\begin{figure}
\centering
\label{fig1}
\includegraphics[width=2cm,scale=6, angle=0,height=6cm]{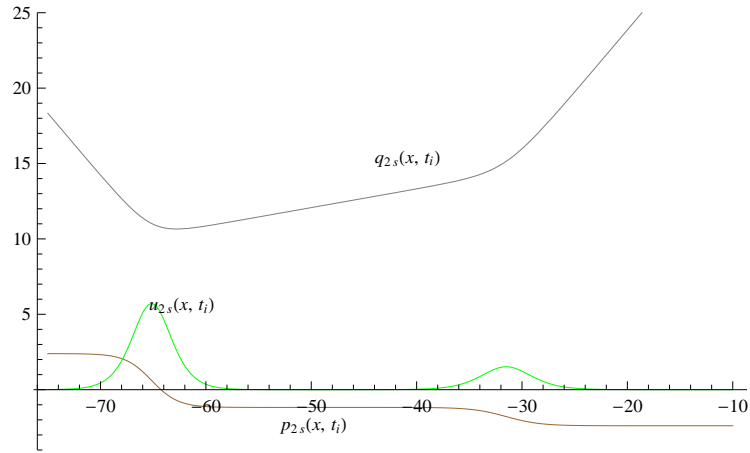} 
\parbox{6in}{\caption{(color online) The initial profiles of the fields $q_{2s}(x,t_i)$, $p_{2s}(x,t_i)$ and the 2-soliton $u_{2s}(x , t_i)$ for initial time $t_i = -17$ and parameter values  $\epsilon_1 =1.2, \epsilon_2=0.9,\,\,b_1=b_2=c_1=c_2=0,\,\, \delta_1=1,\,\,\,\, \delta_2=1, \,\,\, \alpha=1,\, a_1 = 1,\, a_2=0.8,\,\, \,\, k_1 =0.8 $\, and $k_2=0.5. $}}
\end{figure}


\begin{figure}
\centering
\label{fig2}
\includegraphics[width=2cm,scale=6, angle=0,height=6cm]{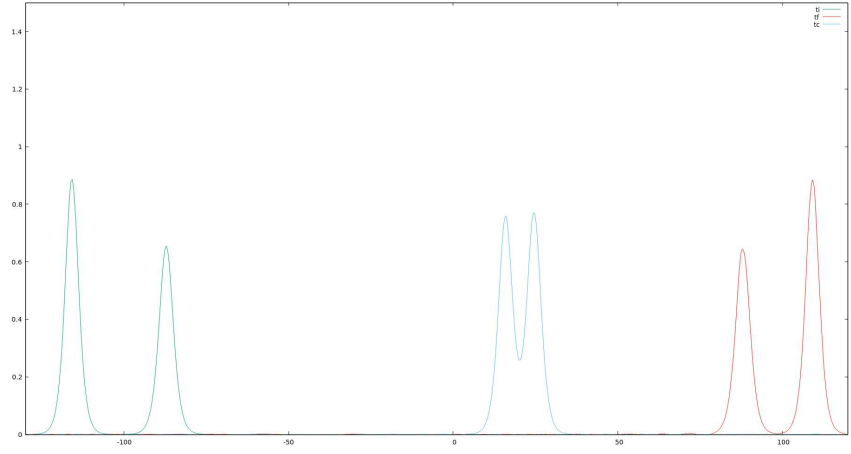} 
\parbox{6in}{\caption{(color online) Numerical simulation of 2-soliton collision for three successive times, $t_i$, before collision (green); $t_c$, collision (blue) and $t_f$, after collision (red); for the parameter values $\epsilon_1 = 1.2, \epsilon_2 = 0.9, \alpha = 4, k_1=0.75, k_2 = 0.71, a_1=a_2=1$, such that the initial condition in (\ref{q2s}) considers $\delta_1 = 87, \delta_2=62, t_i = 0$ and $b_j=c_j=0\, (j=1,2)$.}}
\end{figure}

In the Figs. 2 and 3 we present the plots of the numerical simulations of 2-soliton and 3-soliton collisions for three successive times. In these simulations we have used the LU decomposition method and considered the time steps and spatial grid as $\tau = 0.0025$ and $h=0.14$, and $\tau=0.0045$\,and  $h=0.184$, for 2-soliton and 3-soliton, respectively. Notice that, as an initial configuration, in the case of the 2-soliton we have assumed a linear superposition of two expressions of type (\ref{qii}) (see (\ref{q2s})); similarly, in the case of the 3-soliton we have assumed a linear superposition of three expressions of type (\ref{qii}), i.e. in this case one considers three solitons of the general type (\ref{solgeral}) located far away from each other. So, in the both cases one considers the general solutions (\ref{solgeral}) conveniently located some distance apart from each other. In fact, they turn out to be adequate initial conditions, since there were no visible loss of radiation in the relevant interaction regions.      

\begin{figure}
\centering
\label{fig3}
\includegraphics[width=2cm,scale=6, angle=0,height=6cm]{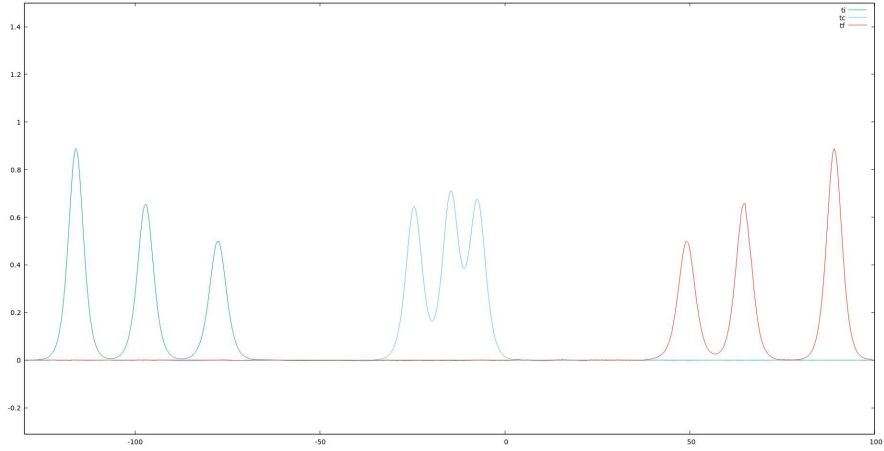} 
\parbox{6in}{\caption{(color online) Numerical simulation of 3-soliton collision for three successive times, $t_i$, before collision (green); $t_c$, collision (blue) and $t_f$, after collision (red); for the parameter values $\epsilon_1 = 1.2, \epsilon_2 = 0.9, \alpha = 4, k_1=0.75, k_2 = 0.71, k_3 = 0.67, a_1=a_2=a_3=1$,  such that the initial condition parameters are:  $\delta_1 = 87, \delta_2=69, \delta_3=52, t_i = 0$ and $b_j=c_j=0 (j=1,2,3)$.}}
\end{figure}

Moreover, in order to check our numerical methods we have verified numerically the relationship (\ref{relation}) between the conserved charges $\widetilde{Q}_{a}^{(2)}$, ${\cal Q}_{a}^{(3)}$ and  $Q_{a}^{(-5)}$  defined in (\ref{qq21}), (\ref{qq31}) and (\ref{qan2}), respectively, for the general 1-soliton solution presented in (\ref{solgeral}). The 1-soliton with the parameter values $\epsilon_1=1.2, \epsilon_2=0.9, \alpha = 3, a =1, k = 0.72$ is allowed to evolve in time, and the relevant charge densities are integrated in $x$. So, one gets $\widetilde{Q}_{a}^{(2)} = 0.116468237, {\cal Q}_{a}^{(3)}=0.184120199$\, and \, $Q_{a} ^{(-5)}=0.0169129905$, which satisfy (\ref{relation}) with good accuracy.

\subsection{2-soliton charges and anomalies}

In the following we present the simulations of four lowest order nontrivial anomalies of the towers of infinite series of quasi-conservation laws, defined in (\ref{bbmn1}), (\ref{bbmn}), (\ref{bbmn2}) and (\ref{hnl}), for the 2-soliton collision of the Fig. 2.  

For the series (\ref{bbmn1}) we will simulate the anomaly for the case $n=3$, the case $n=1$ is trivial anomaly, and for $n=2$ the relevant charge is of the KdV-type in (\ref{n1}). So, the case $n=3$ is new and so, we consider the anomaly in eq. (\ref{mom11}) for $n=3$, with density function taking the form $-6 u u_x v_t$. The Fig. 4 presents the behavior of the anomaly density versus $x-$coordinate for three successive times, before collision, during collision and after the collision of the 2-soliton presented in the Fig. 2. Notice the vanishing of the anomaly and its $t-$integrated anomaly functions of $t$, within numerical accuracy; in fact, the anomaly $\widetilde{\beta}^{(3)}(t) \approx 0$ within the order of $10^{-5}$, whereas the $t-$integrated anomaly vanishes within the order of $10^{-7}$. Therefore, according to (\ref{anoint}) and (\ref{mom11t}) one can argue that the charge $\widetilde{q}^{(3)}_a$ in (\ref{mom11}) is asymptotically conserved, i.e. it satisfies $\widetilde{q}^{(3)}_a(+\widetilde{t}) = \widetilde{q}^{(3)}_a(-\widetilde{t})$ for large time $\widetilde{t}$.

\begin{figure}
\centering
\includegraphics[scale=0.15]{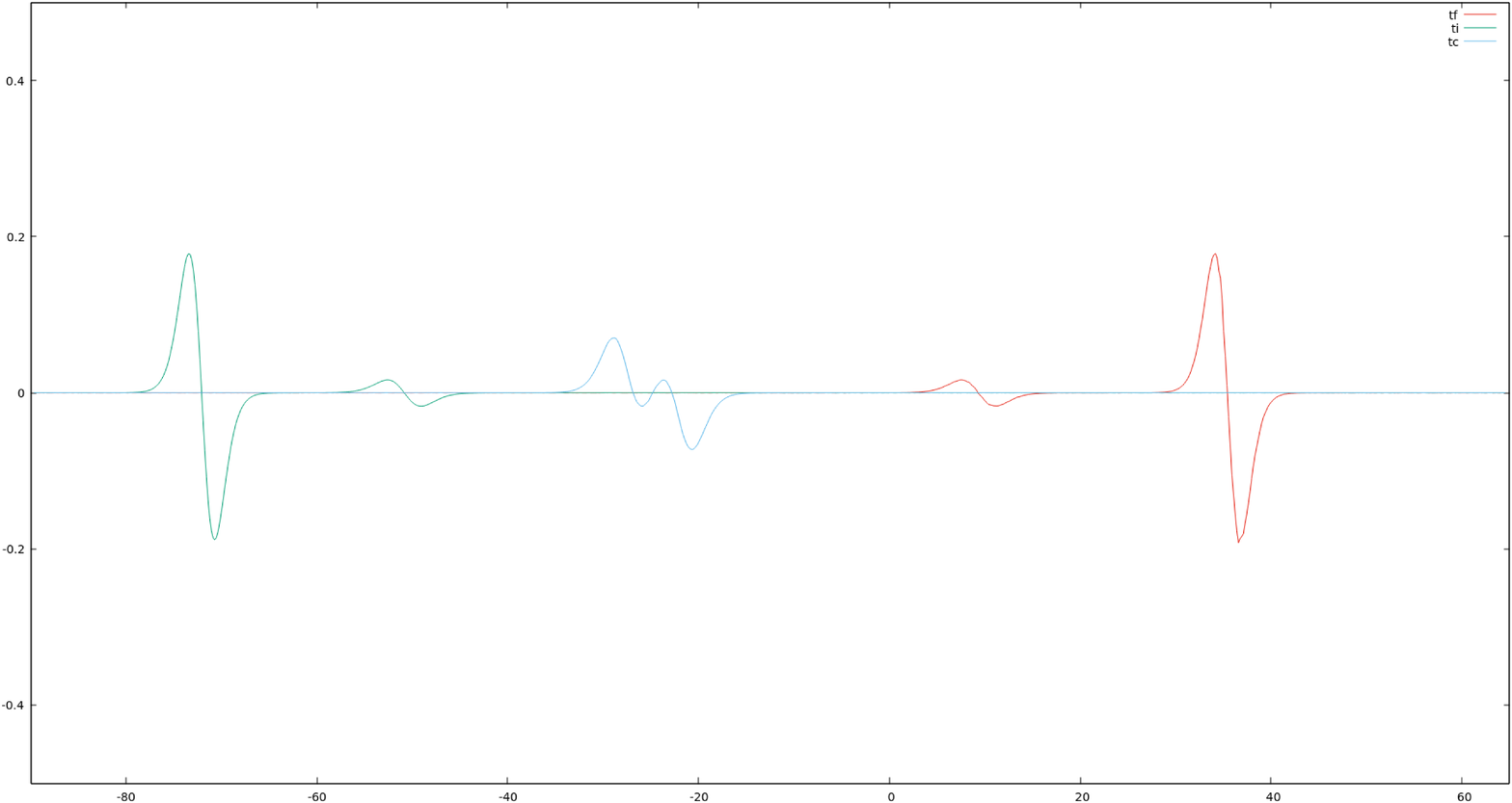}\\
\includegraphics[scale=0.13]{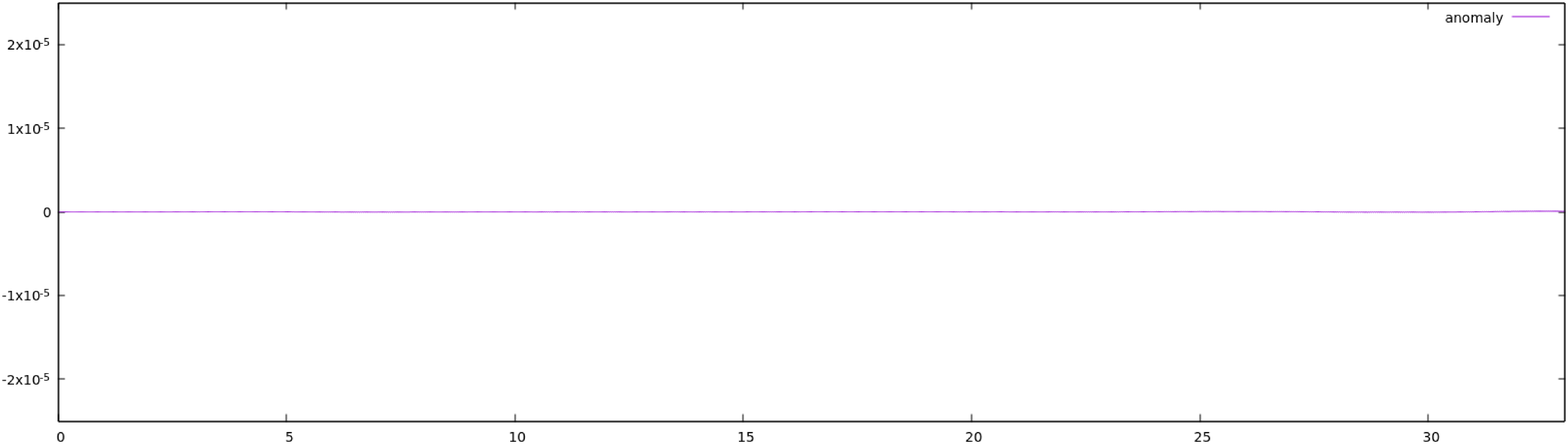}
\includegraphics[scale=0.13]{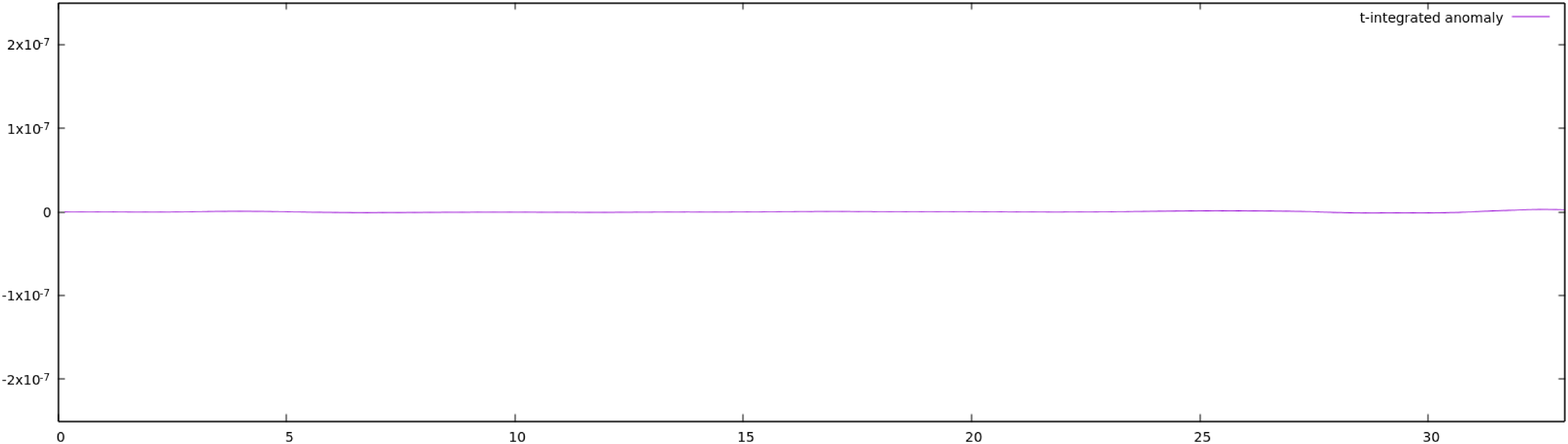}
\caption{Top Fig. shows the anomaly density of $\widetilde{\beta}^{(3)}$ (\ref{mom1}); i.e. the function $(-6 u u_x v_t)\,$ plotted in $x-$coordinate for three successive times, $t_i =$ before collision (green), $t_c=$ collision (blue) and $t_f=$ after collision (red), for the 2-soliton collision of Fig. 2. Bottom Figs. show the plots of the anomaly $\widetilde{\beta}^{(3)}(t) \, \mbox{vs} \,\, t$ and  the $t-$integrated anomaly $\int_{t_i}^{t} \widetilde{\beta}^{(3)}\,  \mbox{vs}\, \, t$, respectively.}
\label{fig4}
\end{figure} 

For the series (\ref{bbmn}) the lowest order quasi-conservation law becomes (\ref{bbm1}). So, we will simulate the anomaly $\widetilde{\alpha}_2$ in (\ref{qq21}) whose corresponding density is $w_x v_t u_x$. The Fig. 5 presents the behavior of $\widetilde{\alpha}_2(t)$ versus $x-$coordinate for three successive times, before collision, during collision and after the collision of the 2-soliton presented in the Fig. 2. Notice the vanishing of the anomaly and its $t-$integrated anomaly functions of $t$, within numerical accuracy; in fact, one has $\widetilde{\alpha}_2(t) \approx 0$ within the order of $10^{-7}$, whereas the $t-$integrated anomaly vanishes within the order of $10^{-9}$. Therefore, according to (\ref{tia}) and (\ref{asym2}) one can argue that the charge $\widetilde{q}^{(3)}_a$ in (\ref{qq21}) is asymptotically conserved, i.e. it satisfies $\widetilde{Q}^{(2)}_a(+\widetilde{t}) = \widetilde{Q}^{(2)}_a(-\widetilde{t})$ for large time $\widetilde{t}$.

\begin{figure}
\centering
\includegraphics[scale=0.15]{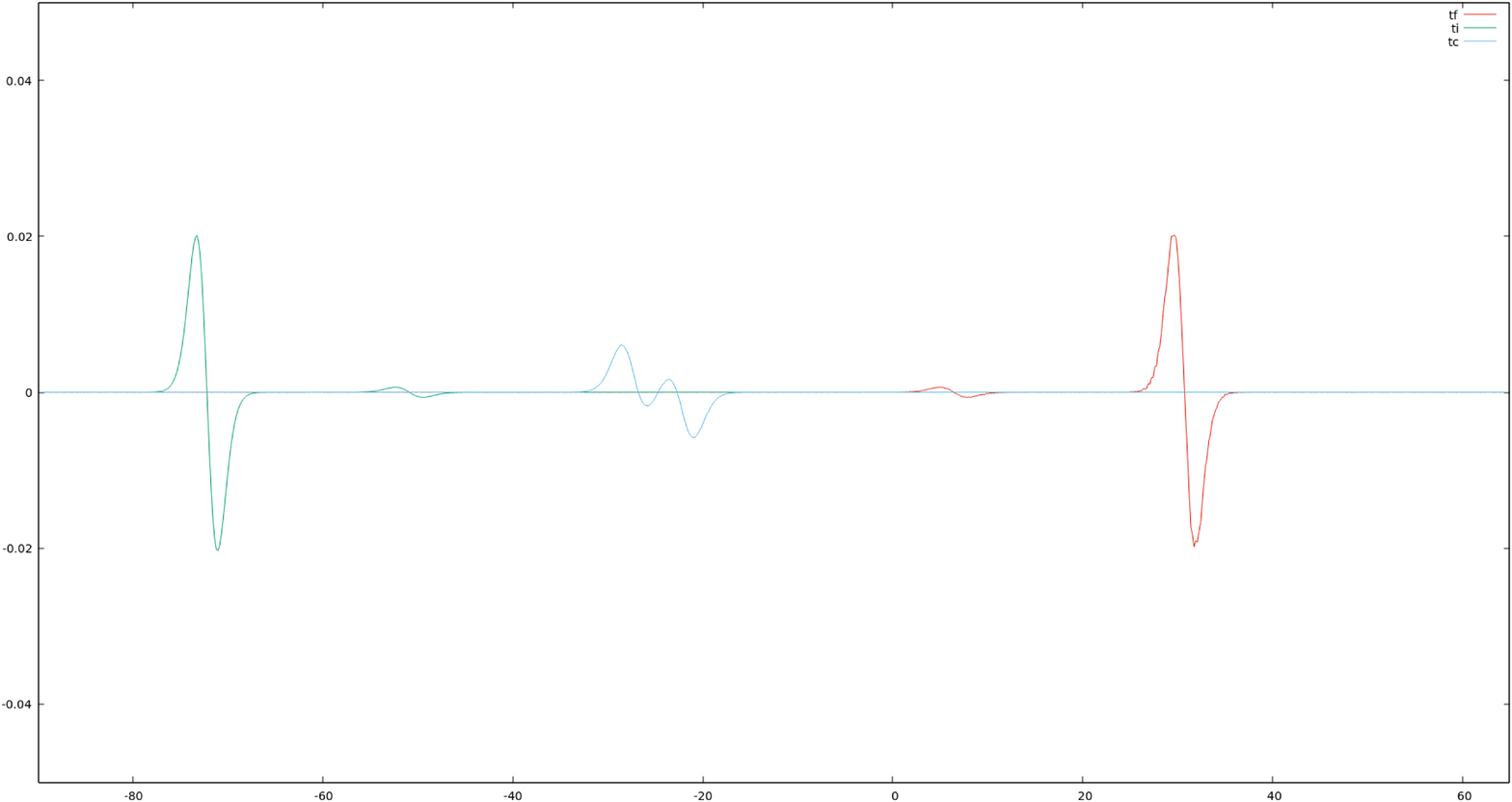}\\
\includegraphics[scale=0.13]{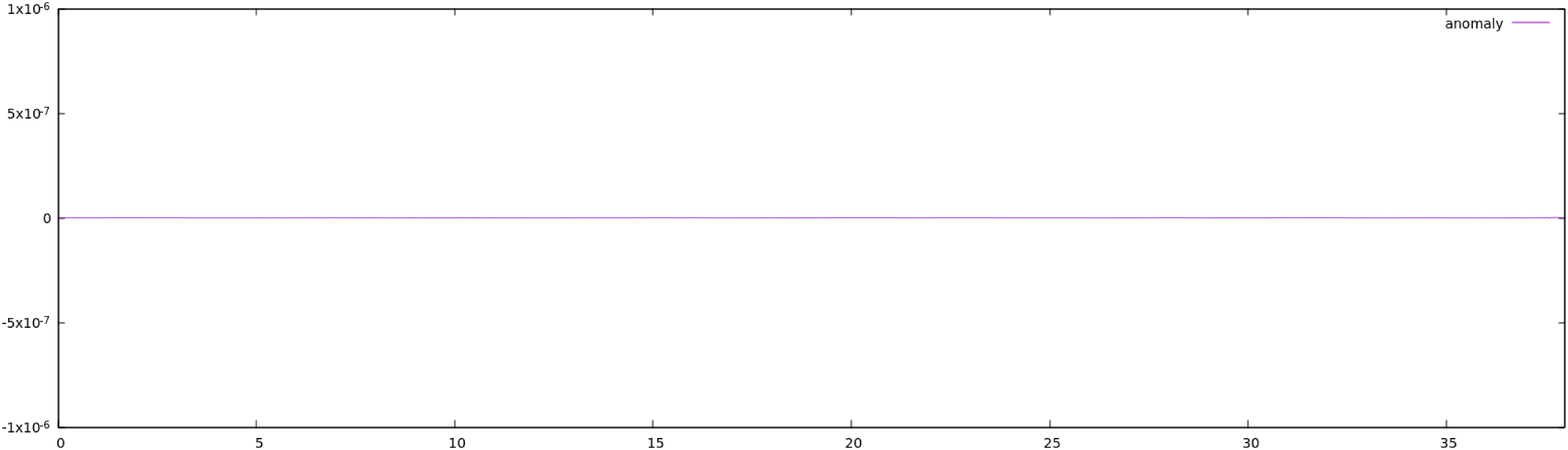}
\includegraphics[scale=0.13]{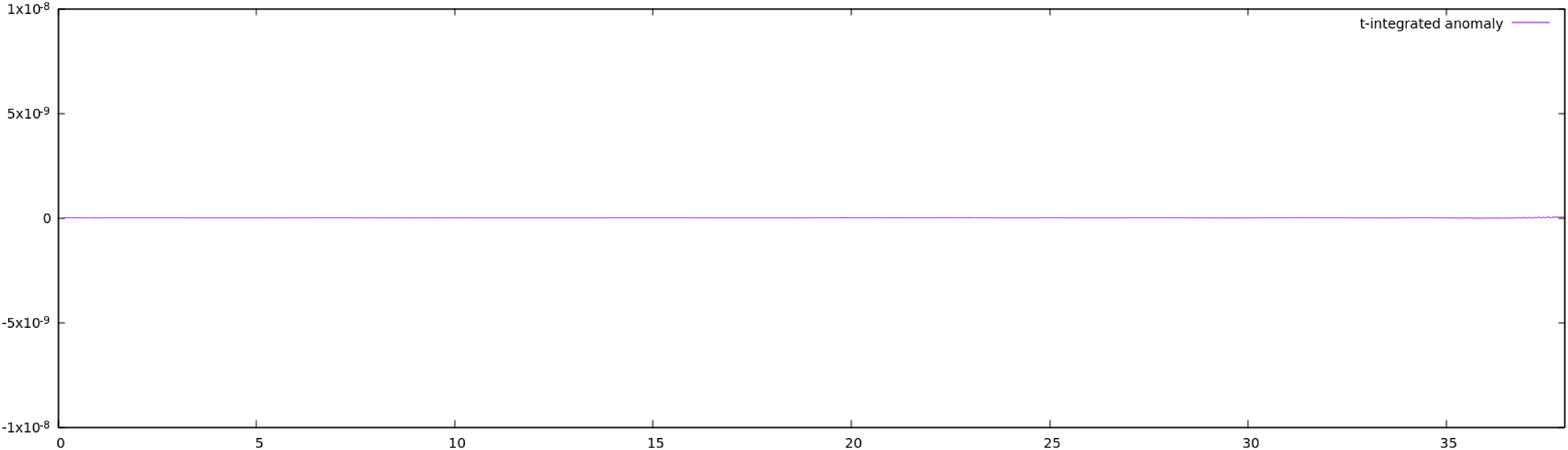}
\caption{Top Fig. shows the anomaly density of $\widetilde{\alpha}_2$ in (\ref{qq21}); i.e. the function $(w_xv_tu_x)\,$ plotted in $x-$coordinate for three successive times, $t_i =$ before collision (green), $t_c=$ collision (blue) and $t_f=$ after collision (red), for the 2-soliton of Fig. 2. Bottom Figs. show the plots of the anomaly $\widetilde{\alpha}_2 \, \mbox{vs} \,\, t$ and  the $t-$integrated anomaly $\int_{t_i}^{t} \widetilde{\alpha}_2\,  \mbox{vs}\, \, t$, respectively.}
\label{fig5}
\end{figure} 
 
Next, for the series (\ref{bbmn2}) the lowest order quasi-conservation law becomes (\ref{evoq1}). So, we will simulate the anomaly $ {\cal \alpha}_{3} $ in (\ref{qq31}) whose corresponding density is $w_x v_t u_t$. The Fig. 6 presents the behavior of $ {\cal \alpha}_{3} $ versus $x-$coordinate for three successive times, before collision, during collision and after the collision of the 2-soliton presented in the Fig. 2. Notice the vanishing of the anomaly and its $t-$integrated anomaly functions of $t$, within numerical accuracy; in fact, one has $ {\cal \alpha}_{3}  \approx 0$ within the order of $10^{-9}$, whereas the $t-$integrated anomaly vanishes within the order of $10^{-10}$. Therefore, according to (\ref{asym21}) one can argue that the charge ${\cal Q}^{(3)}_a$ in (\ref{qq31}) is asymptotically conserved, i.e. it satisfies ${\cal Q}^{(3)}_a(+\widetilde{t}) = {\cal Q}^{(3)}_a(-\widetilde{t})$ for large time $\widetilde{t}$.

\begin{figure}
\centering
\includegraphics[scale=0.15]{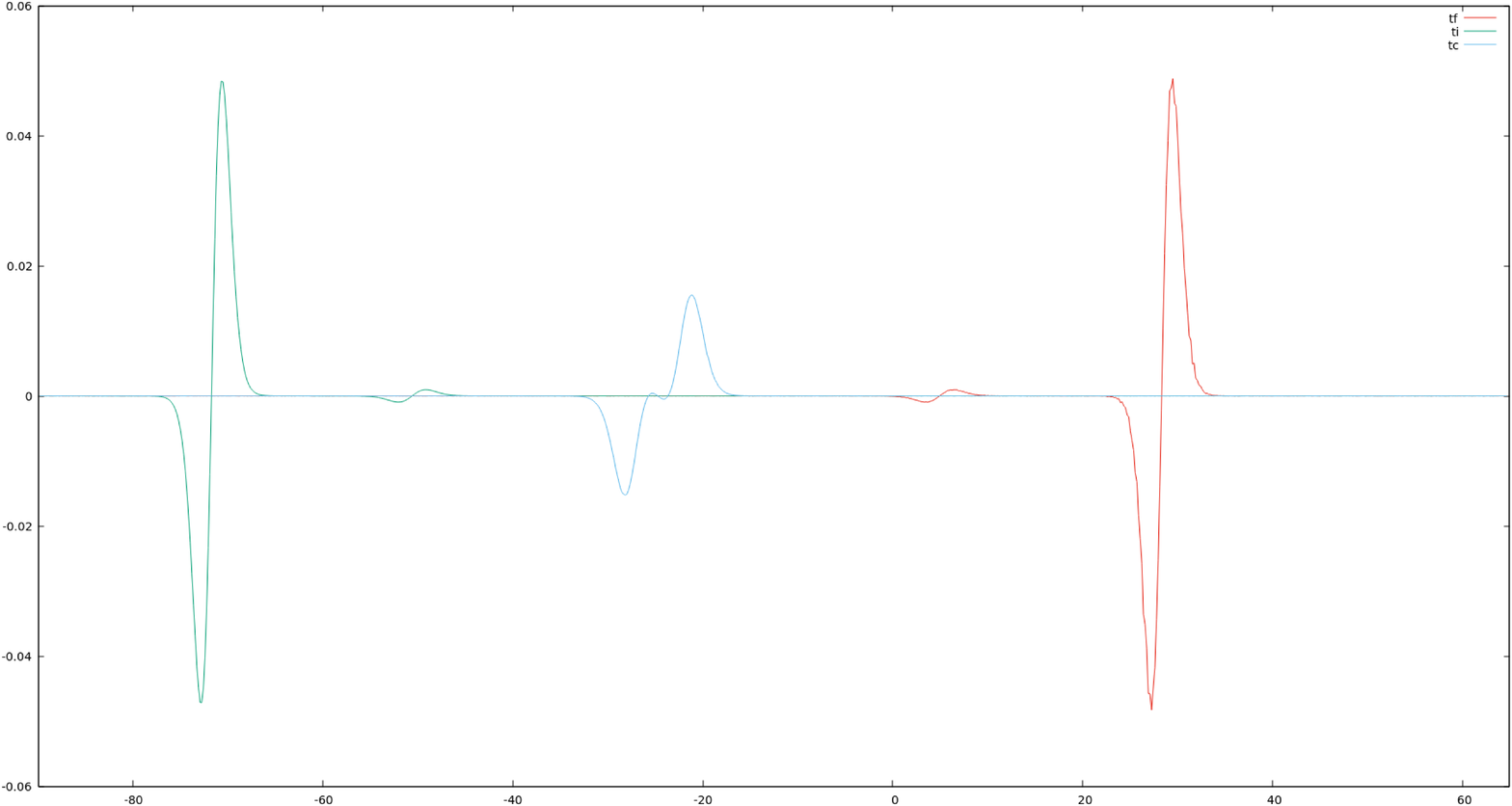}\\
\includegraphics[scale=0.1]{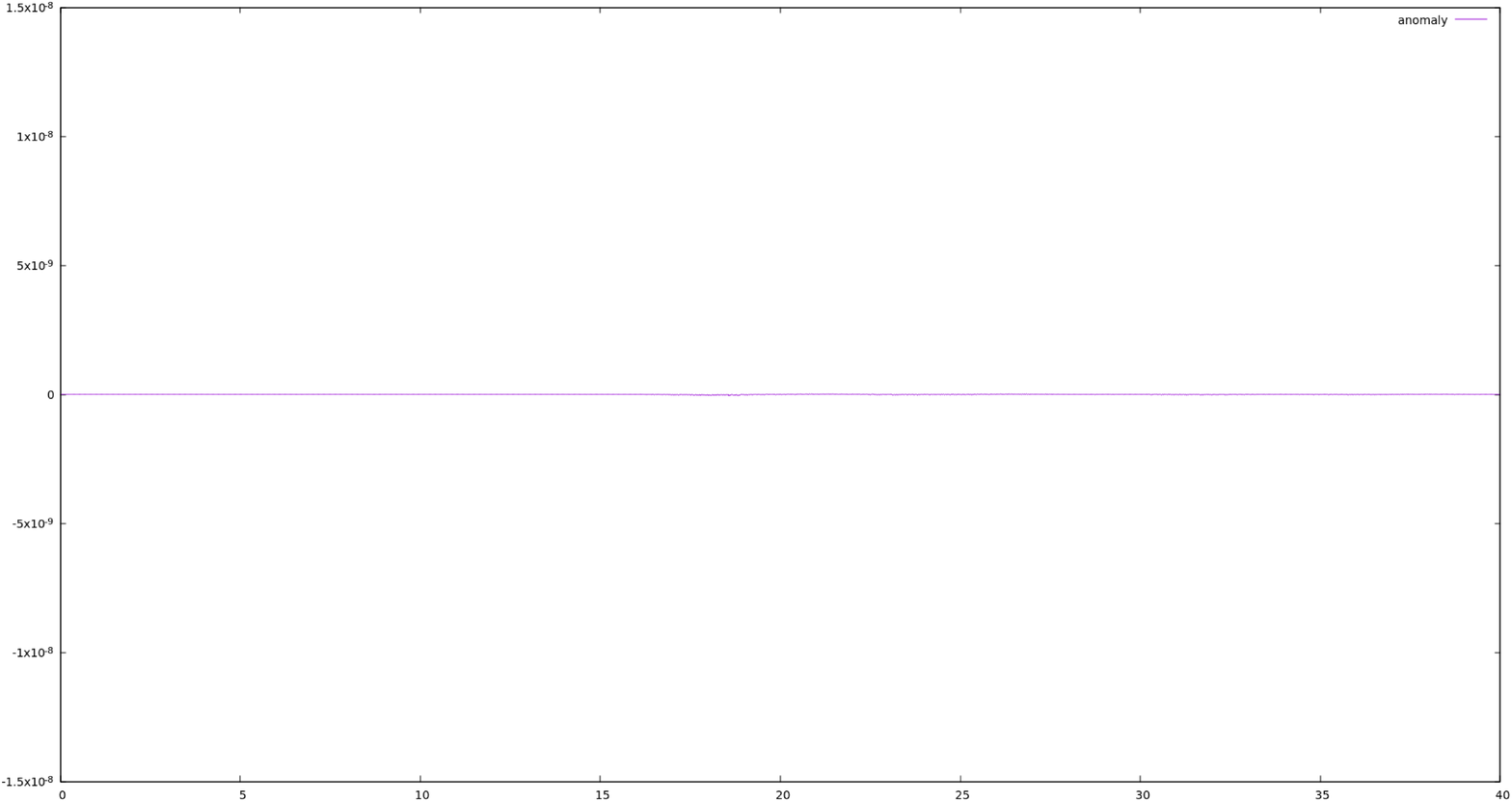}
\includegraphics[scale=0.1]{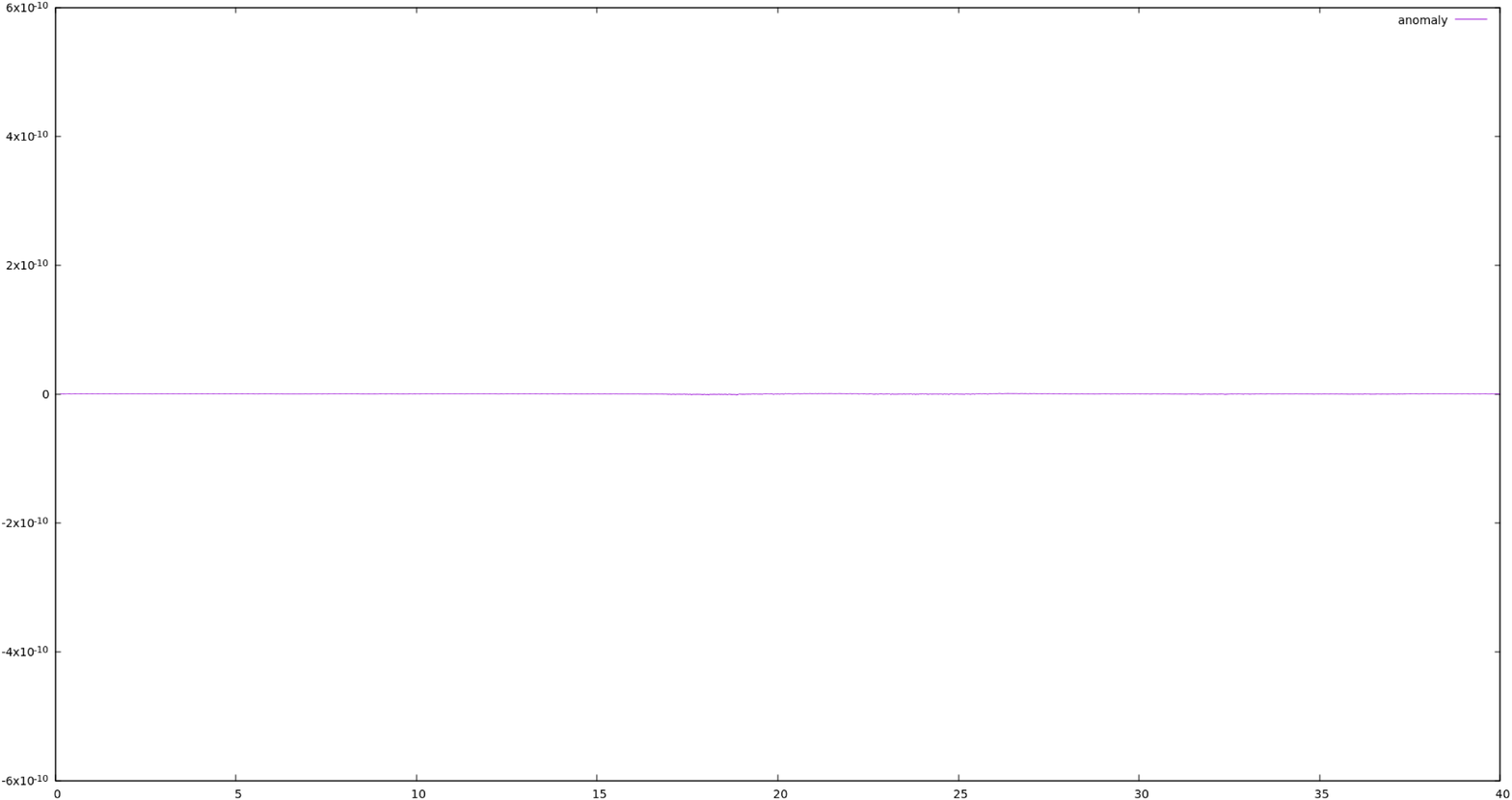}
\caption{Top Fig. shows the anomaly density of ${\cal \alpha}_{3} $ in (\ref{qq31}); i.e. the function $(w_xv_tu_t)\,$ plotted in $x-$coordinate for three successive times, $t_i =$ before collision (green), $t_c=$ collision (blue) and $t_f=$ after collision (red), for the 2-soliton of Fig. 2. Bottom Figs. show the plots of the anomaly ${\cal \alpha}_{3}  \, \mbox{vs} \,\, t$ and  the $t-$integrated anomaly $\int_{t_i}^{t} {\cal \alpha}_{3} \,  \mbox{vs}\, \, t$, respectively.}
\label{fig6}
\end{figure} 
 
Similarly, for the series (\ref{hnl}) the lowest order quasi-conservation law becomes (\ref{Q2}). So, we will simulate the anomaly $ h_{1} $ in (\ref{hh1}) whose corresponding density is $- 4 q_{xxx} \(2 q_{xt}^2 + \epsilon_2 q_{xx} q_{tt}\)$. The Fig. 7 presents the behaviour of $ h_{1} $ versus $x-$coordinate for three successive times, before collision, during collision and after the collision of the 2-soliton presented in the Fig. 2. Notice the vanishing of the anomaly and its $t-$integrated anomaly functions of $t$, within numerical accuracy; in fact, one has $ h_{1}  \approx 0$ within the order of $10^{-8}$, whereas the $t-$integrated anomaly vanishes within the order of $10^{-9}$. Therefore, as in the previous discussions one can argue that the charge $\bar{Q}_2$ in (\ref{hh1}) is asymptotically conserved, i.e. one has  $\bar{Q}_2(+\widetilde{t}) = \bar{Q}_2(-\widetilde{t})$ for large time $\widetilde{t}$.
 
\begin{figure}
\centering
\includegraphics[scale=0.15]{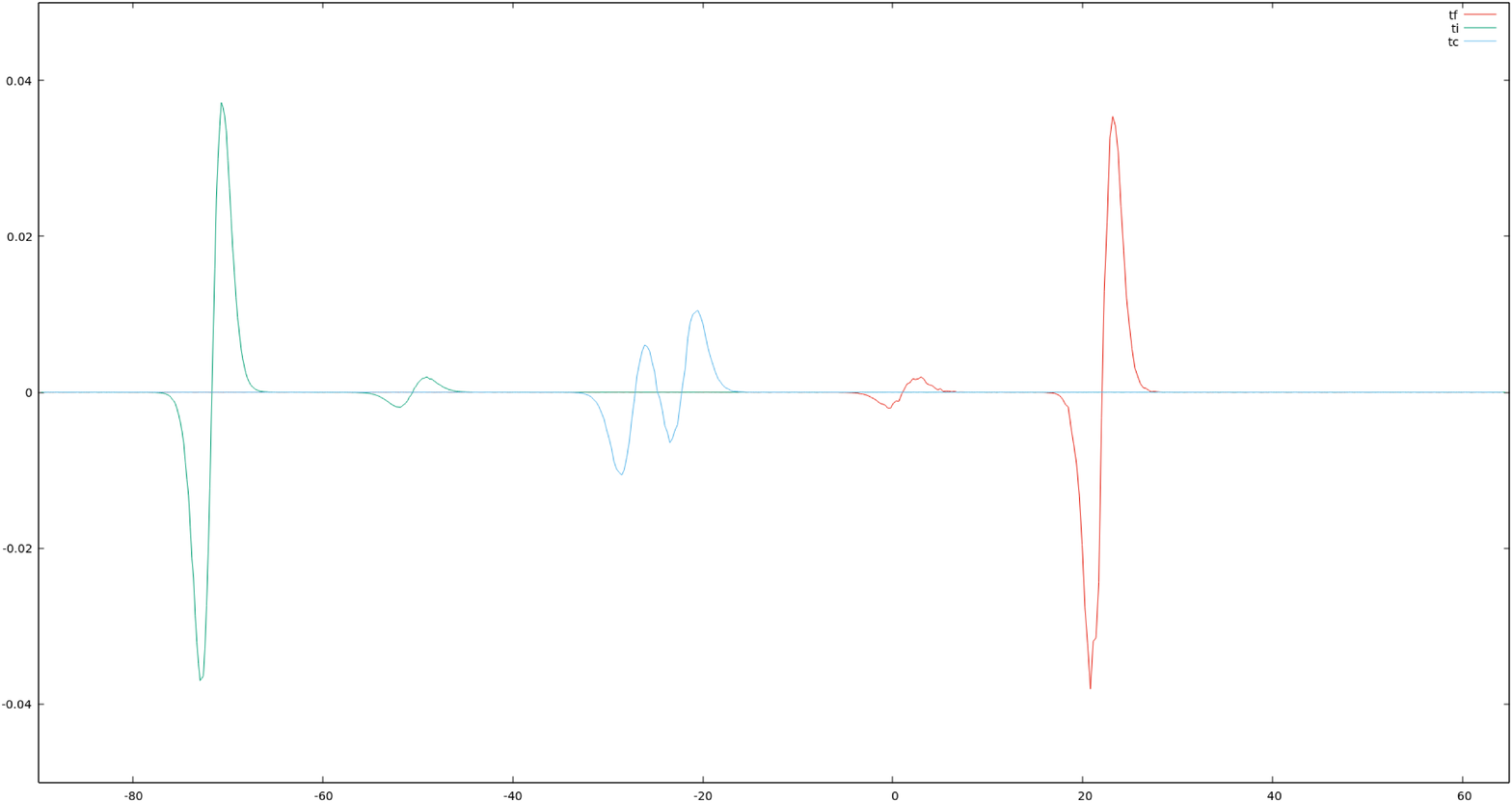}\\
\includegraphics[scale=0.1]{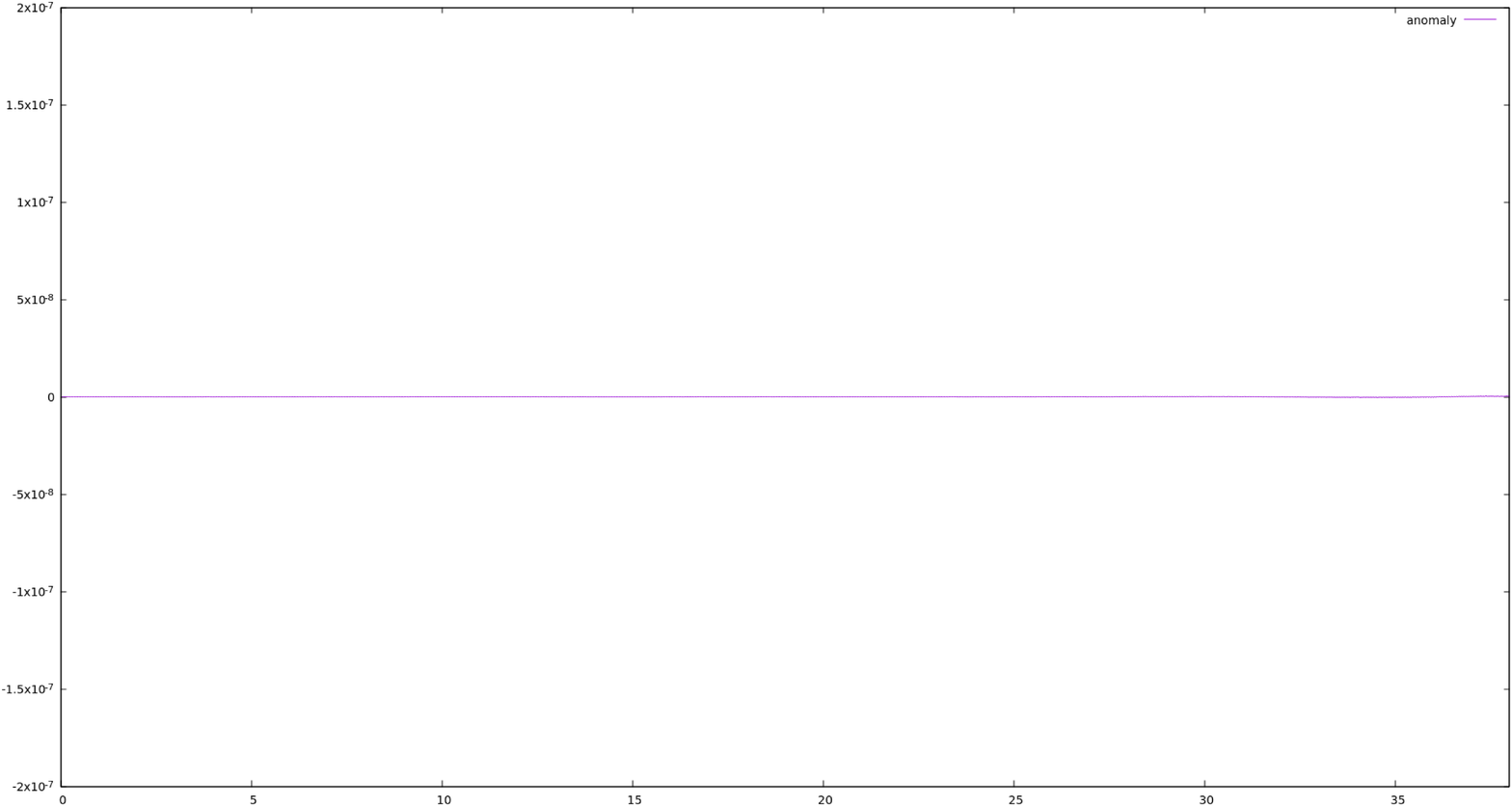}
\includegraphics[scale=0.1]{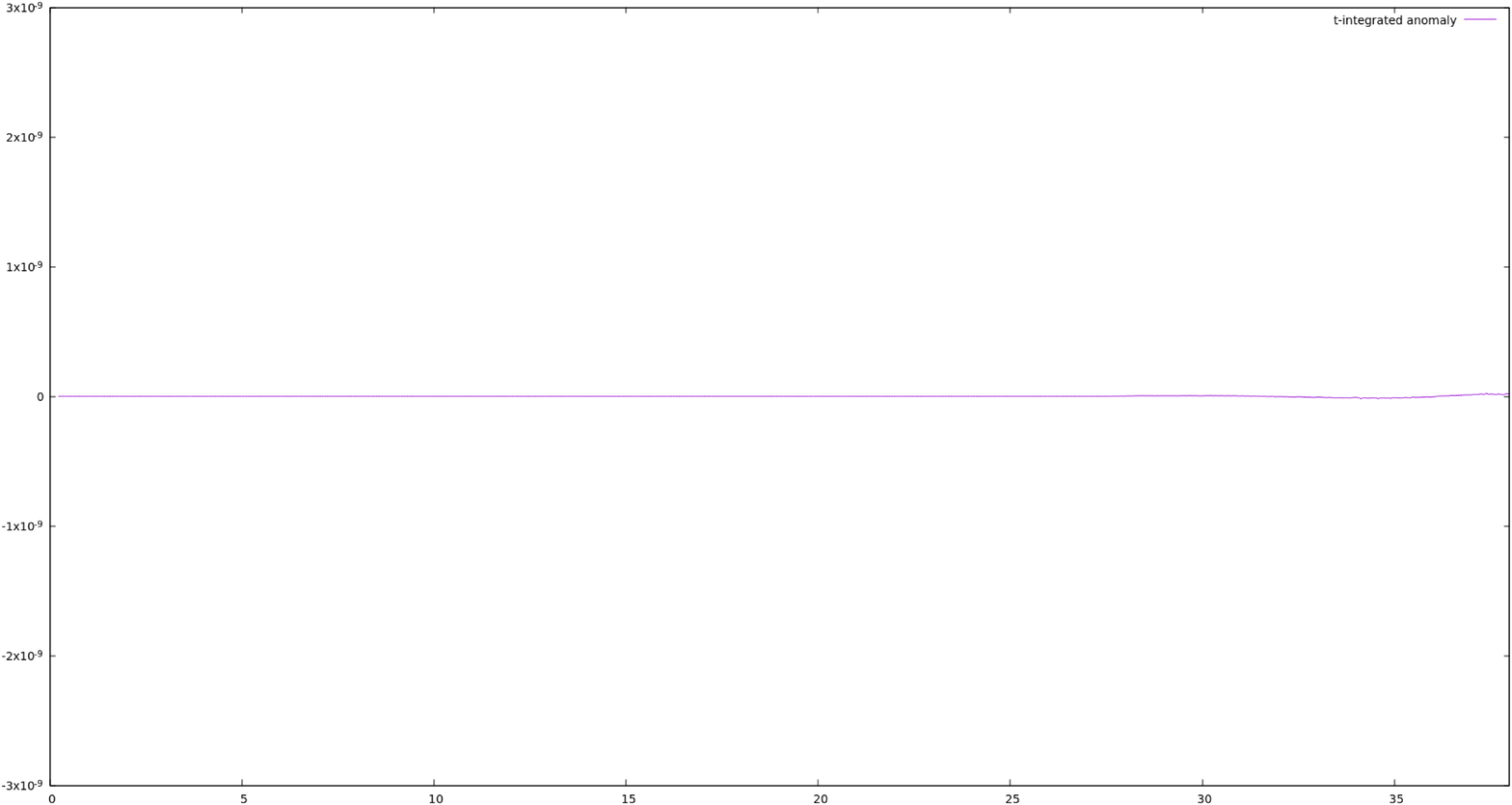}
\caption{Top Fig. shows the anomaly density of $ h_{1} $ in (\ref{hh1}); i.e. the function $- 4 q_{xxx} \(2 q_{xt}^2 + \epsilon_2 q_{xx} q_{tt}\)\,$ plotted in $x-$coordinate for three successive times, $t_i =$ before collision (green), $t_c=$ collision (blue) and $t_f=$ after collision (red), for the 2-soliton of Fig. 2. Bottom Figs. show the plots of the anomaly $ h_{1}  \, \mbox{vs} \,\, t$ and  the $t-$integrated anomaly $\int_{t_i}^{t}  h_{1} \,  \mbox{vs}\, \, t$, respectively.}
\label{fig7}
\end{figure} 

\subsection{3-soliton charges and anomalies}

Likewise, in this subsection we present the simulations of four lowest order nontrivial anomalies of the towers of infinite series of quasi-conservation laws, defined in (\ref{bbmn1}), (\ref{bbmn}), (\ref{bbmn2}) and (\ref{hnl}), for the 3-soliton collision of the Fig. 3.

In the Fig. 8 we consider the anomaly $\widetilde{\beta}^{(3)}$ in eq. (\ref{mom11}), with density function taking the form $-6 u u_x v_t$. We plot the anomaly density versus $x-$coordinate for three successive times, before collision, during collision and after the collision of the 3-soliton presented in the Fig. 3. Notice the vanishing of the anomaly and its $t-$integrated anomaly functions of $t$, within numerical accuracy; in fact, the anomaly $\widetilde{\beta}^{(3)}(t) \approx 0$ within the order of $10^{-6}$, whereas the $t-$integrated anomaly vanishes within the order of $10^{-7}$. Therefore, according to (\ref{anoint}) and (\ref{mom11t}) one has that the charge $\widetilde{q}^{(3)}_a$ in (\ref{mom11}) is asymptotically conserved for the collision of  three solitons, i.e.  $\widetilde{q}^{(3)}_a(+\widetilde{t}) = \widetilde{q}^{(3)}_a(-\widetilde{t})$ for large time $\widetilde{t}$.

\begin{figure}
\centering
\includegraphics[scale=0.15]{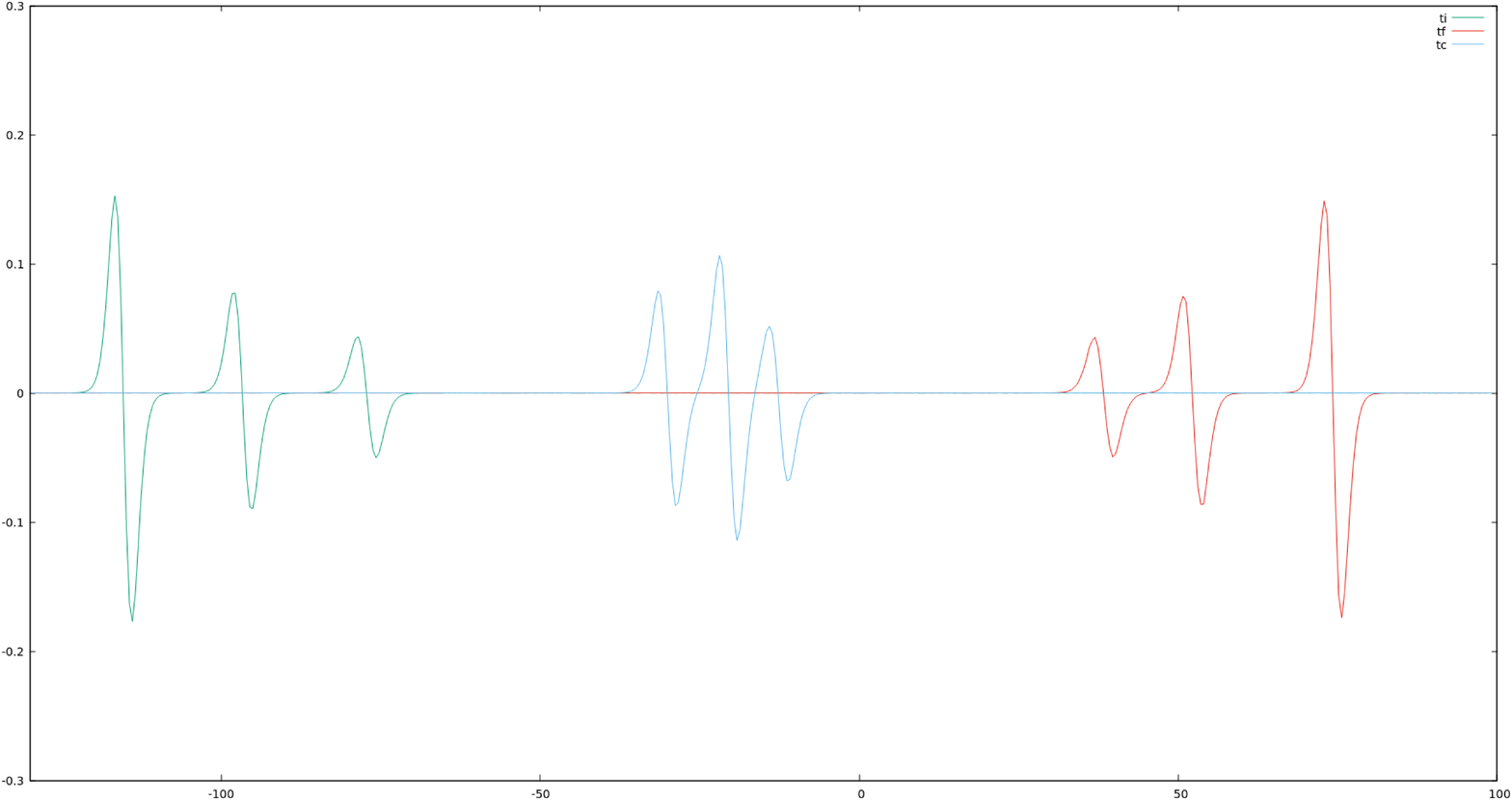}\\
\includegraphics[scale=0.1]{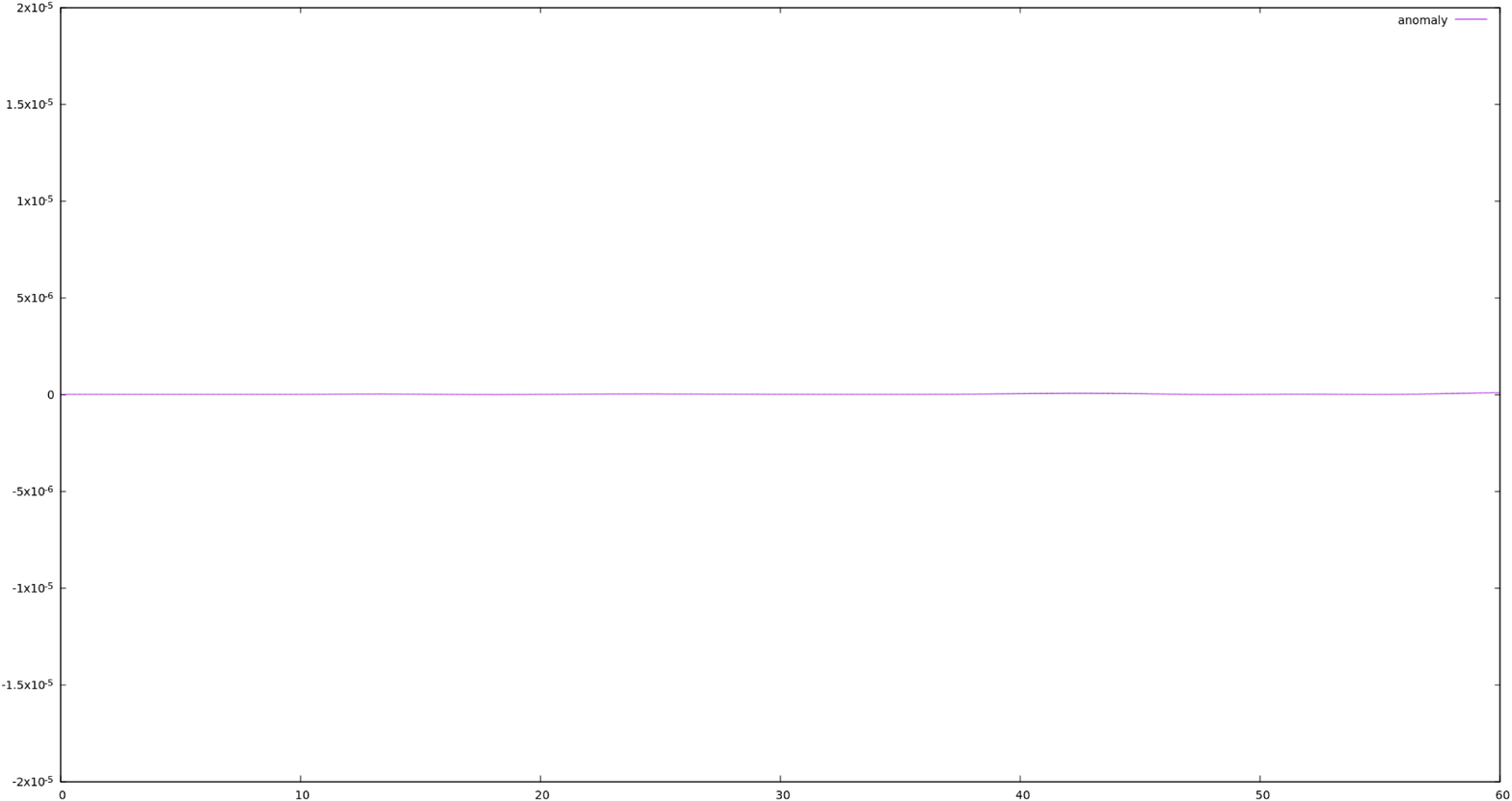}
\includegraphics[scale=0.1]{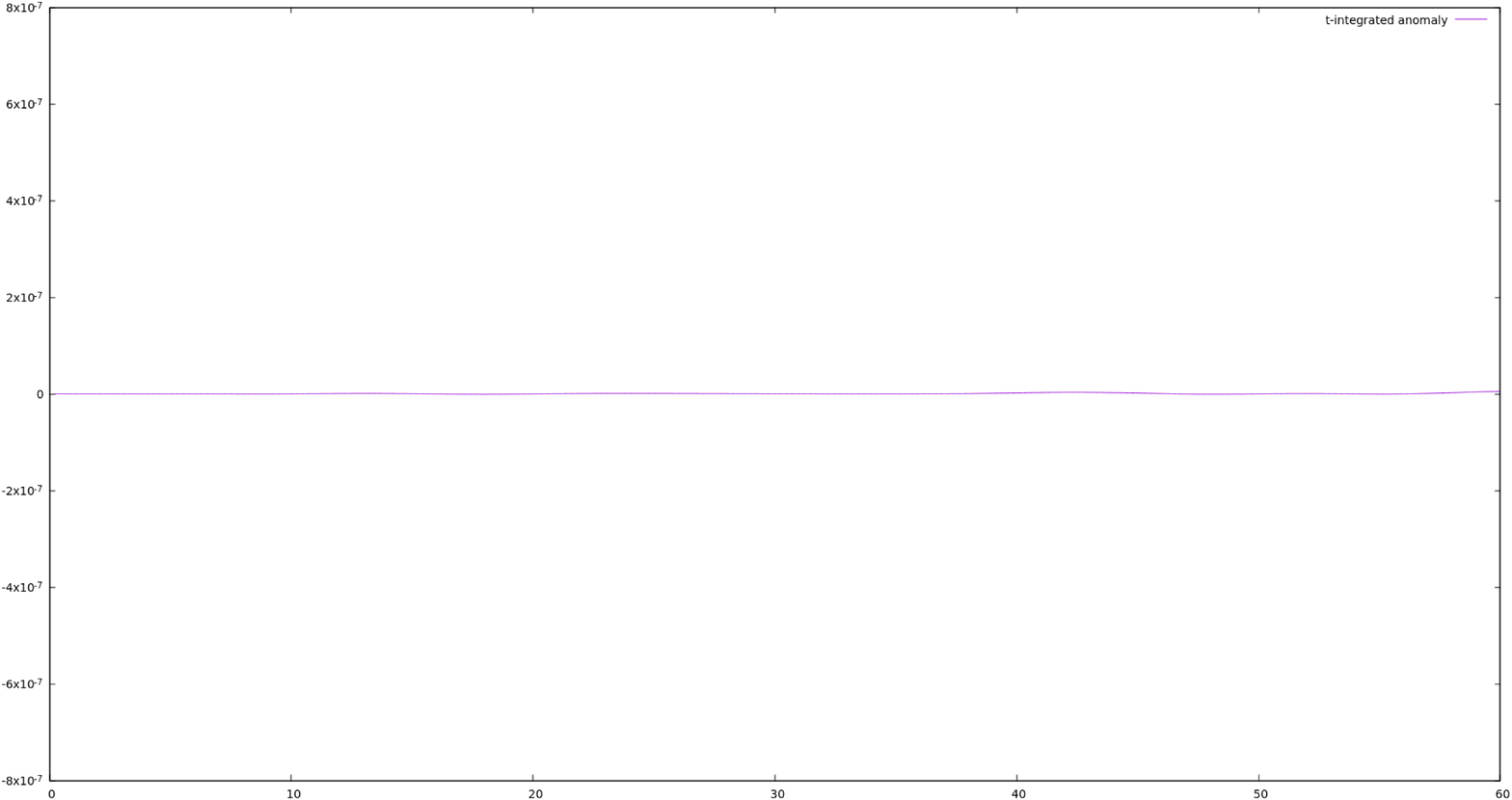}
\caption{Top Fig. shows the anomaly density of $\widetilde{\beta}^{(3)}$ (\ref{mom1}); i.e. the function $(-6 u u_x v_t)\,$ plotted in $x-$coordinate for three successive times, $t_i =$ before collision (green), $t_c=$ collision (blue) and $t_f=$ after collision (red), for the 3-soliton collision of Fig. 3. Bottom Figs. show the plots of the anomaly $\widetilde{\beta}^{(3)}(t) \, \mbox{vs} \,\, t$ and  the $t-$integrated anomaly $\int_{t_i}^{t} \widetilde{\beta}^{(3)}\,  \mbox{vs}\, \, t$, respectively.}
\label{fig8}
\end{figure} 

In Fig. 9 we simulate the anomaly $\widetilde{\alpha}_2$ in (\ref{qq21}) whose corresponding density is $w_x v_t u_x$. It is plotted $\widetilde{\alpha}_2(t)$ versus $x-$coordinate for three successive times, before collision, during collision and after the collision of the 3-soliton presented in the Fig. 3. Notice the vanishing of the anomaly and its $t-$integrated anomaly functions of $t$, within numerical accuracy; in fact, one has $\widetilde{\alpha}_2(t) \approx 0$ within the order of $10^{-6}$, whereas the $t-$integrated anomaly vanishes within the order of $10^{-7}$. Therefore, according to (\ref{tia}) and (\ref{asym2}) the charge $\widetilde{q}^{(3)}_a$ in (\ref{qq21}) is asymptotically conserved for the collision of  three solitons, i.e. $\widetilde{Q}^{(2)}_a(+\widetilde{t}) = \widetilde{Q}^{(2)}_a(-\widetilde{t})$ for large $\widetilde{t}$.

\begin{figure}
\centering
\includegraphics[scale=0.15]{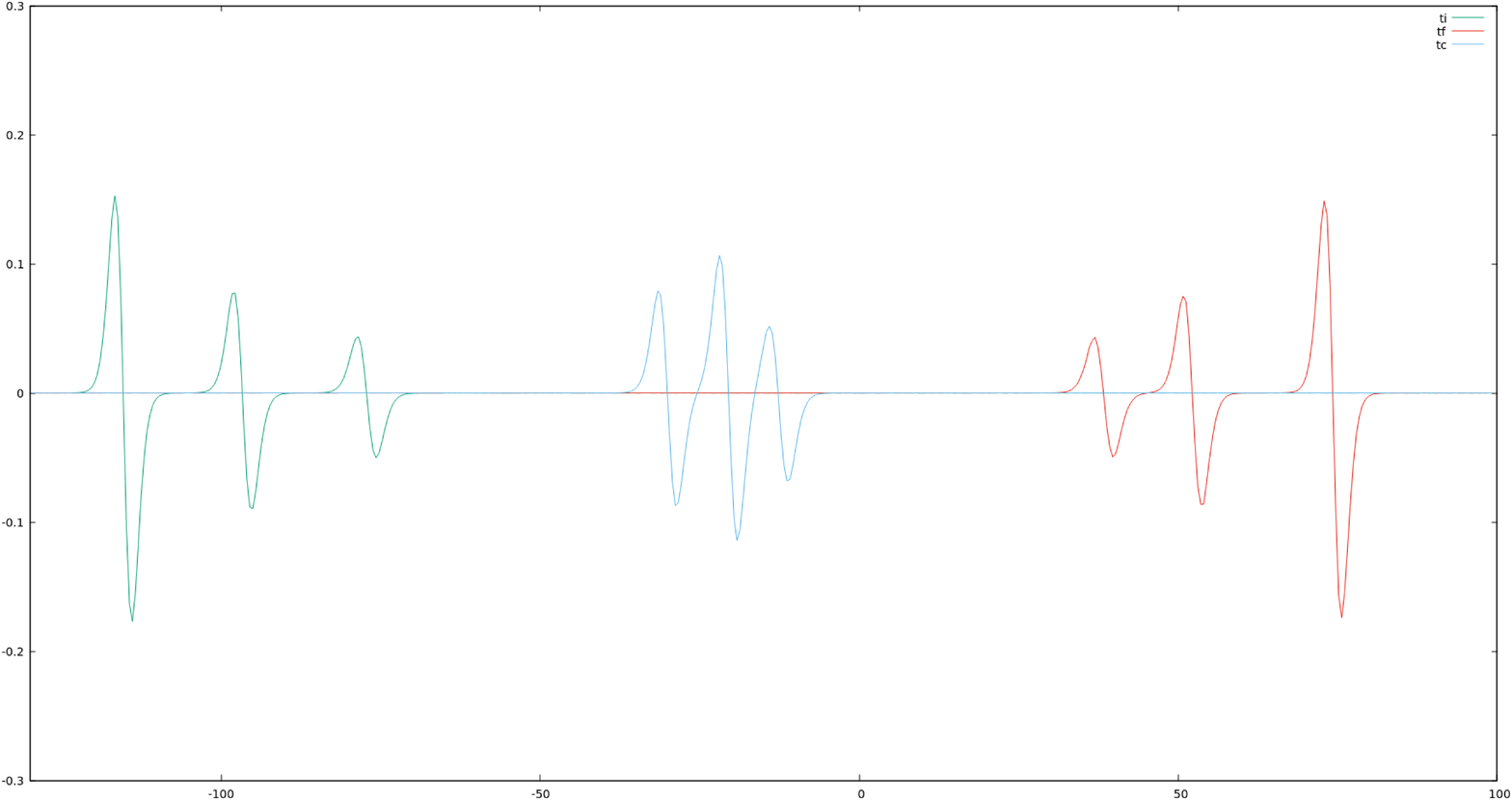}\\
\includegraphics[scale=0.1]{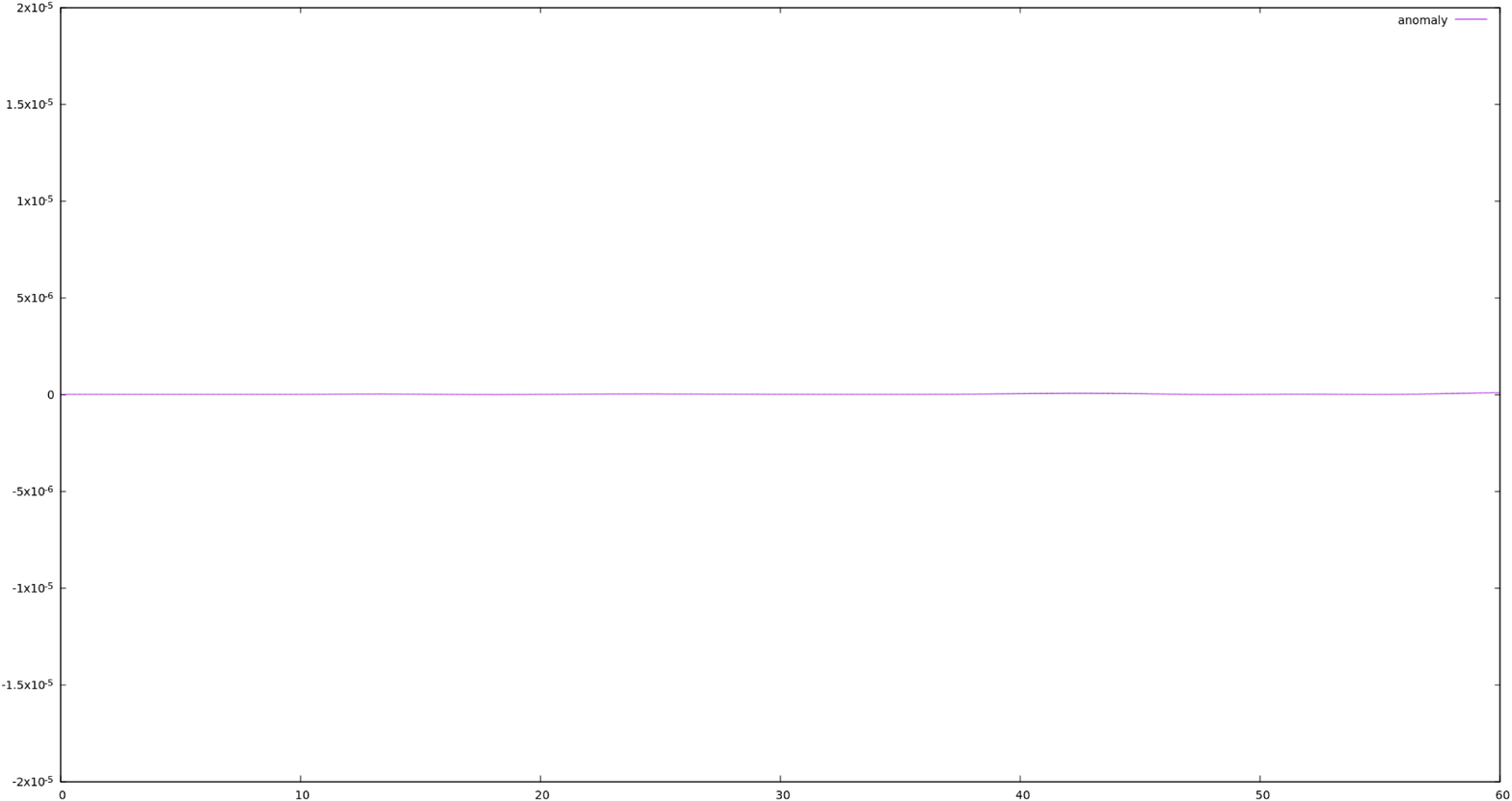}
\includegraphics[scale=0.1]{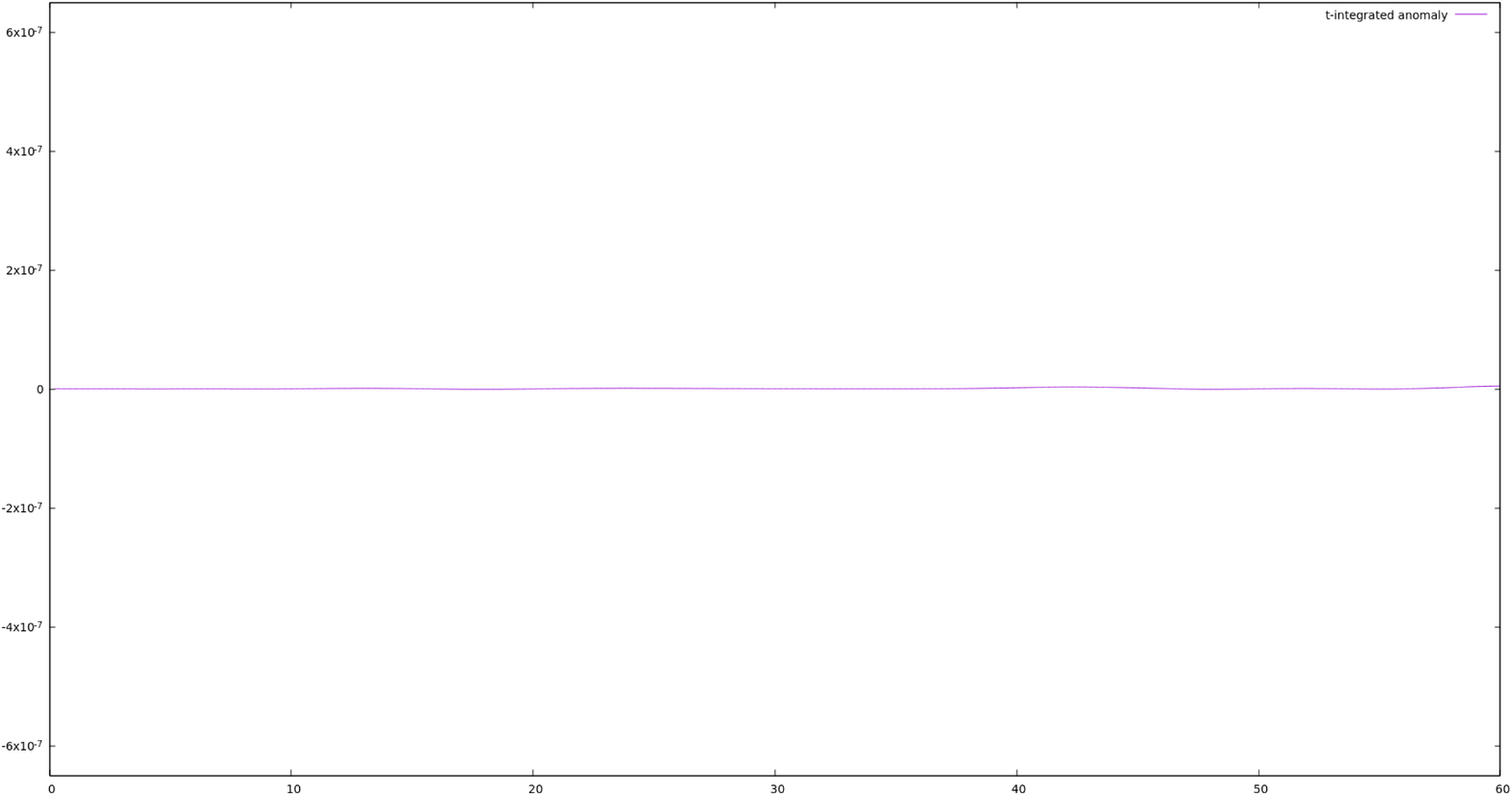}
\caption{Top Fig. shows the anomaly density of $\widetilde{\alpha}_2$ in (\ref{qq21}); i.e. the function $(w_xv_tu_x)\,$ plotted in $x-$coordinate for three successive times, $t_i =$ before collision (green), $t_c=$ collision (blue) and $t_f=$ after collision (red), for the 3-soliton of Fig. 3. Bottom Figs. show the plots of the anomaly $\widetilde{\alpha}_2 \, \mbox{vs} \,\, t$ and  the $t-$integrated anomaly $\int_{t_i}^{t} \widetilde{\alpha}_2\,  \mbox{vs}\, \, t$, respectively. }
\label{fig9}
\end{figure}

Fig. 10 presents the anomaly $ {\cal \alpha}_{3} $ in (\ref{qq31}) whose corresponding density is $w_x v_t u_t$. It presents the behavior of $ {\cal \alpha}_{3} $ versus $x-$coordinate for three successive times, before collision, during collision and after the collision of the 3-soliton presented in the Fig. 3. Notice the vanishing of the anomaly and its $t-$integrated anomaly functions of $t$, within numerical accuracy; in fact, one has $ {\cal \alpha}_{3}  \approx 0$ within the order of $10^{-8}$, whereas the $t-$integrated anomaly vanishes within the order of $10^{-10}$. Therefore, according to (\ref{asym21}), the charge  ${\cal Q}^{(3)}_a$ in (\ref{qq31}) is asymptotically conserved for the collision of  three solitons, i.e. ${\cal Q}^{(3)}_a(+\widetilde{t}) = {\cal Q}^{(3)}_a(-\widetilde{t})$ for large $\widetilde{t}$.

\begin{figure}
\centering
\includegraphics[scale=0.15]{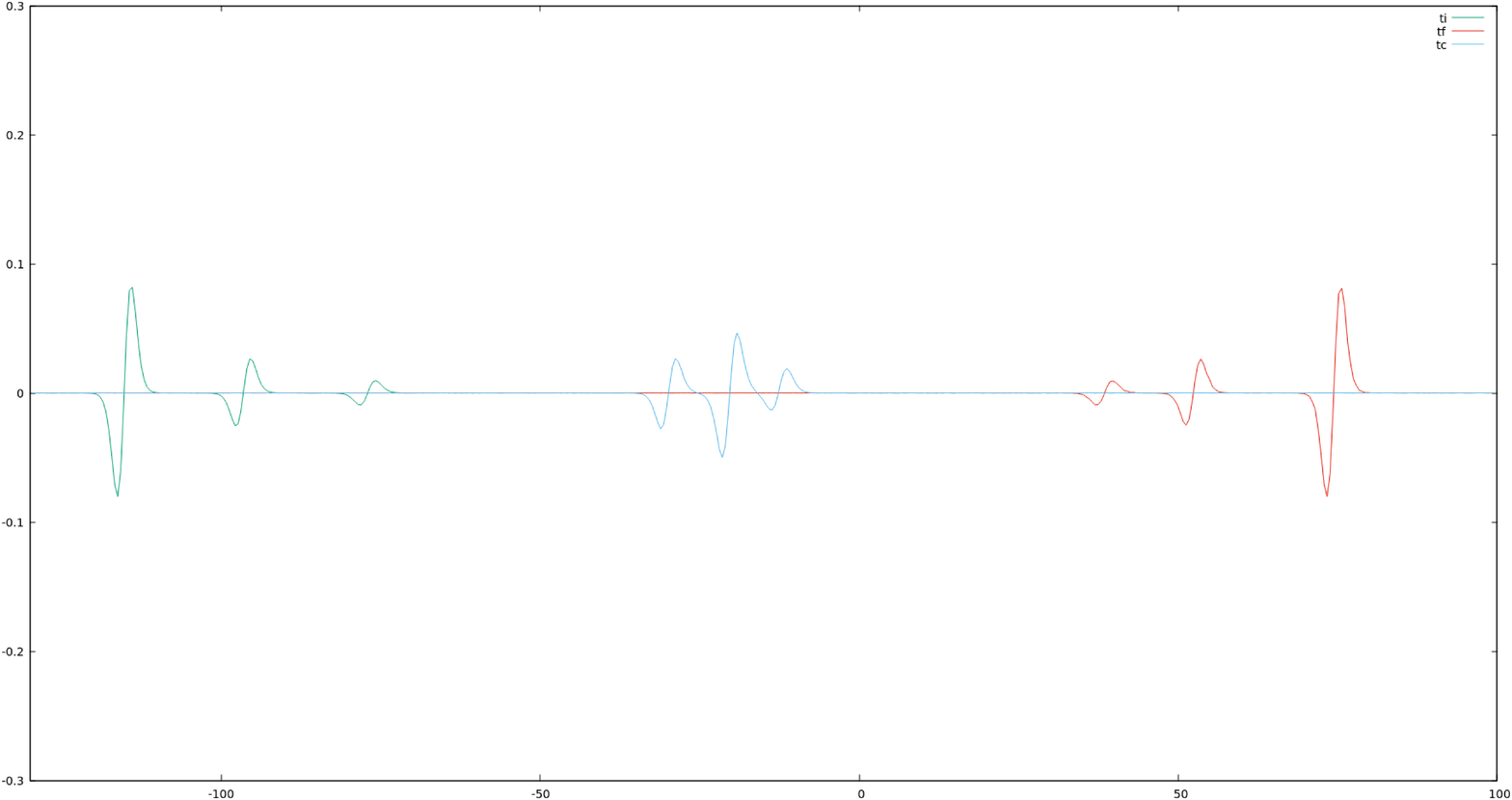}\\
\includegraphics[scale=0.1]{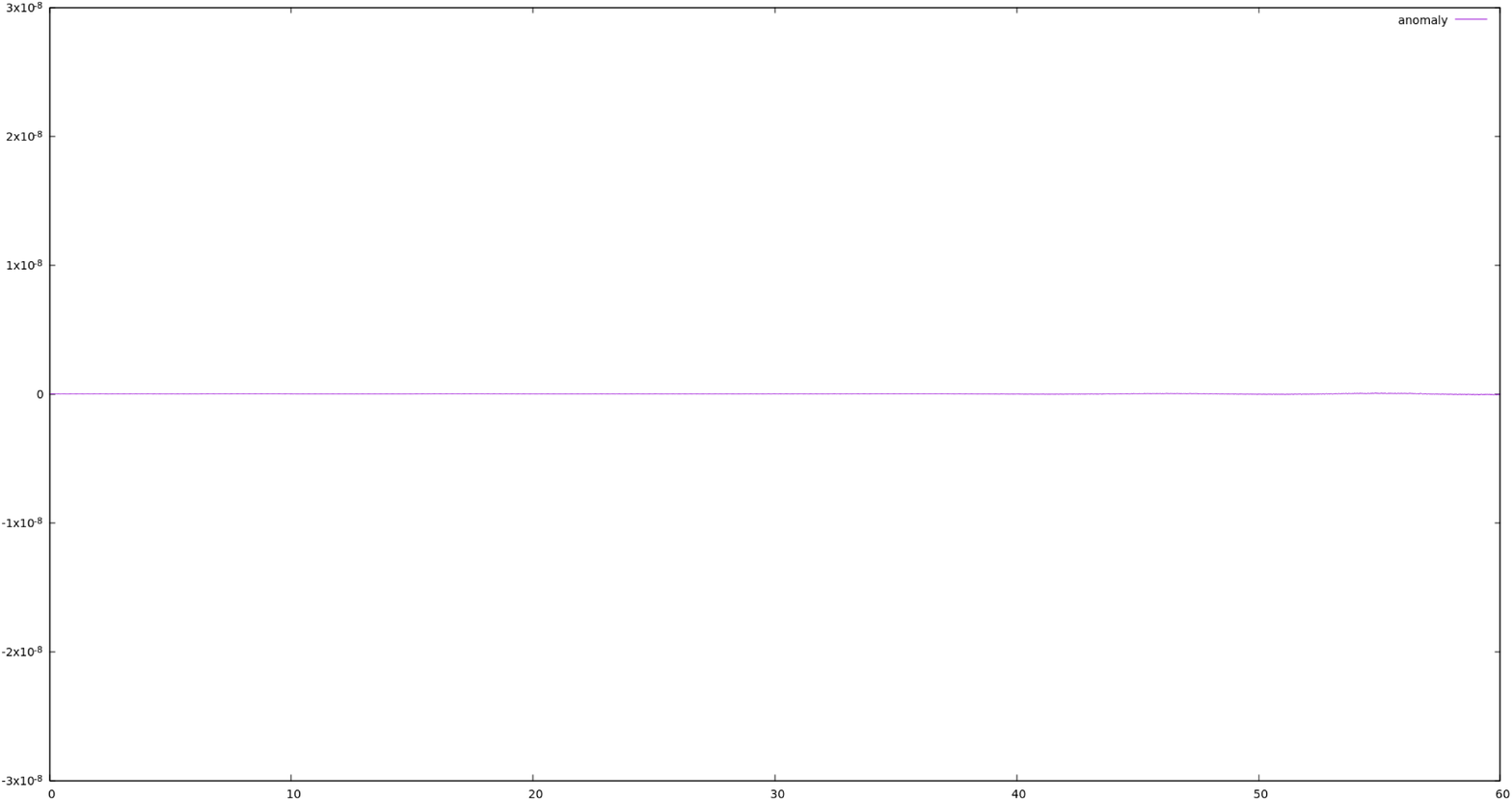}
\includegraphics[scale=0.1]{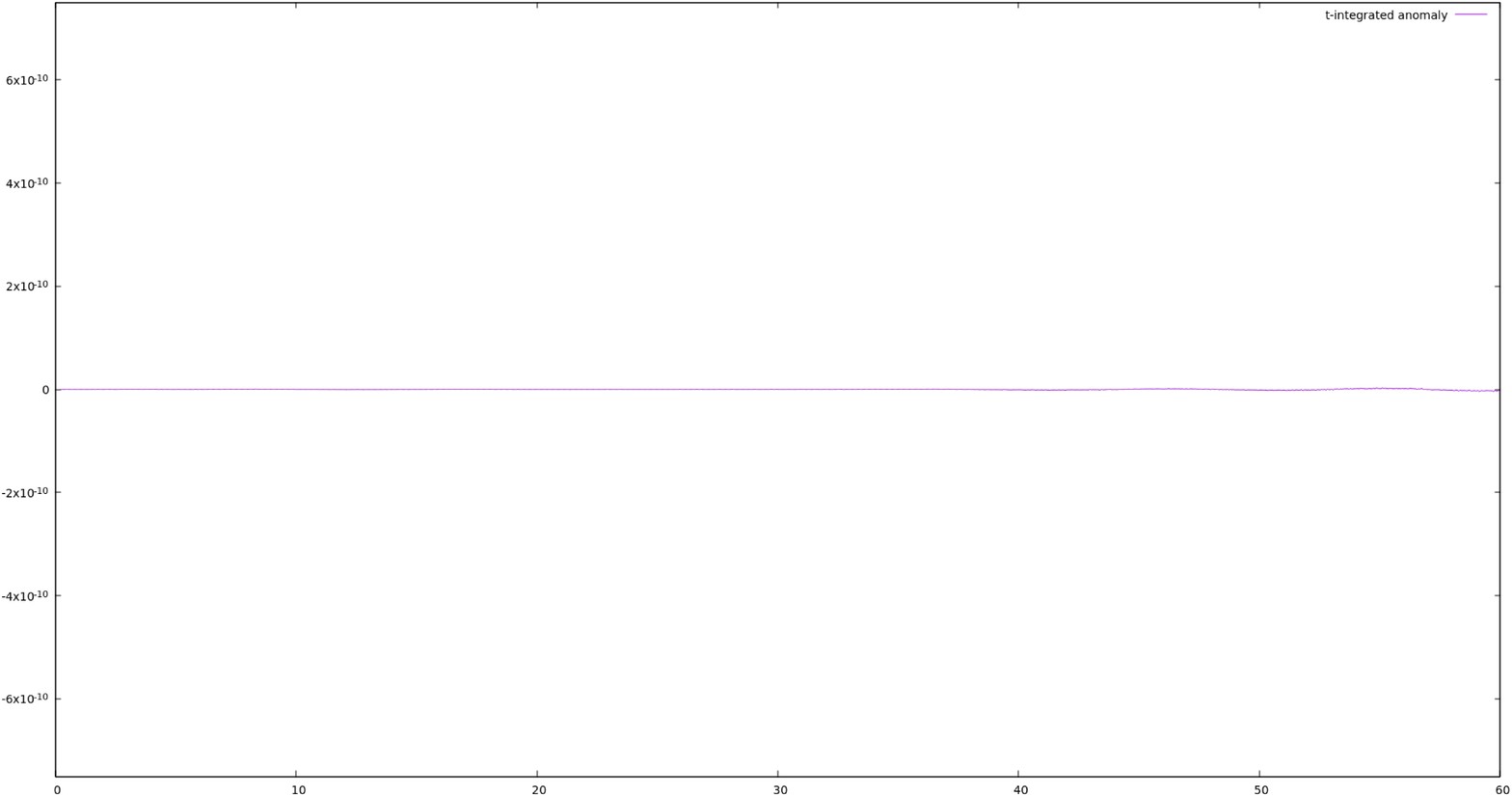}
\caption{Top Fig. shows the anomaly density of ${\cal \alpha}_{3} $ in (\ref{qq31}); i.e. the function $(w_xv_tu_t)\,$ plotted in $x-$coordinate for three successive times, $t_i =$ before collision (green), $t_c=$ collision (blue) and $t_f=$ after collision (red), for the 3-soliton of Fig. 3. Bottom Figs. show the plots of the anomaly ${\cal \alpha}_{3}  \, \mbox{vs} \,\, t$ and  the $t-$integrated anomaly $\int_{t_i}^{t} {\cal \alpha}_{3} \,  \mbox{vs}\, \, t$, respectively.}
\label{fig10}
\end{figure}

Similarly, Fig. 11 shows the simulation of the anomaly $ h_{1} $ in (\ref{hh1}) whose corresponding density is $- 4 q_{xxx} \(2 q_{xt}^2 + \epsilon_2 q_{xx} q_{tt}\)$. It shows the behavior of $ h_{1} $ versus $x-$coordinate for three successive times, before collision, during collision and after the collision of the 3-soliton presented in the Fig. 3. The anomaly and its $t-$integrated anomaly functions of $t$, vanish within numerical accuracy; in fact, one has $ h_{1}  \approx 0$ within the order of $10^{-5}$, whereas the $t-$integrated anomaly vanishes within the order of $10^{-6}$. Therefore, the charge $\bar{Q}_2$ in (\ref{hh1}) is asymptotically conserved for the collision of  three solitons, i.e.  $\bar{Q}_2(+\widetilde{t}) = \bar{Q}_2(-\widetilde{t})$ for large $\widetilde{t}$.
 
\begin{figure}
\centering
\includegraphics[scale=0.15]{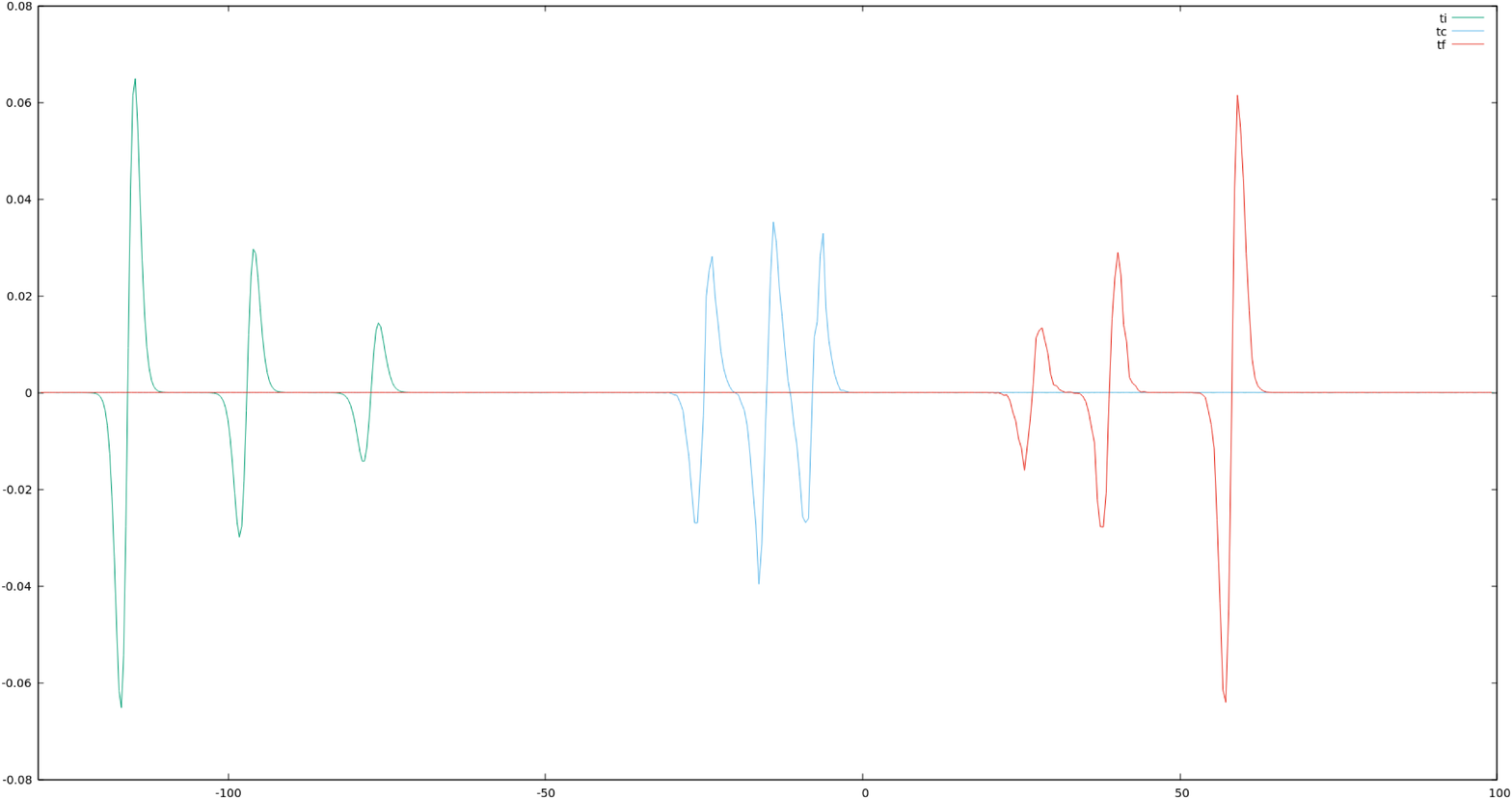}\\
\includegraphics[scale=0.1]{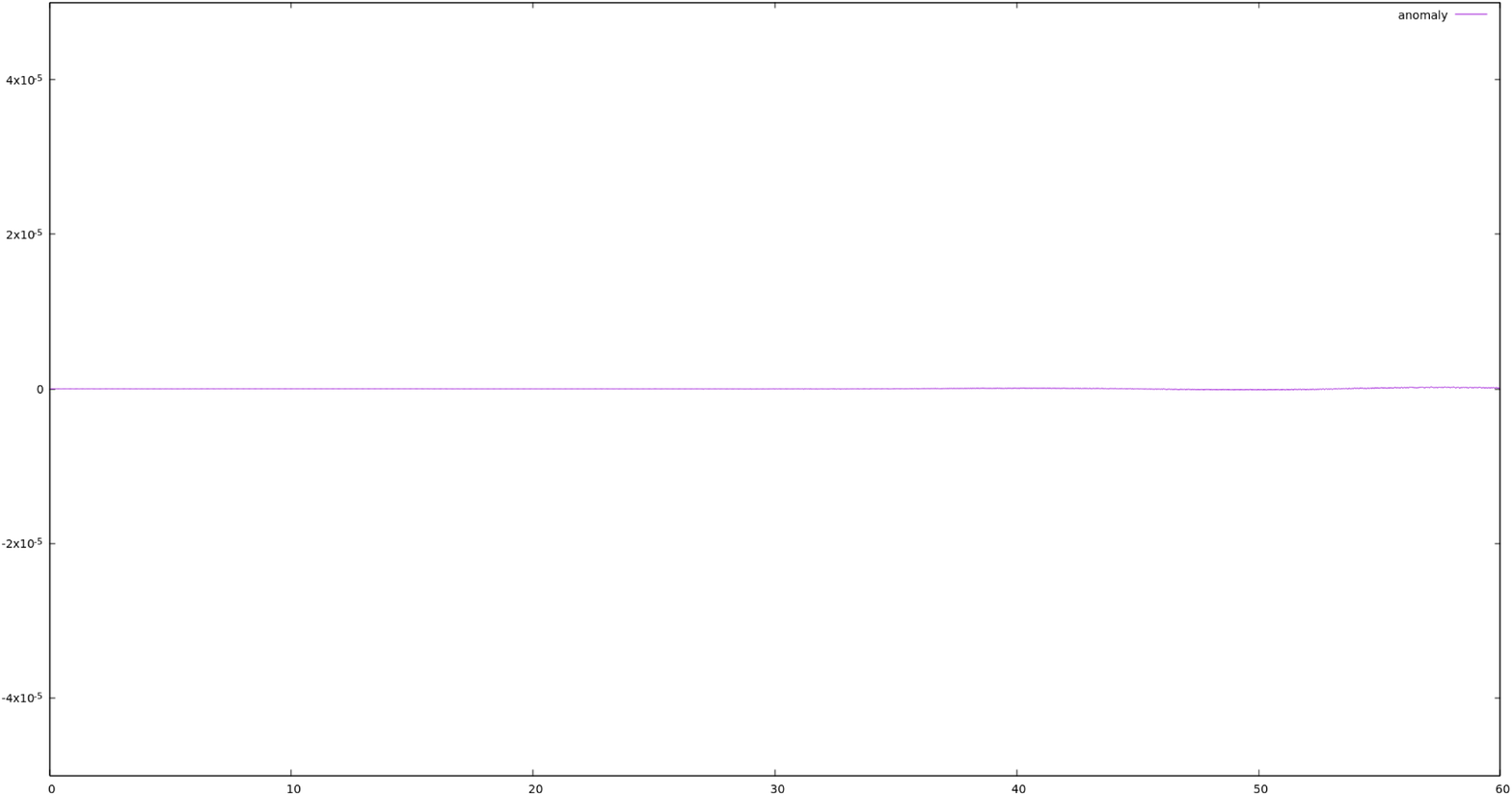}
\includegraphics[scale=0.1]{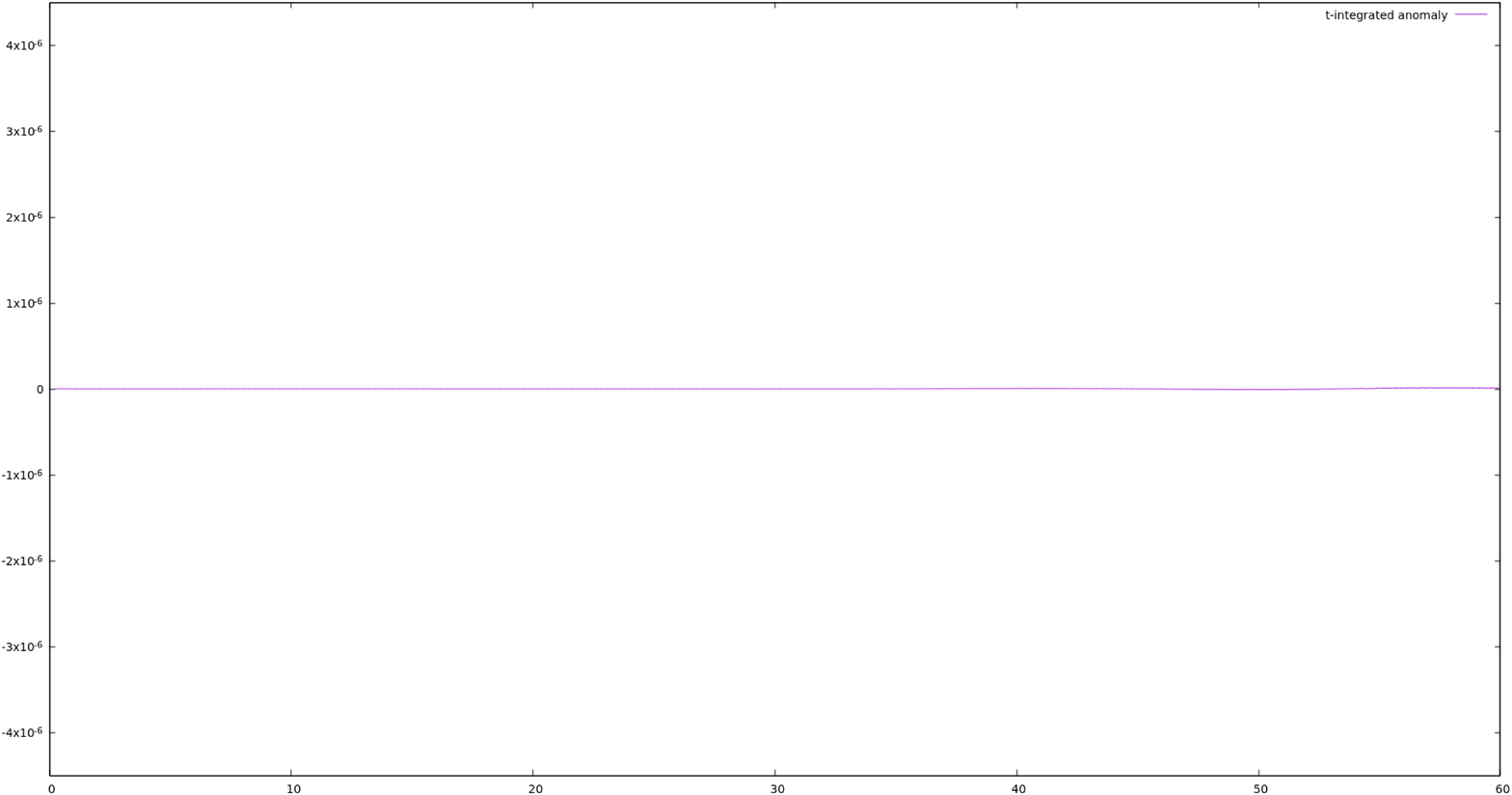}
\caption{Top Fig. shows the anomaly density of $ h_{1} $ in (\ref{hh1}); i.e. the function $- 4 q_{xxx} \(2 q_{xt}^2 + \epsilon_2 q_{xx} q_{tt}\)\,$ plotted in $x-$coordinate for three successive times, $t_i =$ before collision (green), $t_c=$ collision (blue) and $t_f=$ after collision (red), for the 3-soliton of Fig. 3. Bottom Figs. show the plots of the anomaly $ h_{1}  \, \mbox{vs} \,\, t$ and  the $t-$integrated anomaly $\int_{t_i}^{t}  h_{1} \,  \mbox{vs}\, \, t$, respectively.}
\label{fig11}
\end{figure} 

The vanishing of the anomalies and time-integrated anomalies were true for
the lowest order quasi-conservation laws but it was also true for the next order anomalies. However, the expressions for higher order anomalies, such as the anomaly densities in (\ref{qnonl1}) or the expression of $ {\cal H}_n$ in (\ref{ah}), involved more derivatives of the fields and so, our results were more liable to suffer from numerical errors. Thus, although their behavior are consistent with our claims, we have not included them in this paper.

\section{Riccati-type pseudo-potentials and general deformations of KdV}
\label{sec:riccati}

In the search for additional conservation laws and relevant properties next we consider the deformations of the KdV model in the context of the Riccati-type pseudo-potential approach. So, let us consider the Riccati-type system of equations
\br
\label{r1}r_x &=& U + 2 r^2 - 2 \l r,\\
\label{r2}
r_t&=& -4 \l^2 U - 4 U^2- 8( \l^2 +  U) r^2  +2 r (4 \l^3 + 4 \l U - 2 
U_x )+2 \l 
U_x - U_{xx}+Y+\chi,
\er
where the field $U(x,t)$ is a KdV type field, the field $Y(x,t)$ encodes the deformation away from the KdV model, $r(x,t)$ is a Riccati-type pseudo-potential and  $\chi$ is an auxiliary field satisfying
\br
\label{aux1}
\pa_x \chi +  2 (\l - 2 r) \chi = -2 (\l - 2 r) Y.
\er
In the above eqs. $\l$ plays the role of a spectral parameter. Notice that for $Y=0$ and $\chi =0$ one has the Riccati system (\ref{r1})-(\ref{r2}) approach for the usual KdV equation.

Next, through the compatibility condition for the system (\ref{r1})-(\ref{r2}), i.e. $(\pa_t \pa_x r - \pa_x \pa_t r)=0$, and provided that the auxiliary eq. (\ref{aux1}) is taken into account, one has the equation
\br
\label{dkdv2}
U_t + 12 U U_x + U_{xxx} = Y_x.
\er
Since the form of $Y$ can be assumed to be an arbitrary functional of $U$ and its derivatives (containing local and nonlocal terms, as well as some deformation parameters), one can regard the eq. (\ref{dkdv2}) as a general deformation of the KdV model in the pseudo-potential approach. Therefore, the whole problem is transferred to that of the existence of the auxiliary field $\chi$. 

The eq. (\ref{aux1}) is a non-homogeneous ordinary differential equation for $\chi$ in the variable $x$, which can be  integrated  by quadratures. Its general solution becomes 
\br
\nonumber
\chi (x, t) &=& C\, e^{-2 \int^{x}\, [\lambda-2 r(x', t)] dx'} - 2 e^{-2 \int^{x}\, [\lambda-2 r(x', t) ]dx'} \times \\
&& \int^{x} e^{2 \int^{x''}\, [\lambda-2 r(x', t) dx']} Y(x'', t)  \Big[\lambda  - 2   r(x'', t)  \Big]    dx'' .\label{chi11}
\er
Imposing the condition $\chi=0 $ for $Y=0$ to this solution, as it must hold for the usual KdV model, one must set $C=0$. In fact, the contribution of the  homogeneous sector of the differential equation (\ref{aux1}) of the general solution in (\ref{chi11}) must be removed. So, one has 
\br
\chi (x, t) &=&  - 2 e^{-2 \int^{x}\, [\lambda-2 r(x', t) ]dx'} \int^{x} e^{2 \int^{x''}\, [\lambda-2 r(x', t) dx']} Y(x'', t)  \Big[\lambda  - 2   r(x'', t)  \Big]   dx'' .\label{chi111}
\er
The expression for $\chi$ in  (\ref{chi111}) is highly non-local and, once inserted into (\ref{r2}), the new system of  eqs. (\ref{r1}) and (\ref{r2}) will provide a new non-local Riccati-type representation for the dKdV model (\ref{dkdv2}).
 
It is a remarkable fact that in the Riccati-type approach as presented above it is possible to consider more general deformations, since the field $Y$ may depend on arbitrary functions of $U$ and its derivatives, as well as on some auxiliary fields. In the class of non-local deformations, it would be interesting to consider the Alice-Bob (AB) physics recently proposed in \cite{alice} in the framework of quasi-integrability. 
 
The eq. (\ref{dkdv2}) is equivalent to the particular deformation of the KdV equation introduced in (\ref{mrkdv}) as presented in the form (\ref{dx}). In fact, by making the identifications
\br
\label{tr0}
U =\frac{\a}{12} u + \frac{\a}{144},\\
Y = -\frac{1}{2} X,\label{tr01}
\er
one gets the eq. (\ref{dx}). However, in the pseudo-potential approach the variable $Y$ encodes a general deformation, including non-local terms. The deformation can be introduced through a set of parameters $\{\epsilon_i\}$ such that for $\epsilon_i =0$, one recovers the usual KdV model.  As an example, for the particular deformations considered in (\ref{dx}) the field $Y$ can be written as
\br
\label{part2}
Y = -\frac{\a}{12} [\frac{\a}{4} \epsilon_2 w_x v_t -\frac{12 \epsilon_1}{\a} (U_{xt}+U_{xx})].
\er 
For the type of deformations $Y$ satisfying the symmetry ${\cal P}(Y) = Y$, in analogy to $X$ in (\ref{paritys2}), one concludes that the deformed KdV model (\ref{dkdv2}) will be quasi-integrable.   

Next, we examine the anomalous conservation laws in the pseudo-potential approach. Let us consider the relevant (quasi-)conservation law in terms of the pseudo-potential field $r$ and the auxiliary field $\chi$. So, from (\ref{r1})-(\ref{r2}) one can write the next equation\footnote{Different expressions of this type can be written, we adopt the construction below such that the non-homogeneous r.h.s. terms contain the deformation variable $\chi$, such that when $\chi=0 (Y=0)$ one must reconstruct in the l.h.s., order by order in $\lambda^{-1}$, the usual conservation laws of the standard KdV model.}
\br
\label{rr1}
\frac{\pa}{\pa t} r + \frac{\pa}{\pa x}( 4 \l^2 r + 4 U r - 2 \l U + U_x) = Y + \chi.
\er 
From this point forward, we construct the relevant (quasi-)conservation laws in terms of the fields of de deformed KdV model. So, consider the expansions in powers of the $\l$ parameter
\br
\label{pow1}
r &=& \sum_{n=0} c_n \l^{-n-1},\\
\label{pow2}
\chi &=& \sum_{n=1} d_n \l^{-n+1}.
\er 
The first $c_n$'s and $d_n$'s are provided in appendix \ref{cds}.

Notice that setting  $Y=0$ and $\chi =0$ on the r.h.s. of the eq. (\ref{rr1}) it becomes a truly exact conservation law, and then one can construct the infinite tower of exact conservation laws for the usual KdV equation. Next, making use of the power expansions of $r$ (\ref{pow1}) and $\chi$ (\ref{pow2}) on the $\lambda^{-1}$ parameter one gets, substituting them  into the eq. (\ref{rr1}), a polynomial in powers of $\lambda^{-n},\,(n=-1,0,1,2,3...)$. Then, taking into account the $c_n$'s and $d_n$'s expressions in the appendix \ref{cds}, one finds that the first two of this series ($n=-1,0$) provide trivial equations. Likewise, for $n\geq 1$, one can write
\br
\label{setr}
\pa_t \( c_{n-1} \) + \pa_x \(4c_{n+1} + 4 U c_{n-1}\) = d_{n+1},\,\,\,\,\,n=1,2,3....
\er
This is an infinite set of quasi-conservation laws for the deformed KdV model (\ref{dkdv2}) in the Riccati-type pseudo-potential approach.

Moreover, for the particular deformation (\ref{tr0})-(\ref{part2}) from the eq. (\ref{rr1})  one can get the tower of quasi-conservation laws presented in (\ref{qa1})-(\ref{can}). For $n=1,3,5,7$, using the identifications (\ref{tr0}) one can get, up to overall constant factors, the anomalous conservation laws (\ref{n0}), (\ref{n1}),(\ref{n2}) and (\ref{n3}), respectively. We have carefully examined the conservation laws associated to the even order powers $\lambda^{-n},\,n= 2, 4, 6$ and observed that those eqs. are simply the  $x-$derivatives of the relevant conservation laws associated to the orders $n=1,3,5$, respectively; so, they do not exhibit new conservation laws.   

In the next steps we will pursue a linear system of equations associated to the deformed KdV. So, consider the transformation
\br
\label{tr12}
r = - \frac{1}{2} \pa_x \log{\phi}, 
\er
where $\phi$ represents a new pseudo-potential. Next, consider the quasi-conservation law (\ref{rr1}) and integrate that eq. once in $x$. Then one gets 
\br
\label{ss}
s(x,t) = - \frac{1}{2 \phi}\Big[ \phi_t + 4 (\l^2 + U) \phi_x + 2 (2 \l U - U_x)\phi \Big],
\er
where 
\br
\label{ss1}
s(x,t) \equiv \int^x dx'  (Y+\chi).
\er
With the substitution (\ref{tr12}) the Riccati eq. (\ref{r1}) becomes
\br
\label{r1phi}
\phi_{xx} = -2 [\lambda \phi_x + U \phi ].
\er
The auxiliary eq. (\ref{aux1}) upon substitution of (\ref{tr12}) becomes
\br
\label{sxx}
s_{xx} = Y_x - 2 [\lambda + \pa_x \log{\phi}] s_x .  
\er
Substituting (\ref{ss}) into the last eq. and provided that $\phi$ satisfies (\ref{r1phi}) one gets the eq. of motion of the deformed KdV (\ref{dkdv2}). 

Some comments are in order here. First, in the absence of deformations the auxiliary eq. (\ref{sxx}) becomes a trivial one, i.e. $\chi=Y=0$ implies $s=0$. Second, for undeformed KdV and by making the substitution (\ref{tr12}) into the Riccati eq. (\ref{r1}) and the eq. (\ref{rr1}), one can get a linear system of eqs. (\ref{ss}) (set $s\equiv 0$ in the l.h.s.) and (\ref{r1phi}) for the usual KdV model. 

So, following analogous constructions presented in \cite{arxiv2} related to the deformations of the sine-Gordon model, we look for a linear system of eqs. associated to the deformed KdV.  Notice that the function $s$ in (\ref{ss})-(\ref{ss1}) will inherit from $\chi$ in (\ref{aux1}) a highly nonlinear dependence on $r$; then, through the transformation (\ref{tr12}), $s$ will have in general a nonlinear dependence on $\phi$. However, one can argue that the eq. (\ref{ss}) would represent a linear eq. for the pseudo-potential $\phi$ provided that the auxiliary field $s$ is written solely in terms of the fields $U$ and $Y$ and their derivatives. So, let us assume the next Ansatz
\br
\label{lin01}
\pa_x \phi &=& {\cal A}_x \phi,\\
\label{lin02}
\pa_t \phi &=& {\cal A}_t \phi.
\er
The compatibility condition of the above system provides the eq. of motion
\br
\label{eqm}
\pa_{t} {\cal A}_x - \pa_x {\cal A}_t =0.
\er  
Using (\ref{lin01}) into (\ref{r1phi}) one gets the following Riccati eq. for ${\cal A}_x$
\br
\label{riccatiAx}
\pa_x {\cal A}_x = - \Big[2 U + 2 \lambda {\cal A}_x + ({\cal A}_x)^2\Big].
\er
Likewise, replacing (\ref{lin01})-(\ref{lin02}) into  (\ref{ss}) one gets a relationship for the quantity $s$
\br
\label{ss11}
s = \pa_x U- 2 \lambda U - 2 {\cal A}_x (\lambda^2 + U)- \frac{1}{2} {\cal A}_t.
\er
Substituting this form of $s$ into the eq. (\ref{sxx}) and using the eqs. (\ref{riccatiAx}) and (\ref{eqm}) one gets the eq. of motion of the deformed KdV (\ref{dkdv2}). So, the form of $s$ in (\ref{ss11}) is consistent with the dynamics of the deformed model.

Notice that the system of eqs.  (\ref{lin01})-(\ref{lin02}) are defined up to a gauge transformation of the type
\br
\label{gauge1}
\phi &\rightarrow& e^{\Lambda} \phi\\
\label{gauge2}
{\cal A}_x &\rightarrow& {\cal A}_x + \pa_x \Lambda\\
\label{gauge3}
{\cal A}_t &\rightarrow& {\cal A}_t + \pa_t \Lambda,
\er
for an arbitrary function $\Lambda$.

In order to find a linear system it is needed a guesswork out of the eq. (\ref{ss}), and due to the gauge symmetry (\ref{gauge1})-(\ref{gauge3}) a particular choice for the connections ${\cal A}_x$ and ${\cal A}_t$. Let us propose  the following linear system of equations as the linear formulation of the deformed KdV\footnote{Below we will provide a gauge transformation between the system (\ref{sys1})-(\ref{sys2}) and the above system  (\ref{lin01})-(\ref{lin02}).}
\br
\label{sys1}
\pa_t \Phi &=& A_t \Phi\\
\label{sys2}
\pa_x \Phi &=& A_x \Phi\\
A_x &\equiv & \frac{1}{2\lambda^2-U}\Big[U_x -2\lambda U\Big]\\
A_t &\equiv & \frac{1}{2\lambda^2-U}\Big[12 \lambda U^2 -6 U U_x + (2\lambda^2-U) \zeta \Big].
\er  
with
\br
\pa_x \zeta &=& \frac{1}{(2\lambda^2-U)^2} \Big[U_x Y_x + 6 U^2 U_{xx}-U Y_{xx}-U_x U_{xxx}+U U_{xxxx} + 12 \lambda U^2 U_x -\\
&& 2 \lambda^2 (6U_x^2+6UU_{xx}-Y_{xx}+U_{xxxx})-4 \lambda^3 (Y_x-U_{xxx})\Big].
\er
The compatibility condition of the system of eqs. (\ref{sys1})-(\ref{sys2}); i.e. $\pa_t \pa_x (\Phi)-\pa_x \pa_t (\Phi)=0$, furnishes the next expression which is a polynomial in powers of $\lambda$ 
\br
\nonumber
-4 \lambda^3 [U_t + 12 U U_x + U_{xxx} - Y_x]+ 2 \lambda^2 \pa_x[U_t + 12 U U_x + U_{xxx} - Y_x]+\\
\lambda^0 [U_t U_x - U_x Y_x-12 U^2 U_{xx}+U_x U_{xxx}-U (U_{xt} -Y_{xx}+U_{xxxx})]  \equiv 0.\label{exp1}
\er 
Therefore, equating to zero the coefficient of $\lambda^3$ provides the deformed KdV equation of motion (\ref{dkdv2}). The remaining terms in the coefficients of $\lambda^2$ and the zeroth order $\lambda^0$, vanish identically provided that the eq. of motion (\ref{dkdv2}) is assumed. 

Moreover, from the identity $\pa_t [\pa_x \log{\Phi}] - \pa_x [\pa_t \log{\Phi}] \equiv 0$ and the linear system of eqs. (\ref{sys1})-(\ref{sys2}) one can get the conservation law
\br
\label{ab1}
\frac{\pa A_x}{\pa t}   - \frac{\pa A_t}{\pa x} = 0.
\er
Substituting the expressions for $A$ and $A_t$ the last equation turns out to be the same as the eq. (\ref{exp1}). So, the coefficients of the polynomial in powers of $\lambda$ of the conservation law (\ref{ab1}) can directly be verified to vanish by using the equation of motion (\ref{dkdv2}). Since the deformed KdV eq (\ref{dkdv2}) can be written as a conservation law, i.e. $\pa_t [U] + \pa_x [ 6 U^2 + U_{xx} - Y]=0$, the third order term in $\lambda$  of the conservation law (\ref{ab1}), which is the same as the relevant term in (\ref{exp1}), only reproduces the own deformed KdV equation and the ``mass" conservation law (\ref{n0}).  
 
For completeness we provide a gauge transformation between the system (\ref{lin01})-(\ref{lin02}) and the above system  (\ref{sys1})-(\ref{sys2}). So, the gauge transformation  (\ref{gauge1})-(\ref{gauge3}) can be written as
\br
\phi &=& e^{-\Lambda} \Phi\\
A_x &=& {\cal A}_x + \pa_x \Lambda\\
A_t &=& {\cal A}_t + \pa_t \Lambda,
\er
where $\Omega \equiv \pa_x\Lambda$ satisfies the Riccati eq. 
\br
\pa_x \Omega = \Omega^2 - 2 (\frac{2\lambda^3 -3 \lambda U + \pa_x U}{2 \lambda^2-U})\,  \Omega + \frac{(2\lambda^2 -  U) \pa^2_x U + 2[U^3-3\lambda U \pa_x U+(\pa_{x}U)^2]}{(2 \lambda^2-U)^2}.
\er 
Next, for certain deformed models satisfying the parity symmetry (\ref{parity1}) one can rewrite the system (\ref{sys1})-(\ref{sys2}) as
\br
\label{sys11}
\pa_t \widetilde{\Phi} &=& \widetilde{A}_t \widetilde{\Phi}\\
\label{sys21}
\pa_x \widetilde{\Phi} &=& \widetilde{A}_x \widetilde{\Phi}\\
\widetilde{A}_x &\equiv & \frac{1}{2\lambda^2-U}\Big[U_x + 2\lambda U\Big]\\
\widetilde{A}_t &\equiv & -\frac{1}{2\lambda^2-U}\Big[12 \lambda U^2 +6 U U_x + (2\lambda^2-U) \widetilde{\zeta} \Big].
\er  
with
\br
\label{zeta2}
\pa_x \widetilde{\zeta} &=&- \frac{1}{(2\lambda^2-U)^2} \Big[U_x Y_x + 6 U^2 U_{xx}-U Y_{xx}-U_x U_{xxx}+U U_{xxxx} - 12 \lambda U^2 U_x -\\
&& 2 \lambda^2 (6U_x^2+6UU_{xx}-Y_{xx}+U_{xxxx})+4 \lambda^3 (Y_x-U_{xxx})\Big].
\er
Thus, it is a second linear representation of the deformed KdV model, such that the soliton solutions satisfy the parity symmetry (\ref{parity1}).
  
\subsection{Infinite set of non-local conserved charges}
\label{sec:nonlocal}

For linear systems as above it is possible to construct a set of non-local conserved charges. So, let us construct a set of infinite number of non-local conservation laws using the iterative approach introduced by Br\'ezin et.al. \cite{brezin}. In fact, the system (\ref{sys1})-(\ref{sys2}) satisfies  the properties: i) $(A_x, A_t)$ is a ``pure gauge"; i.e. $A_{\mu} = \pa_{\mu} \Phi \Phi^{-1}, \mu = x, t$; ii) $J_{\mu} =(A_x, A_t)$ defines a conserved current according to (\ref{ab1}). So, one can construct an infinite set of non-local 
conserved currents through an inductive procedure following \cite{brezin}. Let us define the currents 
\br
J_{\mu}^{(n)} &=& \frac{\pa}{\pa x_\mu} \chi^{(n)},\,\,\,x_\mu \equiv x, t;\,\,\,\,n=0,1,2,...\\
d \chi^{(1)} &=& A_{\mu} dx_{\mu}\\
&\equiv& A_x dx + A_t dt ,\\
J_{\mu}^{(n+1)} &=& \frac{\pa}{\pa x_\mu} \chi^{(n)}-A_{\mu} \chi^{(n)};\,\,\,\,\,\chi^{(0)}=1,
\er
 
Then one can show by an inductive procedure that the  (non-local) currents $J_{\mu}^{(n)}$ are conserved
\br
\label{nlcl}
\pa_{t} J^{(n)}_{t} - \pa_{x} J^{(n)}_{x} =0,\,\,\,\,n=1,2,3,...
\er
The first non-trivial  current becomes $J_{\mu}^{(1)}=(A_x, A_t)$ whose conservation law $\pa_t A - \pa_x A_t=0$  reproduces  the eq. (\ref{ab1}), and then provides the ``mass" conservation law. The second  order current becomes $J_{\mu}^{(2)}=(A_x - A_x\chi^{(1)},A_t-A_t \chi^{(1)})$, and from the conservation law (\ref{nlcl}), using the first order conservation law (\ref{ab1}), one gets
\br
\label{j2}
\pa_t [A_x \chi^{(1)}] - \pa_x [A_t\chi^{(1)}]=0.
\er
The third order current becomes $J_{\mu}^{(3)}=(\frac{\pa}{\pa x}\chi^{(2)}-A_x\chi^{(2)}, \frac{\pa}{\pa t}\chi^{(2)}-A_t\chi^{(2)})$. The conservation law  (\ref{nlcl}), upon using the first and second order conservation laws, can be written as 
\br
\label{j3}
\pa_t [A_x \chi^{(2)}] - \pa_x [A_t \chi^{(2)}]=0.
\er  
where
\br
\pa_ x \chi^{(2)}= A_x - A_x\chi^{(1)},\,\,\,\,\pa_ t \chi^{(2)}= A_t-A_t\chi^{(1)}.
\er
In summary, one can write the infinite tower of non-local conservation laws as 
\br 
\pa_t [ A_x \chi^{(1)} ] - \pa_x [ A_t \chi^{(1)} ] &=& 0,\\
\pa_t [ A_x \chi^{(n)} ] - \pa_x [ A_t\chi^{(n)} ] &=&0,\,\,\,\,\,n=2,3,4,...\\
\pa_x \chi^{(n)} &=& A_x - A_x \chi^{(n-1)},\,\,\,\,\,\,\pa_t \chi^{(n)} = A_t - A_t\chi^{(n-1)}.
\er 

The construction of analogous  linear systems and their associated non-local charges have recently been performed for some deformations of the sine-Gordon model \cite{arxiv2}. So, it would be interesting to search for the classical Yangian as a Poisson-Hopf type algebra related to those set of non-local currents and charges \cite{mackay} for the deformations of the known integrable models.  The non-local conserved charges, as in the non-linear $\sigma-$model, are relevant at the quantum level and they imply absence of particle production and the first non-trivial charge fixes almost completely the on-shell dynamics of the model (see e.g. \cite{abdalla, luscher}).

Moreover, in view of the ubiquitous presence of the KdV-type models in the various areas of nonlinear science it would be interesting to investigate the relevance and physical consequences of the various towers of asymptotically conserved charges discovered above. For example, it is known the relationship between gravitation in three-dimensional space-times and two-dimensional integrable systems. In particular, the KdV-type and KdV-Gardner models have recently been uncovered as describing the dynamics of the boundary degrees of freedom of General Relativity on AdS$_3$ (see e.g. \cite{grav} and references therein). We will postpone those important issues and some relevant applications for a future work.
    
\section{Discussions and some conclusions}
\label{sec:discuss}

We have studied the quasi-integrability properties of certain deformations of the KdV model. The 
charges introduced in \cite{npb}, in the anomalous zero-curvature approach, were carefully examined. The non-homogeneous (anomalous) conservation laws (\ref{curr1}) were considered and verified that they give rise to vanishing trivial charges, $q^{(-2n-1)}=0,\, n=1,2,...$, provided that the anomalies are rewritten conveniently such that the quasi-conservation laws are expressed as exact conservation laws. Our computations considered the first four cases for $n=0,1,2,3$. The first  charge $Q^{(-1)}, $ for $n=0$, becomes a non-trivial exactly conserved charge, which maintains the same form as in the usual KdV model. 

By direct construction in sec. \ref{sec:newasy}, we have obtained additional towers of quasi-conservation laws with true anomalies such that each of their densities exhibit the special space-time  symmetry (\ref{parity1})-(\ref{paritys1}) for definite parity $N-$soliton configurations. We have considered the exact conservation laws of the deformed model, and constructed a tower of quasi-conservation laws as extensions of them with higher order derivatives. So, for each exact conservation law of the deformed model it has been constructed a tower of related family of higher order infinite number of quasi-conservation laws.

In sec. \ref{sec:mrlw1} it has been performed an analytical and not only numerical, proof, of the quasi-integrability of a well known non-integrable theory. We have showed that the 2-soliton solution  of the mRLW theory, written as a ${\cal P}$ invariant solution,  was analytically quasi-integrable. The proof of this result holds for all the anomalous charges from 
the various towers of infinite number of quasi-conservation laws presented in sec. \ref{sec:newasy} .

Moreover, we showed, in sec. \ref{sec:standard}, that even the standard KdV model possesses some towers of infinite number of anomalous conservation laws. Subsequently, we showed analytically the vanishing of each anomaly and then the quasi-conservation of the infinite tower of anomalous charges for $N-$soliton solution satisfying the special parity symmetry (\ref{parity1})-(\ref{paritys1}). So, our results show the first example of an analytical, and not only numerical, demonstration of the vanishing of the anomalies associated to the quasi-conservation laws in an integrable system. These kind of anomalous charges also appear in the standard sine-Gordon model \cite{arxiv2}, and they are expected to appear in the other integrable systems and their quasi-integrable deformations.   
   
We have checked through numerical simulations of soliton collisions, in sec. \ref{sec:num}, the  conservation properties of the lowest order  charges appearing in the towers of quasi-conservation laws defined in (\ref{bbmn1}), (\ref{bbmn}), (\ref{bbmn2}) and (\ref{hnl}), for the 2-soliton and 3-soliton collisions. We have used, as a particular example, the model of Ferreira et. al. \cite{npb}, which depends on two deformation parameters $\{\epsilon_1, \epsilon_2\}$ (such that  for $\epsilon_1-\epsilon_2=0$ it reduces to the usual KdV model) and possesses a general soliton solution (for any real set $\{\epsilon_1, \epsilon_2\}$). We have studied these models numerically and computed the anomalies $\widetilde{\beta}^{(3)}$ in eq. (\ref{mom11}),
$\widetilde{\alpha}_2$ in (\ref{qq21}), 
${\cal \alpha}_{3} $ in (\ref{qq31}), and 
$ h_{1} $ in (\ref{hh1}), for various N-soliton ($N=2,3$) configurations. In our numerical simulations presented in the Figs 4-11 we have observed that the non-trivial lowest order anomalies, and their $t-$integrated anomalies, of the various towers of quasi-conservation laws, vanish for the 2-soliton and 3-soliton collisions. So, our numerical simulations allow us to argue that for  2-soliton and 3-soliton configurations the relevant charges are exactly conserved, within numerical accuracy. 
 
We have applied the pseudo-potential approach to deformations of KdV model in sec. \ref{sec:riccati}. We showed that when the Riccati-type pseudo-potential equations are deformed, away from the KdV model, one can construct infinite towers of quasi-conservation laws associated to general deformations of KdV as in (\ref{dkdv2}). It has been constructed an infinite set of quasi-conservation laws for the deformed KdV model (\ref{dkdv2}) in this approach (\ref{setr}). This construction reproduced the KdV-type quasi-conservation laws presented in the anomalous zero-curvature approach of \cite{npb}.

In the framework of the Riccati-type pseudo-potential approach we have constructed a couple of  linear systems of equations, (\ref{sys1})-(\ref{sys2}) and  (\ref{sys11})-(\ref{sys21}), whose relevant compatibility conditions furnish the deformed KdV model (\ref{dkdv2}). The second system of linear eqs. (\ref{sys11})-(\ref{sys21}) holds for certain deformed models satisfying the parity symmetry (\ref{parity1}). In subsection \ref{sec:nonlocal} we have constructed an infinite set of non-local charges associated to the linear formulation of the deformed model. The study of the properties of these linear systems deserves more careful consideration; in particular, the relation of their associated non-local currents with  the so-called classical Yangians \cite{mackay}. 

In view of the recent results, on deformations of sine-Gordon \cite{arxiv2}, and the present paper on deformations of KdV, one can inquire about the non-local properties of the quasi-integrable systems studied in the literature, such as the deformations of the non-linear Schr\"odinger, Bullough-Dodd, Toda and SUSY sine-Gordon systems \cite{jhep3, jhep4, jhep5, jhep6, toda, susy}, and more specific structures, such as an infinite number of (non-local) exact conservation laws and new towers of quasi-conservation laws. So, they deserve careful considerations in the lines discussed above.  

Finally, it would be an interesting issue to analyze, in the context of the quasi-integrable KdV models, the behavior of the so-called statistical moments defined by the integrals of the type (\ref{smom}), which would be relevant to the study of certain structures in (quasi-)integrable systems, such as soliton turbulence, soliton gas dynamics and rogue waves \cite{pla1, prlgas}.

\section{Acknowledgments}

HB thanks the Science Faculties at UNI (Lima-Per\'u) and UNASAM (Huaraz-Per\'u), respectively, for kind hospitality during his sabbatical year.

\appendix

\section{The first few current components}
\label{cc1}
For completeness we record the first few current and anomaly components as provided in \cite{npb}
\br
a^{(-1)}_x &=& \frac{\a}{2^2 3} u,\nonumber \\
a^{(-3)}_x &=&\frac{\a^2}{2^5 3^2}  u^2,\nonumber\\
a^{(-5)}_x &=&\frac{\a^3}{2^7 3^3}  u^3 + \frac{\a^2}{2^7 3^2}  u u_{xx}, \label{axn}\\
a^{(-7)}_x &=&\frac{ 5 \a^4}{2^{11} 3^4}  u^4 + \frac{\a^3}{2^7 3^3}  u^2  u_{xx} + \frac{1}{2^9 3^2} \( \a^3 u  u_{x}^2 + \a^2 u u_{xxxx}\).\nonumber
\er
Let us define the potential $\G^{(-2n-1)}$ such that  
\br
\label{gdef}
\g^{(-2n-1)} = - \pa_x \G^{(-2n-1)},\,\,\,n=0,1,2,3,...;\er
where
\br
\G^{(-1)} &=&0,\nonumber\\
\G^{(-3)} &=& \frac{\a}{2^3 3 }  u, \nonumber\\
\G^{(-5)} &=& \frac{\a^2}{2^6 3 }  u^2 + \frac{\a}{2^5 3 }  u_{xx} ,\label{gn}\\
\G^{(-7)} &=& \frac{5 \a^3}{2^8 3^3 }  u^3 + \frac{5 \a^2}{2^8 3^2}  u_{x}^2 + \frac{5 \a^2}{2^7 3^2 }  u u_{xx}+ \frac{ \a}{2^7 3}  u_{xxxx}  .\nonumber
\er

\section{Numerical methods}
\label{app:num}

The suitable form of the model (\ref{mrkdv}), in order to undertake a numerical simulation of its soliton solutions, is provided by the eq (\ref{eqq}). Next, taking into account the eq. (\ref{pdef}) one can rewrite the eq. (\ref{eqq}) as
\br
\label{eqqp}
p_{t} + p_{x} - 4 p^2_{x} - 2 \epsilon_2  q_{xx} p_{t} + p_{xxx} - \epsilon_1 (p_{xxt} + p_{xxx} ) = 0.
\er

We will apply a finite difference method in order to perform numerical simulations of the equation (\ref{eqqp}). Notice that the two and three-soliton solutions for RLW ($\epsilon_1 = 1, \epsilon_2 = 0$ case of eq. ( \ref{eqqp}) ) are only known numerically and were obtained in  \cite{eil1,eil2}. We follow the methods discussed by J.C. Eilbeck and G.R. McGuire \cite{eil1, eil2} and Ferreira et. al. \cite{npb}. The equation requires the introduction of implicit methods and we will use the LU method, in order to solve for the vector $P$ of a linear system $A P = D$ with tri-diagonal  matrix $A$. 

So, let us discretize the coordinates  $x$ and $t$ by the set of points $x_0, x_1, ..., x_N$ and $t_0, t_1, ..., t_K$. Next, we will use  the notation $p_{j}^{m} \equiv p(j h, m \tau)$ and $q_{j}^{m} \equiv q(j h, m \tau)$, where $h$ and $\tau$ denote the step size in space 
and time, respectively. In order to denote the relevant approximations to  $q_{j}^{m} $ and $p_{j}^{m}$  we will use the same notations, respectively. 
 
Applying the central finite difference operators on $p_j^m$ and $q_j^m$ one can write the following system of equations
\br\label{disc1}
\widetilde{b}_2^{m}\, p_2^{m+1} + \delta \, p_{3}^{m+1} &=& \widetilde{d}_2^{m}- 2 \tau \,\delta \, \widetilde{d}_1^{m},\\
\label{disc2}
\delta\, p_{j-1}^{m+1}+\widetilde{b}_j^{m} \,p_j^{m+1} + \delta\, p_{j+1}^{m+1} &=& \widetilde{d}_j^{m},\,\,\,\,\,\,\,\,\,\,\,\,\,\,\,\,\,\,\,\,\,\,\,\,\,\,\,j= 3,....,N-3\\
\delta\, p_{N-3}^{m+1}+\widetilde{b}_{N-2}^{m} \,p_{N-2}^{m+1}  &=& \widetilde{d}_{N-2}^{m} - 2 \tau \, \delta \,d_{N-1}^{m},\label{disc3}
\er
where
\br
\label{delta}
\delta &\equiv & - \frac{\epsilon_1}{2 h^2 \tau},\\
\widetilde{b}_j^{m} &\equiv & \frac{1}{2 \tau}- \epsilon_2 \frac{q_{j+1}^{m}-2 q_{j}^{m}+q_{j-1}^{m} }{h^2 \tau} +\epsilon_1 \frac{1}{h^2 \tau},
\\
\nonumber
\widetilde{d}_j^{m} & \equiv & \frac{p_{j}^{m-1}}{2\tau}-\frac{p_{j+1}^{m}-p_{j-1}^{m}}{2 h} - \epsilon_1 \frac{p_{j+1}^{m-1}-2p_{j}^{m-1}+p_{j-1}^{m-1}}{2 h^2 \tau}\\
&& - \epsilon_2\( \frac{q_{j+1}^{m}-2q_{j}^{m}+q_{j-1}^{m}}{h^2 \tau}\) p_{j}^{m-1}
\label{dt}\\
&&+ \frac{(p_{j+1}^{m}-p_{j-1}^{m})^2}{h^2} \nonumber\\
&&-(1-\epsilon_1) \frac{p_{j+2}^{m}-2p_{j+1}^{m}+2p_{j-1}^{m}-p_{j-2}^{m}}{2 h^3},\,\,\,\,\,\,j=2,3,...,N-2,
\nonumber 
\er
and the next boundary conditions will be imposed
\br
\widetilde{d}_0^{m} & \equiv & \frac{1}{2\tau} \, p_0^{m+1},\,\,\,\, \widetilde{d}_N^{m}  \equiv  \frac{1}{2\tau}\, p_{N}^{m+1} \label{boun1}. \\
 \widetilde{d}_1^{m} &=& \frac{1}{2\tau}\, p_1^{m+1},\,\,\,\,\widetilde{d}_{N-1}^{m} = \frac{1}{2\tau}\, p_{N-1}^{m+1}.\label{boun2}  
\er
The boundary conditions (\ref{boun1})-(\ref{boun2}) are consistent with the behaviours of the fields $q(x,t)$ and $p(x,t)$ at the both ends of the interval  $x \in [-75,\, 75]$, as presented in Fig. 1 for $t_i$. For any fixed time $t$ and for regions far away from the solitons, represented by the field $u(x,t)$, one has approximately a linear behavior of $q(x, t)$ and an approximately  constant function behaviour of $p(x,t)$. Notice that the term containing $p_{xxx}$ implied the appearance of the two relationships per boundary in (\ref{boun1})-(\ref{boun2}) as the relevant boundary conditions. 

Next, our problem reduces to solving the linear system of equations (\ref{disc1})-(\ref{disc3}) for the unknown variables $p_{j}^{m+1}$ for each time step provided that the  $p_{j}^{m}$'s are known. In fact, one has the matrix equation 
\br
A P = D,
\er
where  A is a tridiagonal matrix with relevant components provided by the l.h.s. of the system of eqs. (\ref{disc1})-(\ref{disc3}). As mentioned above, we have used the LU method in order to solve for the vector $P$. Moreover, we realized that, alternatively to the LU method, the tridiagonal matrix algorithm (Thomas algorithm) would be useful; however, we have mainly used the LU method. These will provide us the values of $p_{j}^{m}$ at the next time level $t_{m+1}$. Therefore, equation (\ref{pdef}) allows us to determine all the values of $q_{j}^{m}$ at the next time level $t_{m+1}$ through the formula
\br
p_{j}^m = \frac{q^{m+1}_{j} - q_{j}^{m-1}}{2 \tau} \Rightarrow q^{m+1}_{j}  = 
2 \tau p_{j}^m + q_{j}^{m-1}. 
\er
So, the algorithm above allows us to determine the numerical time evolution of the system by repeating the procedure for many time steps.
  
\section{The first $c_n$'s and $d_n$'s}
\label{cds}
Substituting (\ref{pow1}) into (\ref{r1}) one can get the first six of the $c_n$'s as
\br
c_0 &=& \frac{1}{2} U,\\
c_1 &=& -\frac{1}{4} U_x,\\
c_2 &=& \frac{1}{4} (U^2 +\frac{1}{2} U_{xx}),\\
c_3 &=& -\frac{1}{2} (U U_x + \frac{1}{8} U_{xxx}),\\
c_4 &=&  \frac{1}{32} (8 U^3+10 U_x^2+12 U U_{xx}+U_{xxxx}),\\
c_5 &=& - \frac{1}{64} (64 U^2 U_x + 36 U_x U_{xx} + 16 U U_{xxx} + U_{xxxxx}),\\
c_6 &=& \frac{1}{128}[40 U^4+120 U^2 U_{xx}+38 U_{xx}^2+56 U_x U_{xxx}+20 U(10U_x^2+U_{xxxx})+U_{xxxxxx}].
\er 
Further, by substituting these results into (\ref{aux1}) and the expansion (\ref{pow2}) one gets the first six of the $d_n$'s as
\br
d_1&=& -Y,\\
d_2 &=& \frac{1}{2} Y_x,\\
d_3 &=& -\frac{1}{4} Y_{xx},\\
d_4 &=& \frac{1}{8} (4 U Y_x +Y_{xxx}),\\
d_5 &=& -\frac{1}{16} (8 U_x Y_x+ 8 U Y_{xx} + Y_{xxxx}),\\
d_6 &=& \frac{1}{32} (24 U^2 Y_x +12 U_{xx} Y_x + 20 U_x Y_{xx}+ 12 U Y_{xxx} + Y_{xxxxx}),\\
d_7 &=& -\frac{1}{64} [64 U^2 Y_{xx} +36 U_{xx} Y_{xx} +16 U_{xxx} Y_x+36 U_{x}Y_{xxx}+16 U(8U_x Y_x + Y_{xxxx}) +Y_{xxxxxx}].
\er


\begin{thebibliography}{**}
\bibitem{jhep1}
L.A. Ferreira and Wojtek J. Zakrzewski, \JHEP{05}{2011}{130}.
\bibitem{jhep2}
L.A. Ferreira and Wojtek J. Zakrzewski, \JHEP{01}{2014}{058}\\
 L. A. Ferreira and W. J. Zakrzewski, \Nonl{29}{2016}{1622}.
\bibitem{hietarinta}
J. Hietarinta, {\sl Hirota's bilinear method and partial integrability}, in {\sl Partially Integrable Equations in Physics}, R. Conte and N. Boccara eds., NATO ASI Series C310, Les Houches France March 21-30 1989.
\bibitem{malomed}
Y. S. Kivshar, Boris A. Malomed, Dynamics of solitons in nearly integrable systems, \RMP{61(4)}{1989}{763}.
\bibitem{arnaudon}
 A. Arnaudon, On a Lagrangian reduction and a deformation of completely integrable systems, \JNS{26}{2016}{1133}. 
\bibitem{jhep3}
L.A. Ferreira, G. Luchini  and Wojtek J. Zakrzewski, \JHEP{09}{2012}{103}. 
\bibitem{jhep6}
V.H. Aurichio and L.A. Ferreira, \JHEP{03}{2015}{152}.  
\bibitem{toda}
L.A. Ferreira, P. Klimas and Wojtek J. Zakrzewskic, \JHEP{05}{2016}{065}.
\bibitem{npb}
F. ter Braak, L. A. Ferreira and W. J. Zakrzewski, \NPB{939}{2019}{49}.
\bibitem{susy}
K. Abhinav and P. Guha, \EPL{116}{2016}{10004}. 
\bibitem{arxiv2}
H. Blas, H. F. Callisaya and J.P.R. Campos, \NPB{950}{2020}{114852}.
\bibitem{jhep4} 
H. Blas and M. Zambrano, \JHEP{03}{2016}{005}.\\
 H Blas and M Zambrano, 2018 J. Phys.: Conf. Ser. 1143 012004.
\bibitem{jhep5}
H. Blas,  A.C.R. do Bonfim and A.M. Vilela, JHEP05(2017)106.
\bibitem{cnsns}
H. Blas and H. F. Callisaya, \CNSNS{55}{2018}{105}.\\
see also the Research Highlight: {\sl``An exploration of kinks/anti-kinks and breathers in deformed sine-Gordon models"} in Advances in Engineering, https://advanceseng.com/kinks-anti-kinks-breathers-deformed-sine-gordon-models/.  
\bibitem{frantzeskakis}
D. J. Frantzeskakis, \JPA{43}{2010}{213001}.
\bibitem{tanaka1}
A. Gurevich and V. M. Vinokur, \PRL{90}{2003}{047004}.\\
Y. Tanaka, \PRL{88}{2002}{017002}.
\bibitem{grav}
E. Ojeda and A. Perez, \JHEP{08}{2019}{079}.
\bibitem{pla1}
E.N. Pelinovsky et al. \PLA{377}{2013}{272}.\\
E. N. Pelinovsky and E. G. Shurgalina, \RQE{57}{2015}{737}.
\bibitem{prlgas}
I. Redor, E. Barthelemy, H. Michallet, M. Onorato, and N. Mordant,\PRL{122}{2019}{214502}.
\bibitem{alice}
 S.Y. Lou and  F. Huang, \ScR{7}{2017}{869}.\\
Man Jia and Sen Yue Lou, \PLA{382}{2018}{1157}.
\bibitem{das}
A. Das, {\sl Integrable Models}, World Scientific, 1989. 
\bibitem{faddeev}
L. D. Faddeev and L. A. Takhtajan, {\sl Hamiltonian Methods in the Theory of Solitons}, Springer, Berlin, 2007, Translated from the 1986 Russian original by Alexey G. Reyman.
\bibitem{babelon}
O. Babelon and D. Bernard, \PLB{317}{1993}{363}.\\
O. Babelon, D. Bernard and F.A. Smirnov, \CMP{182}{1996}{319}.
\bibitem{tao}
G. Staffilani, G., ``KdV and almost conservation laws'', in {\sl Harmonic analysis at
Mount Holyoke} (South Hadley, MA, 2001), 367-381, Contemp. Math.,
320, Amer. Math. Soc., Providence, RI, 2003.\\
J. Colliander, M. Keel, G. Staﬃlani, H. Takaoka, T. Tao, ``Sharp
global well-posedness for KdV and modified KdV on $\IR$ and $T$ , J. Amer.
Math. Soc. 16 (2003), no. 3, 705-749.
\bibitem{eil1}
J.C. Eilbeck, G.R. McGuire, Numerical study of the regularized longe-wave equation I: numerical methods, \JCP{19}{1975}{43}.
\bibitem{eil2}
J.C. Eilbeck, G.R. McGuire, Numerical study of the regularized long-wave equation II: interaction of solitary waves, \JCP{23}{1977}{63}. 
\bibitem{mme}
S. Israwi and H, Kalisch, \PLA{383}{2019}{854}.
\bibitem{physd}
E.G. Didenkulova (Shurgalina), \PHSD{399}{2019}{35}.
\bibitem{pla2}
D. Dutykh and E. Pelinovsky, \PLA{378}{2014}{3102}. 
\bibitem{chai}
Z. Chai, N. He, Z. Guo and B. Shi, \PRE{97}{2018}{013304}.
\bibitem{hereman}
W. Hereman and U. Koktas, {\sl Integrability Tests for Nonlinear Evolution Equations}, in Computer Algebra Systems: A Practical Guide (Ed. M. Wester), 1999, Wiley and Sons,New York
\bibitem{pla3}
 D.   Dutykh,   T.   Katsaounis,   D.   Mitsotakis, \IJNMF{71}{2013}{717}.
\bibitem{bbm}
 T.B.   Benjamin,   J.L.   Bona,   J.J.   Mahony,   Model   equations   for   long   waves   in   nonlinear   dispersive   systems,   Philos.   Trans.   R.   Soc.   Lond.   Ser.   A,   Math.   Phys.   Sci.   272  
(1972) 47.
\bibitem{junior}
V.A.S. Junior, \CNSNS{69}{2019}{73}.
\bibitem{hirota}
 R. Hirota, \PRL{27}{1971}{1192}.
\bibitem{gibbon}
J.D. Gibbon, J.C. Eilbeck, R.K. Dodd, \JPA{9}{1976}{L127}.
\bibitem{olver}
P.J. Olver, Proc. Camb. Philos. Soc. 85 (1979) 143160. 
\bibitem{brezin}
E. Brezin, C. Itzykson, J. Zinn-Justin and J.-B. Zuber, \PLB{82}{1979}{442}. 
\bibitem{mackay}
N. J. Mackay, \IJMPA{30}{2005}{7189}.
\bibitem{abdalla}
 Abdalla, E., Abadalla, M.C.B., Rothe, K.: Non-perturbative methods in two-dimensional 
quantum field theory. Singapore: World Scientific, 2nd Ed. 2001.
\bibitem{luscher}
M. L\"uscher, \NPB{135}{1978}{1}.
\end{thebibliography}
\end{document}